\documentclass[longauth]{aa}
\usepackage{graphicx}
\usepackage{tabularx}
\usepackage{multirow}
\usepackage[colorlinks=true, linkcolor=blue, citecolor=blue, draft=False]{hyperref}
\usepackage{gensymb}
\usepackage{color}
\usepackage{txfonts}
\usepackage{amsmath}
\usepackage{ulem}
\usepackage[flushleft]{threeparttable}

\defcitealias{Franco2018}{F18}
\defcitealias{Franco2020}{F20a}
\begin{document} 
\title{GOODS-ALMA: The slow downfall of star formation in $z$\,=\,2-3 massive galaxies}
\author{M.~Franco\inst{\ref{inst1},\ref{inst2}}\thanks{Email: m.franco@herts.ac.uk}
\and D.~Elbaz\inst{\ref{inst1}} 
\and L.~Zhou\inst{\ref{inst1},\ref{inst3},\ref{inst4}}
\and B.~Magnelli\inst{\ref{inst5}}
\and C.~Schreiber\inst{\ref{inst6}}
\and L.~Ciesla\inst{\ref{inst1},\ref{inst7}}
\and M.~Dickinson\inst{\ref{inst8}}
\and N.~Nagar\inst{\ref{inst9},\ref{inst10}}
\and G.~Magdis\inst{\ref{inst11},\ref{inst12},\ref{inst13},\ref{inst14}}
\and D.~M.~Alexander\inst{\ref{inst15}}
\and M.~B\'ethermin\inst{\ref{inst7}}
\and R.~Demarco\inst{\ref{inst9},\ref{inst10}}
\and E.~Daddi\inst{\ref{inst1}}
\and T.~Wang\inst{\ref{inst1},\ref{inst16}}
\and J.~Mullaney\inst{\ref{inst17}}
\and M.~Sargent\inst{\ref{inst18}}
\and H.~Inami\inst{\ref{inst19},\ref{inst20}}
\and X.~Shu\inst{\ref{inst21}}
\and F.~Bournaud\inst{\ref{inst1}}
\and R.~Chary\inst{\ref{inst22}}
\and R.~T.~Coogan\inst{\ref{inst23}}
\and H.~Ferguson\inst{\ref{inst24}}
\and S.~L.~Finkelstein\inst{\ref{inst25}}
\and M.~Giavalisco\inst{\ref{inst26}}
\and C.~G\'omez-Guijarro\inst{\ref{inst1}}
\and D.~Iono\inst{\ref{inst27},\ref{inst28}}
\and S.~Juneau\inst{\ref{inst1},\ref{inst8}}
\and G.~Lagache\inst{\ref{inst7}}
\and L.~Lin\inst{\ref{inst29}}
\and K.~Motohara\inst{\ref{inst30}}
\and K.~Okumura\inst{\ref{inst1}}
\and M.~Pannella\inst{\ref{inst31},\ref{inst32}}
\and C.~Papovich\inst{\ref{inst33},\ref{inst34}}
\and A.~Pope\inst{\ref{inst26}}
\and W.~Rujopakarn\inst{\ref{inst35},\ref{inst36},\ref{inst37}}
\and J.~Silverman\inst{\ref{inst37}}
\and M.~Xiao\inst{\ref{inst1},\ref{inst3}}}

\institute{AIM, CEA, CNRS, Universit\'{e} Paris-Saclay, Universit\'{e} Paris Diderot, Sorbonne Paris Cit\'{e}, F-91191 Gif-sur-Yvette, France\label{inst1}
\and Centre for Astrophysics Research, University of Hertfordshire, Hatfield, AL10 9AB, UK\label{inst2}
\and School of Astronomy and Space Science, Nanjing University, Nanjing 210093, China\label{inst3}
\and Key Laboratory of Modern Astronomy and Astrophysics (Nanjing University), Ministry of Education, Nanjing 210093, China\label{inst4}
\and Argelander-Institut f\"{u}r Astronomie, Universit\"{a}t Bonn, Auf dem H\"{u}gel 71, D-53121 Bonn, Germany\label{inst5}
\and Department of Physics, University of Oxford, Keble Road, Oxford OX1 3RH, UK\label{inst6}
\and Aix Marseille Universit\'{e}, CNRS, LAM, Laboratoire d'Astrophysique de Marseille, Marseille, France\label{inst7}
\and Community Science and Data Center/NSF’s NOIRLab, 950 N. Cherry Ave., Tucson, AZ 85719, USA \label{inst8}
\and Department of Astronomy, Universidad de Concepci\'{o}n, Casilla 160-C Concepci\'{o}n, Chile \label{inst9}
\and Departamento de Astronom\'ia, Facultad de Ciencias F\'isicas y Matem\'aticas,
Universidad de Concepci\'on, Concepci\'on, Chile \label{inst10}
\and Cosmic Dawn Center at the Niels Bohr Institute, University of Copenhagen and DTU-Space, Technical University of Denmark\label{inst11}
\and DTU Space, National Space Institute, Technical University of Denmark, Elektrovej 327, DK-2800 Kgs. Lyngby, Denmark\label{inst12}
\and Niels Bohr Institute, University of Copenhagen, DK-2100 Copenhagen, Denmark\label{inst13}
\and Institute for Astronomy, Astrophysics, Space Applications and Remote Sensing, National Observatory of Athens, 15236, Athens, Greece\label{inst14}
\and Centre for Extragalactic Astronomy, Department of Physics, Durham University, Durham DH1 3LE, UK\label{inst15}
\and Institute of Astronomy, University of Tokyo, 2-21-1 Osawa, Mitaka, Tokyo 181-0015, Japan\label{inst16}
\and Department of Physics and Astronomy, The University of Sheffield, Hounsfield Road, Sheffield S3 7RH, UK\label{inst17}
\and Astronomy Centre, Department of Physics and Astronomy, University of Sussex, Brighton, BN1 9QH, UK\label{inst18}
\and Univ Lyon, Univ Lyon1, ENS de Lyon, CNRS, Centre de Recherche Astrophysique de Lyon (CRAL) UMR5574, F-69230, Saint-Genis-Laval, France\label{inst19}
\and Hiroshima Astrophysical Science Center, Hiroshima University, 1-3-1 Kagamiyama, Higashi-Hiroshima, Hiroshima 739-8526, Japan\label{inst20}
\and Department of Physics, Anhui Normal University, Wuhu, Anhui, 241000, China\label{inst21}
\and Infrared Processing and Analysis Center, MS314-6, California Institute of Technology, Pasadena, CA 91125, USA\label{inst22}
\and Max-Planck-Institut f\"{u}r Extraterrestrische Physik (MPE), Giessenbachstr.1, 85748 Garching, Germany \label{inst23}
\and Space Telescope Science Institute, 3700 San Martin Drive, Baltimore, MD 21218, USA\label{inst24}
\and Department of Astronomy, The University of Texas at Austin, Austin, TX 78712, USA\label{inst25}
\and Astronomy Department, University of Massachusetts, Amherst, MA 01003, USA\label{inst26}
\and National Astronomical Observatory of Japan, National Institutes of Natural Sciences, 2-21-1 Osawa, Mitaka, Tokyo 181-8588, Japan\label{inst27}
\and SOKENDAI (The Graduate University for Advanced Studies), 2-21-1 Osawa, Mitaka, Tokyo 181-8588, Japan\label{inst28}
\and Institute of Astronomy \& Astrophysics, Academia Sinica, Taipei 10617, Taiwan\label{inst29}
\and Institute of Astronomy, Graduate School of Science, The University of Tokyo, 2-21-1 Osawa, Mitaka, Tokyo 181-0015, Japan\label{inst30}
\and Astronomy Unit, Department of Physics, University of Trieste, via Tiepolo 11, I-34131 Trieste, Italy\label{inst31}
\and Fakult\"{a}t f\"{u}r Physik der Ludwig-Maximilians-Universit\"{a}t, D-81679 M\"{u}nchen, Germany\label{inst32}
\and Department of Physics and Astronomy, Texas A\&M University, College Station, TX, 77843-4242, USA\label{inst33}
\and George P. and Cynthia Woods Mitchell Institute for Fundamental Physics and Astronomy, Texas A\&M University, College Station, TX, 77843-4242, USA\label{inst34}
\newpage
\and Department of Physics, Faculty of Science, Chulalongkorn University, 254 Phayathai Road, Pathumwan, Bangkok 10330, Thailand\label{inst35}
\and National Astronomical Research Institute of Thailand (Public Organization), Donkaew, Maerim, Chiangmai 50180, Thailand\label{inst36}
\and Kavli Institute for the Physics and Mathematics of the Universe (WPI), The University of Tokyo Institutes for Advanced Study, The University of Tokyo, Kashiwa, Chiba 277-8583, Japan \label{inst37}
}
\date{Received: 30 April 2020; accepted 31: August 2020}

\abstract{
We investigate the properties of a sample of 35 galaxies, detected with the Atacama Large Millimeter/Submillimeter Array (ALMA) at 1.1\,mm in the GOODS-ALMA field (area of 69\,arcmin$^2$, resolution\,=\,0.60\arcsec, RMS\,$\simeq$\,0.18\,mJy\,beam$^{-1}$).
Using the ultraviolet-to-radio deep multiwavelength coverage of the GOODS--South field, we fit the spectral energy distributions of these galaxies to derive their key physical properties.
The galaxies detected by ALMA are among the most massive at $z$\,=\,2-4 (M$_{\star,med}$\,=\,8.5$\,\times$\,10$^{10}$\,M$_\odot$) and they are either starburst or located in the upper part of the galaxy star-forming main sequence. A significant portion of our galaxy population ($\sim$\,40\%), located at $z\sim$2.5-3, exhibits abnormally low gas fractions. The sizes of these galaxies, measured with ALMA, are compatible with the trend between the rest-frame 5000\,$\AA$ size and stellar mass observed for $z\sim2$ elliptical galaxies, suggesting that they are building compact bulges. We show that there is a strong link between star formation surface density (at 1.1\,mm) and gas depletion time: The more compact a galaxy's star-forming region is, the shorter its lifetime will be (without gas replenishment). 
The identified compact sources associated with relatively short depletion timescales ($\sim$100\,Myr) are the ideal candidates to be the progenitors of compact elliptical galaxies at $z$\,$\sim$\,2.
}


    \maketitle
%

\section{Introduction}

Over the last 8 billion years, the cosmic star formation density has decreased by a factor of $\sim$\,10 \citep[e.g.,][]{Madau2014}. One of the major key questions in astrophysics is to understand why the Universe's star-forming activity reaches a peak around $z$\,=\,$2$ and why it is now so ineffective at generating stars.

Due to the lack of infrared (IR) surveys able to detect ``typical'' star-forming galaxies at $z$\,$>$\,2, the actual contribution of dust-obscured galaxies to the cosmic star formation history at these redshifts remains largely unknown, especially at high masses where galaxies are known to be metal-rich \citep[e.g.,][]{Tremonti2004} and dust-rich \citep[e.g.,][]{Boissier2004, Reddy2010}. The star-formation rates (SFRs) of these high redshift galaxies are mostly estimated from ultraviolet (UV) measurements emitted by short-lived massive stars \citep[e.g.,][]{Kennicutt2012}. This UV emission is strongly affected by the presence of dust in the interstellar medium (ISM), which absorbs part of this emission that is to be re-emitted in IR. Therefore, to correctly estimate the SFR, a dust correction needs to be applied. This approach has proved effective for distant galaxies up to epochs close to reionization \citep[e.g.,][]{Oesch2015, Bouwens2015, McLeod2015}, but it suffers from caveats due to uncertainties on the attenuation law and the difficulties to constrain it \citep[e.g.,][]{Cowie1996, Pannella2009}. For this reason, constraining galaxy IR emission is essential to obtaining a robust star formation estimate. 

The rest-frame peak of a galaxy's spectral energy distribution (SED) with a dust temperature between 30 and 50 K can vary between 72 and 125 $\mu$m \citep[e.g.,][]{Casey2014}, corresponding to an observed peak between $\sim$280 and 500 $\mu$m at $z$\,=\,3. To constrain the shape of the IR SED, at least one measurement must be carried out beyond this peak in the FIR part of the spectrum. This is why (sub)millimeter observations are necessary to constrain the IR luminosity of a galaxy. 
Thanks to the negative K-correction, submillimeter observations of galaxies are not affected by the flux decrease with increasing redshift over 2\,$<$\,$z$\,$<$\,10 \citep{Blain2002}. With the Atacama Large Millimeter/Submillimeter Array (ALMA), it is now possible to detect galaxies with continuum emission below 1\,mJy and angular resolution lower than 1\arcsec, which makes it possible to overcome the limit of confusion. 

The study of distant and massive galaxies is essential to understanding our models of galaxy formation and evolution, as they are the ideal candidate progenitors of compact quiescent galaxies at $z$\,$\sim$\,2 (\citealt{Barro2013, Williams2014, Van_der_Wel2014, Kocevski2017}, see also \citealt{Elbaz2018}) and of present-day elliptical galaxies (\citealt{Swinbank2006, Michalowski2010, Ricciardelli2010,Fu2013}) that represent 60\% of the total stellar mass in the local Universe \citep[e.g.,][]{Fukugita1998, Hogg1998, Bell2003}. In particular, one of the most critical questions about the growth of galaxies concerns the evolution of the gas fraction over cosmic time and of the efficiency of galaxies to transform this gas into stars \citep[e.g.,][]{Somerville2015, Schinnerer2016, Tacconi2018}. Therefore, the study of massive and distant galaxies is of the utmost importance to constrain galaxy evolution models.

The Great Observatories Origins Deep Survey--South (GOODS--South) field benefits from deep and ultra-deep surveys over a large range of wavelengths (the Cosmic Assembly Near-infrared Deep Extragalactic Legacy Survey, CANDELS (\citealt{Koekemoer2011, Grogin2011}, PIs: S. Faber, \hbox{H. Ferguson}), the \textit{Spitzer} Extended Deep Survey \citep{Ashby2013}, the GOODS--\textit{Herschel} Survey \citep{Elbaz2011}, the \textit{Chandra} Deep Field-South \citep{Luo2017}, ultra-deep radio imaging with the Karl G. Jansky Very Large Array (VLA) \citep{Rujopakarn2016} and the \textit{Hubble} Ultra Deep Field, HUDF).
This large effort allows us to study the whole SED of massive and distant galaxies by securing the cross-identification of ALMA detected galaxies thanks to its high angular resolution.

The GOODS-ALMA large survey covers 69\,arcmin$^2$ in the deepest region of CANDELS, with a depth of 0.18\,mJy, in which 20 sources were blindly detected (\citealt{Franco2018}, hereafter \citetalias{Franco2018}). A detailed description of this survey, detection techniques, first results, and the presentation of optically-dark galaxies revealed by ALMA are presented in \citetalias{Franco2018}. Going further in the analysis of these data, we used \textit{Spitzer}/IRAC and VLA to extend our catalog to 16 additional sources detected down to 3.5$\sigma$ (see \citealt{Franco2020}, hereafter \citetalias{Franco2020}).

Beyond cosmic noon ($z\gtrsim2$) most studies on the evolution of star formation density are based on Lyman break galaxy (LBG) samples \citep[e.g.,][]{Steidel1999, Alvarez2016, Bouwens2016A}. Already, evidence exists that above a stellar mass of typically 5\,$\times$\,10$^{10}$\,M$_\odot$ the LBG technique misses the majority of massive dusty galaxies, because of their faintness and the redness of their UV slope \citep{vanDokkum2006, Bian2013, Wang2016, Wang2019}. Furthermore, recent studies with ALMA of a population of galaxies previously undetected by the \textit{Hubble} Space Telescope (HST) has shed new light on our understanding of the origin and formation of massive galaxies \citep{Wang2016, Fujimoto2016, Elbaz2018, Schreiber2018,Williams2019,Wang2019}. 
These optically dark galaxies constitute 20\% of the sources detected in GOODS-ALMA \citepalias{Franco2018}, 17\% if we include the sources detected down to 3.5$\sigma$ (see \citetalias{Franco2020}). Despite the fact that they are undetected by the HST (at 5$\sigma$ limiting depth \textit{H}\,=\,28.2 AB at 1.6\,$\mu$m), they are detectable through their thermal dust emission thanks to the depth and capabilities of ALMA. 
The systematic study of massive galaxies (M$_\star$\,$>$\,5\,$\times$\,10$^{10}$\,M$_\odot$) during this period of rapid transition between star-forming and quenched galaxies \citep{Muzzin2013} is crucial to understand the mechanism by which star formation ceases in these galaxies. 

Several surveys of the GOODS--South field have been carried on with ALMA around 1\,mm, resulting in a ``wedding cake'' distribution of surveys. A deep survey in the \textit{Hubble} Ultra Deep Field (HUDF, 20\,h, 4.5\,arcmin$^2$, RMS\,=\,35\,$\mu$Jy, $\lambda$\,=\,1.3\,mm, \citealt{Dunlop2017}), a wider, shallower survey  encompassing the HUDF, the ALMA 26\,arcmin$^2$ survey of GOODS-S at one millimeter (ASAGAO, 45\,h, 26\,arcmin$^2$, RMS\,=\,61\,$\mu$Jy, $\lambda$\,=\,1.2\,mm, \citealt{Hatsukade2018}) and finally the GOODS-ALMA survey itself encompassing both fields and covering the full area of GOODS--South with the deepest WFC3/$H$-band coverage (PI: D. Elbaz, 20\,h, 69\,arcmin$^2$, RMS\,=\,182\,$\mu$Jy, $\lambda$\,=\,1.1\,mm, \citetalias{Franco2018}). In addition, two spectroscopic surveys (the ALMA Spectroscopic Surveys; ASPECS), a pilot \citep{Walter2016} and large program \citep{Gonzalez2019}, were performed with ALMA over an area of $\sim$1\,arcmin$^2$ and $\sim$3\,arcmin$^2$ respectively, inside the HUDF. This ``wedding cake'' approach allows us to both to collect information on extreme and rare galaxies in mapping large regions and also to have a precise view of more common and abundant galaxies with deep observations on a small area. Interestingly, extending the survey area allows the detection of more distant galaxies than deep observations of a smaller area. Indeed, while only $\sim$13\% of the galaxies detected with ALMA have a redshift $\ge$ 3 in the deep ALMA survey of HUDF \citep{Dunlop2017}, $\sim$40\% of the \citetalias{Franco2018} galaxies are at $z\gtrsim3$. In addition, ALMA surveys over large areas allow the detection of particularly massive and dusty galaxies that are rare in terms of surface density.

This paper is organized as follows: in $\S$\ref{sec:Data}, we describe the data used in this paper. In $\S$\ref{sec:SED_fitting}, we describe how we took advantage of our large multiwavelength coverage to fit the spectral energy distributions (SEDs) of the galaxies detected in GOODS-ALMA. In $\S$\ref{sec:derived_parameter} we explain how we derived the main parameters of the galaxies (M$_\text{dust}$, M$_\text{gas}$, SFR, depletion time). 
In $\S$\ref{sec:slow_death}, we discuss the results and interpret them as the evidence for a slow downfall of star formation in $z\sim2-3$ massive galaxies.

Throughout this paper, we adopt a spatially flat $\Lambda$CDM cosmological model with H$_0$\,=\,70 km\,s$^{-1}$Mpc$^{-1}$, $\Omega_m$\,=\,0.3 and $\Omega_{\Lambda}$\,=\,0.7. We assume a Salpeter \citep{Salpeter1955} Initial Mass Function (IMF). We use the conversion factor of M$_\star$ (\citealt{Salpeter1955} IMF)\,=\,1.7\,$\times$\,M$_\star$ (\citealt{Chabrier2003} IMF) as suggested by \citealt{Reddy2006, Nordon2010, Caputi2015, Riechers2013, Seko2016, Elbaz2018}. All magnitudes are quoted in the AB system \citep{Oke1983}.

\section{Data}
\label{sec:Data}
\subsection{ALMA data}
This paper uses the ALMA observations obtained between August and September 2016 (Project ID: 2015.1.00543.S; PI: D. Elbaz), extending over an effective area of 69\arcmin$^2$, covering the deepest part of the CANDELS field -- in the GOODS--South field -- centered at $\alpha=3^{\rm h}$\,32$^{\rm m}$\,30.0$^{\rm s}$, $\delta=-27$\degree\,48\arcmin\,00\arcsec (J2000). We perform this analysis in a 0.60\arcsec-tapered mosaic reaching a RMS $\simeq 0.18$\,mJy\,beam$^{-1}$. The complete description of this survey and the data reduction are presented in detail in \citetalias{Franco2018}, where the properties of 23 bright ALMA sources discovered as the result of the blind survey in this field are discussed and cataloged. Sources that were most likely false (indicated by an * in Table 2 in \citetalias{Franco2018} and AGS22) are not taken into account in the rest of this paper. We consider these sources spurious because they have no counterparts at other wavelengths including mid-infrared. Among the 4 flagged sources, the 3 (AGS14, AGS16 and AGS19) which are in the 100 arcmin$^2$ of the SUPER-GOODS field of view have not been detected down to an RMS $\sim$0.56\,mJy at 850\,$\mu$m with SCUBA-2 (see the complete explanation of the rejection of these galaxies in \citetalias{Franco2020}).

In addition, this catalog has been enriched with 16 galaxies, detected with a lower S/N, using the VLA and \textit{Spitzer}/IRAC counterparts (see \citetalias{Franco2020} for more details). In this work, we analyze a sample of 35 galaxies with redshifts between 0.6 and 4.7 (z$_{med}$\,=\,2.7) and stellar masses ranging from 10$^{10.3}$ to 10$^{11.5}$\,M$_\odot$ (M$_{\star,med}$\,=\,10$^{10.93}$M$_\odot$ ). The stellar mass and redshift distributions of these 35 galaxies are visible in Fig.~10 of \citetalias{Franco2020}.

\subsection{Multiwavelength coverage}

We take advantage of the excellent multiwavelength coverage of the GOODS--South field to derive the physical properties of our galaxies. Below, we reproduce the list of the bands and filters used to observe this field (see \citealt{Guo2013}; \citealt{Straatman2016}), to fit the spectral energy distribution of these galaxies from the radio to the UV. We use the following bands: In UV, optical and near infrared (NIR), we used filters from VLT/VIMOS (U and R; \citealt{Nonino2009}), from ESO/MPG/WFI (U$_{38}$, V, R$_c$; \citealt{Baade1999,Hildebrandt2006}), from HST/ACS (F435W, F606W, F775W, F814W, F850LP; \citealt{Giavalisco2004,Wuyts2008, Guo2013}), from HST/WFC3 (F098M, F105W, F125W, F160W; \citealt{Grogin2011, Koekemoer2011}), from Subaru Suprime-Cam (IA484, IA527, IA550, IA574, IA598, IA624, IA651, IA679,  IA738,  IA797, IA856; \citealt{Cardamone2010}), from CFHT/WIRCAM (K; \citealt{Hsieh2012}) from Magellan/FourStar (J1, J2, J3, Hs, Hl; \citealt{Straatman2016}), \textit{Spitzer}/IRAC, channel 1 to 4 \citep{Fazio2004}, and \textit{Spitzer}/MIPS \citep{Rieke2004} filters. In far infrared (FIR), PACS (70$\mu$m, 100$\mu$m, 160$\mu$m; \citealt{Poglitsch2010}) and SPIRE (250$\mu$m, 350$\mu$m, 500$\mu$m;  \citealt{Griffin2010}). In radio, 5 and 10 cm images (\citealt{Rujopakarn2016}, Rujopakarn et al., in prep). In addition, where possible, we add the (sub)millimeter flux from previous pointings of the GOODS--South field \cite[e.g.,][]{Zhou2020, Elbaz2018,Cowie2018, Barro2017, Talia2018} or during previous surveys \cite[e.g.,][]{Hodge2013,Aravena2016,Dunlop2017,Hatsukade2018}.

As the SPIRE beam is very large (18.1\arcsec, 24.9\arcsec, and 36.6\arcsec at 250\,$\mu$m, 350\,$\mu$m, and 500\,$\mu$m, respectively) and yielding a high confusion limit, we use the catalog of Wang et al. (in prep.), which is built with a state-of-the-art de-blending method. MIPS, radio and ALMA sources have been used to define prior positions to perform source extractions on the SPIRE images. Moving from shorter to longer wavelengths, predictions were made for the galaxy fluxes at longer wavelengths based on the redshift and photometry information at all available shorter wavelengths. The faint priors at longer wavelengths were then removed , which helps to break blending degeneracies and achieve deeper detection limits. Similar methods have been used to produce \textit{Herschel}/SPIRE catalogs in GOODS--North \citep{Liu2018}.

\section{SED-fitting}
\label{sec:SED_fitting}
\subsection{Method}

\begin{figure}
   \centering
   \includegraphics[width=\hsize]{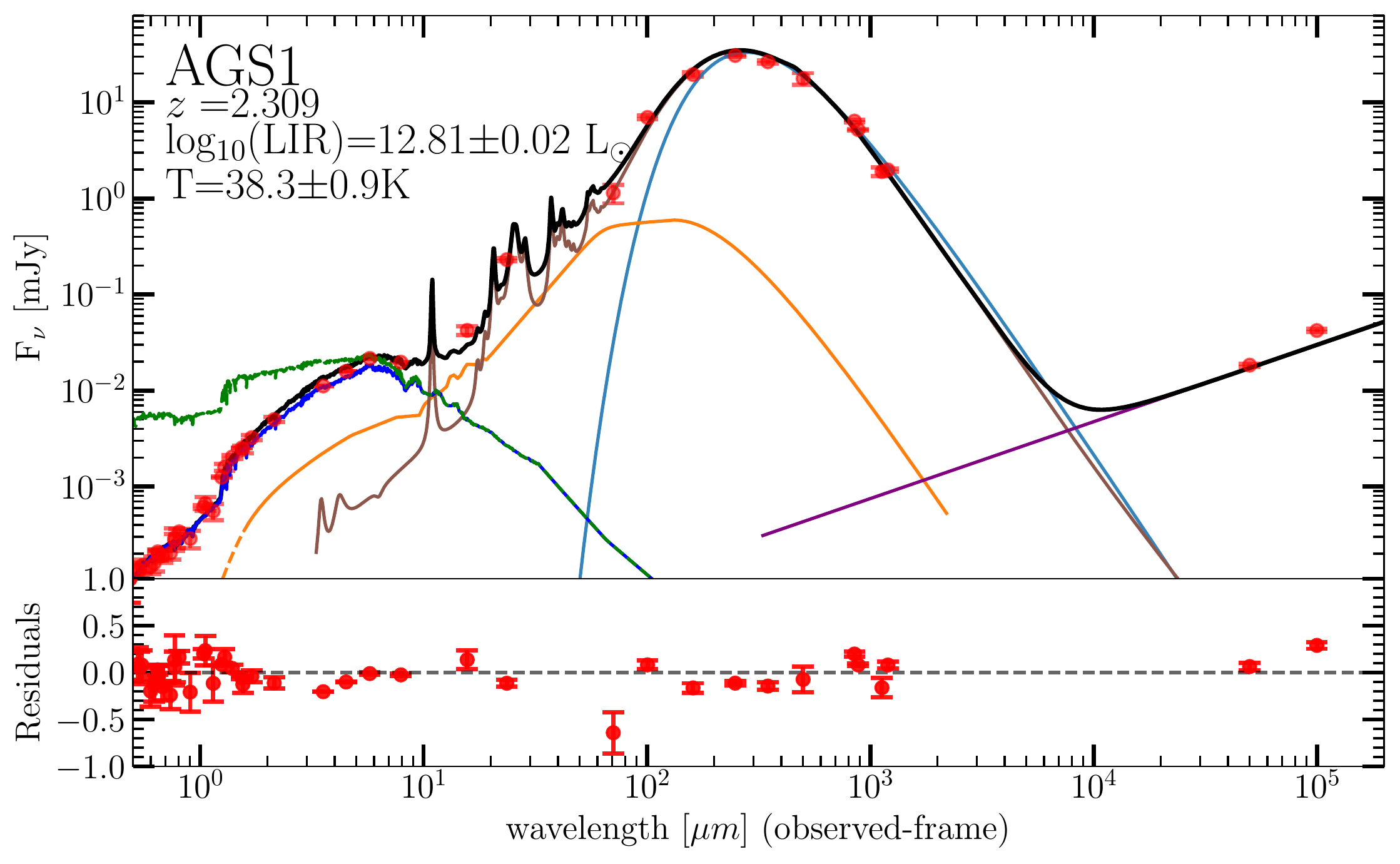}
      \caption{SED of AGS1 shown as an example. The solid black line represents the best fit, which can be decomposed into the IR dust contribution (brown line), a stellar component uncorrected for dust attenuation (dark blue line), synchrotron emission (purple line) and the AGN contribution (orange line). An extrapolation of the AGN model of \cite{Kirkpatrick2015} to shorter wavelengths is displayed with an orange dashed line. In addition, we show the best fit of a modified black body, with $\beta=1.5$ (light blue line). The corrected UV emission is also shown in green. The bottom panel shows the residuals: (observation - model)/observation. The 34 other SEDs are given in the Appendix (see Fig.~\ref{SEDs})}
         \label{SED_AGS1_bis}
   \end{figure}

We fit the spectral energy distributions using two different methods, depending on whether or not the galaxy has a \textit{Herschel} counterpart. For galaxies that have a far-IR flux density measured by the \textit{Herschel} space observatory, we employ the SED-fitting code \texttt{CIGALE} \footnote{Publicly available at \url{http://cigale.lam.fr}} (Code Investigating Galaxies Emission; \citealt{Boquien2019}). We use the stellar population models of \cite{Bruzual2003} and the attenuation law of \cite{Calzetti2000}. The IR SED fitting was performed using the dust infrared emission model given by \cite{Draine2014}. We independently fit the wavelengths from the UV up to 16 $\mu$m, and from 24 $\mu$m up to the millimeter wavelengths respectively (see Fig.~\ref{SED_AGS1_bis} for an example and Fig.~\ref{SEDs} for the full sample). The radio portion has been added after the fitting process, using the FIR/radio correlation, with a constant ratio between FIR and radio luminosity of 2.34 \citep{Yun2001}. The parameters used in \texttt{CIGALE} were given by \cite{Ciesla2018} and are shown in Table~\ref{table:CIGALE_parameter}.

In contrast, if the galaxy has no \textit{Herschel} infrared counterpart, we fit the data with the dust spectral energy distribution library\footnote{Publicly available at \url{http://cschreib.github.io/s17- irlib/}} presented in \cite{Schreiber2017}, and normalized to the ALMA flux density at 1.13\,mm in the SED. We proceed iteratively. After fitting the galaxy with a star formation main sequence (MS; \citealt{Noeske2007, Rodighiero2011, Elbaz2011}) SED, we compute the distance to the main sequence (R$_{SB}$\,=\,SFR/SFR$_{MS}$) using the output IR luminosity (8-1000 $\mu$m) and the redshift. The R$_{SB}$ and the redshift of the galaxy can be used to calculate the dust temperature (T$_{dust}$) and IR8 (L$_\text{IR}$/L$_8$) from Eq.~18 and 19 of \cite{Schreiber2017}. IR8 can be used as an indication of the compactness of distant galaxies \citep{Elbaz2011}. T$_{dust}$ and IR8 are therefore set to these newly calculated values in the SED-fitting process, and an updated SED is generated. The differences in the SED fitting derived quantities between the two methods are discussed in Sect.~\ref{sec:dust_mass}.

\begin{table}
\caption{Parameters used in the SED fitting procedures by \texttt{CIGALE}.}
\centering          
\begin{tabularx}{0.5\textwidth}{l X}
\hline
Parameter & Value \\
\hline
\hline
 & Delayed SFH \\
age [Gyr]&  100, 250, 500, 1000, 2000, 3000 \\
$\tau_{\text{main}}$ [Gyr] & 100, 500, 1000, 3000, 5000, 8000, 10000 \\
\hline
 & Dust attenuation \\
E(B-V)$_*$& 0.01, 0.1 , 0.3 , 0.4 , 0.6 , 0.7 , 0.9 , 1.  , 1.3  \\
\hline
 & Dust emission \\
$U_{min}$ & 1, 5,  12, 15, 20, 25, 30, 35, 40, 50  \\
$\alpha$ &  1, 1.5, 2, 2.5  \\
$\gamma$ & 1.e-04, 1.e-03, 1.e-02, 1.e-01, 5.e-01, 1.  \\
\hline                  
\end{tabularx}
\label{table:CIGALE_parameter}      
\end{table}

\subsection{AGN subtraction}
To fit an SED with an AGN component, we used the code $\texttt{decompIR}$ by \cite{Mullaney2011}. This code proposes to fit an AGN according to the spectrum of a sample of host galaxies representative of galaxies with an AGN. We consider that a galaxy hosts an AGN when one of the five $\texttt{decompIR}$ models converges on a reasonable solution. The contribution of the AGN to the IR luminosity can lead to an overestimation of the dust infrared emission and therefore an overestimation of the SFR. The AGN SED used in $\texttt{decompIR}$ does not include the wavelengths below 3\,$\mu$m. To better characterize the contribution of AGN to the total infrared luminosity of galaxies, we need to know their behavior at rest-frame wavelengths lower than 3\,$\mu$m, corresponding to the domain where the contribution of AGN is most important. Since this AGN model is only defined for wavelengths $>$\,3\,$\mu$m, we therefore used another AGN model for wavelengths shorter than 3\,$\mu$m. We extrapolate AGN emission to shorter wavelengths, using an AGN model from \cite{Kirkpatrick2015}, by fitting the flux of the AGN model from \cite{Kirkpatrick2015} to the one from $\texttt{decompIR}$ at 5\,$\mu$m (the part least polluted by PAHs). The subtraction of the AGN contribution from the optical part of the galaxy spectrum remains highly uncertain, so we have chosen not to modify the stellar masses of galaxies hosting an AGN, whilst keeping in mind that they could be overestimated. The SED fitting is done in two iterations; during the first iteration, we remove the contribution of the AGN. We then apply the \texttt{CIGALE} code in order to derive the properties of the galaxy having eliminated the bias produced by the AGN.


\subsection{Dust temperature}
For the sake of simplicity and comparison with previous studies, we measure the dust temperature by fitting a modified black body (MBB) model, following:
\begin{equation}
  S_\nu \propto \frac{\nu^{3+\beta}}{\exp(\frac{h \nu}{k_B T_{dust}}) - 1},
\end{equation}
where $k_B$ is Boltzmann's constant, $h$ is the Planck's constant, $\beta$ is the dust emissivity spectral index, $T_{dust}$ is the dust temperature, and S$_\nu$ is the flux density. We have assumed a spectral index $\beta=1.5$ \citep[e.g.,][]{Kovacs2006,Gordon2010}. We fit the flux densities at $\lambda_{rest} \ge 0.55 \lambda_{peak}$ using the MBB model as suggested by \cite{Hwang2010}, and exclude the synchrotron contribution. The criteria we have defined to select the points to be modeled with a MBB are as follows: (i) at least one data point between 0.55\,$\times\,\lambda_{peak}$ and $\lambda_{peak}$ and (ii) at least one data point beyond $\lambda_{peak}$, with a wavelength lower than or equal to 3\,mm.

Galaxies selected in (sub)millimeter flux density are expected to be biased toward low dust temperatures \cite[e.g.,][]{Chapman2003, Magdis2010, Casey2014b,  McAlpine2019}. Indeed, at fixed redshift and IR brightness, the (sub)millimeter flux of a galaxy with a colder dust temperature will be higher than that of a galaxy with a warmer dust temperature. We were interested to know whether the galaxies detected in this survey had unusual temperature characteristics, and if a possible temperature discrepancy could explain the properties of our sample. Indeed, the computation of both the dust mass and the gas mass (with the method we used), is temperature dependent.

\begin{figure*}
   \centering
      \begin{minipage}[t]{1.0\textwidth}
\resizebox{\hsize}{!} {
\includegraphics[width=\hsize]{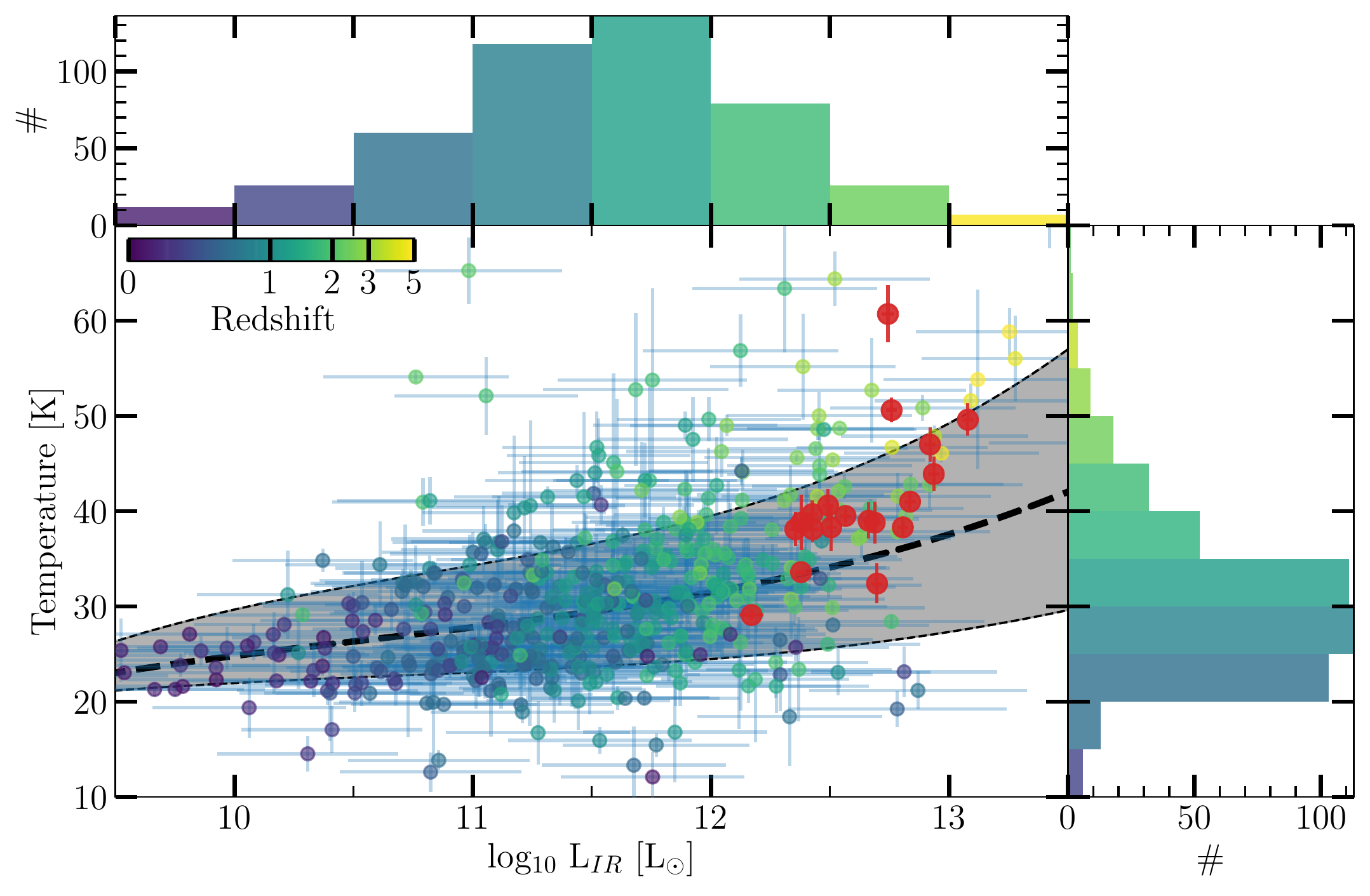}\\
\includegraphics[width=\hsize]{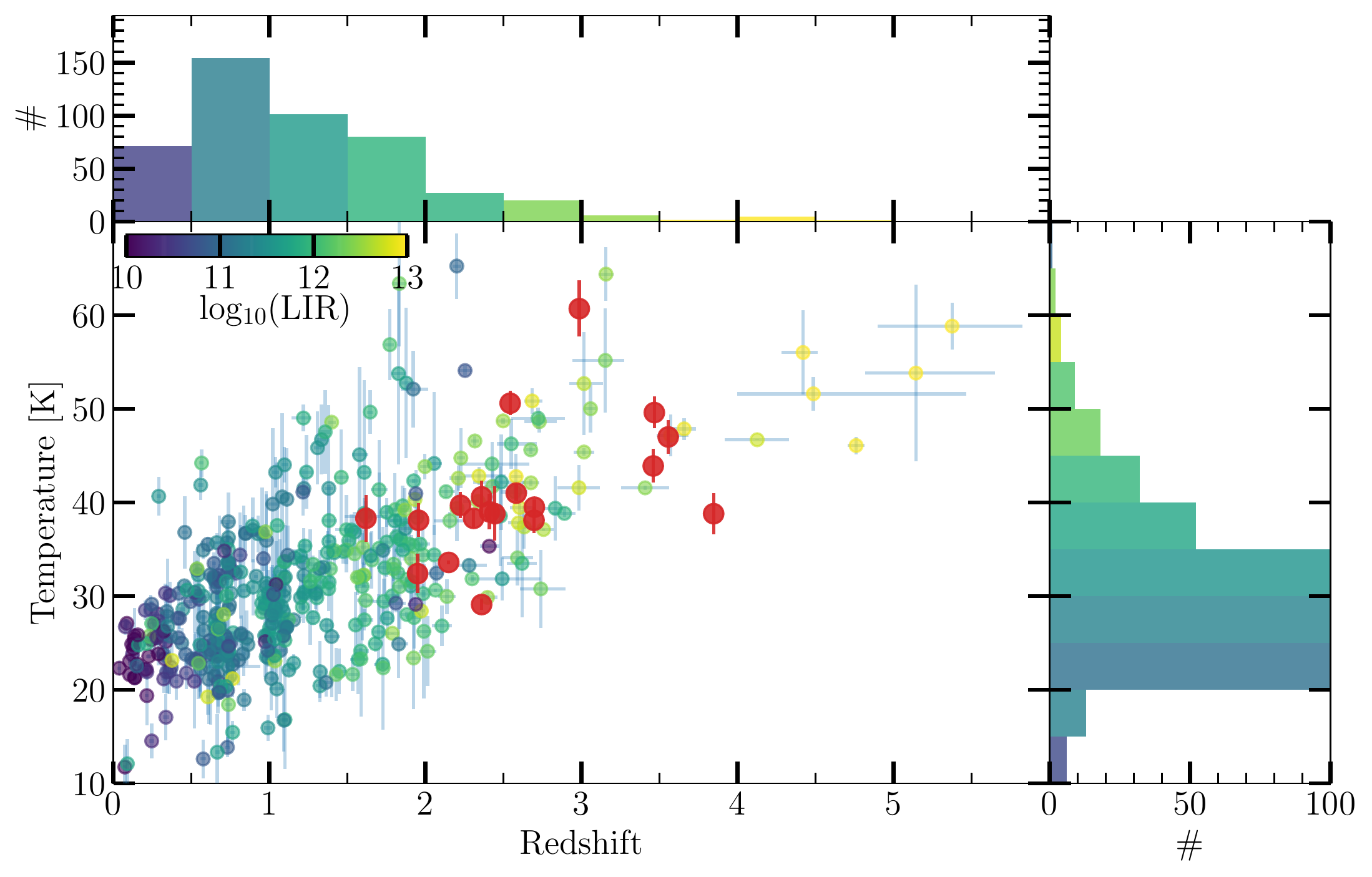}
}
\end{minipage}
      \caption{Evolution of the dust temperature as a function of IR luminosity (left panel) and redshift (right panel) for the galaxies with a \textit{Herschel} counterpart. The ALMA detections are shown in red. By comparison, we also plot the dust temperatures of all the galaxies within the GOODS-ALMA field, color-coded by redshift (left panel) or IR luminosity (right panel). We show the sliding median (and the 1$\sigma$ error) in black.}
         \label{T_dust}
   \end{figure*}

We investigated where the galaxies detected in the GOODS-ALMA survey are located in the IR Luminosity-Temperature plane (Fig.~\ref{T_dust}, left panel) and in the Redshift-Temperature plane (Fig.~\ref{T_dust}, right panel). For comparison, we also plot the dust temperature of all the galaxies located in GOODS-ALMA with an MBB fit, as described above. We find that the galaxies detected by ALMA do not exhibit a systematic offset in the luminosity-temperature and redshift-temperature plane compared to those undetected by ALMA.   
For an SMG, the dust temperature is correlated with the IR luminosity \cite[e.g.,][]{Wardlow2011}. We have found a median dust temperature of 40\,K for our sample. However, we note that the spectral index $\beta$ has an influence on the temperature. We chose to fix it, at $\beta$\,=\,1.5 in order to have fewer free parameters in our fit and to compare all galaxies consistently. If we had taken $\beta$\,=\,2, on the other hand, the MBB temperatures would have been slightly lower (1 - 4\,K lower). We note that we do not use this T$_{dust}$ temperature to determine dust masses (see Sect.~\ref{sec:dust_mass}).

\section{Derived parameters}
\label{sec:derived_parameter}
\subsection{Dust mass}
\label{sec:dust_mass}
 Following \cite{Draine2007}, we adopt the maximum starlight intensity relative to the local interstellar radiation field $U_{\max}$\,=\,$10^6$\,U$_{\odot}$, and the power-law index $\alpha$\,=\,2 in Eq.~\ref{radiationM_dust}. For the galaxies with a \textit{Herschel} counterpart, the dust mass is estimated from the \texttt{CIGALE} fit (see Sect.~\ref{sec:SED_fitting}) using the formula from \cite{Draine2007}:
\begin{equation}
\label{radiationM_dust}
\frac{dM_{dust}}{dU}\,=\,(1\,-\,\gamma)M_{dust}\delta(U\,-\,U_{min})\,+\,\gamma
M_{dust} \frac{\alpha\,-\,1}{U_{min}^{1-\alpha}\,-\,U_{max}^{1-\alpha} } U^{-\alpha},
\end{equation}
\noindent where $U_{min} \le U_{max}$, $\alpha \ne 1$ is the exponent of the power law describing the intensity distribution of the interstellar radiation field, and $\gamma$ is the relative fraction of dust heated by each source. \cite{Draine2007} showed that $\alpha=2$ and $U_{max}=10^6$ provided a good fit to a large sample of nearby galaxies from the \textit{Spitzer} SINGS program \citep{Kennicutt2003}.

We compared the dust masses derived using \texttt{CIGALE} for galaxies with \textit{Herschel} counterparts to those derived using the dust spectral energy distribution library presented in \cite{Schreiber2017}. We find a systematic offset of approximately 0.15 dex between the two methods, where dust masses derived using \texttt{CIGALE} are systematically larger. This may be caused by a slight underestimation of the ALMA fluxes (see \citetalias{Franco2020}) directly affecting the dust measurement with the \cite{Schreiber2017} dust library but compensated by the \textit{Herschel} measurements when we use the full spectrum of the galaxy. In order to avoid biasing our study, and because we did not want to derive two physical quantities (the mass of dust and the mass of gas) from a single data point, we do not indicate the points without \textit{Herschel} counterparts in the different figures and in the analysis. The differences however being small, we have presented the dust masses for galaxies with no \textit{Herschel} counterpart in Table~\ref{derived_properties} as an indication.

\subsection{Gas mass}

As we discuss in Sect.~\ref{sec:SFR_IR}, the ALMA detected galaxies are located in the SB region or in the upper part of the MS. To understand if their position is due to an increased star formation efficiency (SFE\,$\equiv$\,SFR/$M_{\text{gas}}$) or a large gas reservoir compared to normal MS galaxies, we computed their gas mass $M_{gas}$ as well as their gas fraction $f_{\text{gas}}$, defined by:
\begin{equation}
\label{eq:f_gas}
f_{\text{gas}}= \frac{M_{\text{gas}}}{M_{\text{gas}} + M_\star}.
\end{equation}
To compute the gas mass, we assume a gas-to-dust ratio ($\delta_{GDR}$) depending only on metallicity. This method of derivation of the gas mass, in comparison with using the CO-to-H$_2$ factor, which is observationally time consuming, can be used for a large survey. However, this method is based by a wide range of assumptions and limitations, such as the precise measurement of the dust mass, the assumption that the properties of the dust grains remain unchanged with redshift, that metallicity can be estimated at high redshift or that this relationship remains valid regardless of the distance to the main sequence \cite[e.g.,][]{Magdis2011, Magdis2012, Santini2014, Berta2016, Magdis2017}.

This ratio was directly derived by \cite{Leroy2011} in the local Universe, and can be applied to our sample, assuming that this relation is valid at all redshifts \cite[e.g.,][]{Ivison2011, Casey2014b, Tan2014, Swinbank2014}:
\begin{equation}
\log_{10}(\delta_{GDR})\,=\,\frac{M_{gas}}{M_{dust}}=\,(9.4 \pm 1.1)\,-\,(0.85 \pm 0.13) [12 + \log(O/H)],
  \label{eq:gdr}
\end{equation}
\noindent where $M_{gas}\,=\,M(H_2) + M(H_I)$.
At the redshifts of this study, the atomic hydrogen can be considered negligible compared to the molecular form \citep[e.g.,][]{Leroy2008, Obreschkow2009, Daddi2010}.

We note that recent studies have found evidence for a steep increase in the gas-to-dust ratio of subsolar metallicity galaxies at z$\sim$2 compared with this local relation \citep{Coogan2019}, but we do not expect this effect to be significant for our more massive, enriched galaxies.
As we do not have direct metallicity measurements for our galaxies, we use the equation given by \cite{Genzel2012} to compute the metallicity:
\begin{multline}
   12+\log(O/H)\,=\, -4.51 + 2.18 \text{log}_{10}(\text{M}_\star/1.7) \\
   - 0.0896 \big[\text{log}_{10}(\text{M}_\star/1.7)\big]^2 .
\end{multline}

In this equation, we include a conversion factor (1/1.7) to transform the original formula from a Chabrier IMF to a Salpeter IMF. However, the metallicity can be underestimated for galaxies above the main sequence \cite[e.g.,][]{Silverman2015}, which could artificially increase the proportion of gas and conversely underestimate the gas depletion time \citep{Elbaz2018}. 
We compared our calculated metallicities to the metallicities obtained using the fundamental metallicity relation (FMR) of \cite{Mannucci2010}:
\begin{multline}
\label{eq:FMR}
 12+\log_{10}(O/H)\,=\,8.90 + 0.37m\,-\,0.14s\,-\,0.19m^2 \\
 + 0.12ms\,-\,0.054s^2,
\end{multline}

\noindent with $m$\,=\,log$_{10}$(M$_{\star}$/1.7)-10, and $s$\,=\,log$_{10}$(SFR/1.7).
We applied an average correction factor of -0.25 $\pm$ 0.02 to convert from the FMR derived using the \cite{Kewley2002} metallicity calibration to the calibration of \cite{Pettini2004}, as given in \cite{Kewley2008}. The median metallicity ratio between these two methods is 1.03\,$\pm$\,0.01, where the uncertainty corresponds to the standard deviation. For our galaxy sample, both methods are, therefore, in good agreement. However, the metallicities of these ALMA detected galaxies remain uncertain. Indeed, the metallicity evolution is poorly constrained for galaxies at high redshift, as well as for starburst galaxies and galaxies with AGN \cite[e.g.,][]{Tan2013, Kewley2013}. We keep in mind that the uncertainties on the determination of $M_{gas}$ are large, taking into account all of the assumptions used.


We also verified that the mass of gas derived by the method described above was in agreement with that derived using the method of \cite{Scoville2016}. The \cite{Scoville2016} method is based on the assumption that continuum measurements of the Rayleigh-Jeans tail can be used to estimate the mass of dust and therefore, the mass of gas. Since this method is based on the Rayleigh-Jeans tail, it can only be used at long wavelengths ($\lambda$\,$>$\,250 $\mu$m). However, if the dust emission is optically thin, the \cite{Scoville2016} method may underestimate the gas mass \citep{Miettinen2017}. At 1.13\,mm, the estimate of the gas mass can be written, according to equations Eq.~6 and Eq.~16 of \cite{Scoville2016}, as:
\begin{equation}
    M_{mol}\ [M_{\odot}] \,=\,S_\nu \times 5.12\times10^{10}\times(1+z)^{-4.8}\times(d_L)^2\frac{\Gamma_{RJ}^0}{\Gamma_{RJ}^z},
\end{equation}
\noindent with S$_\nu$ the flux at 1.13\,mm in mJy and $\Gamma_{RJ}^z$ the correction for departure in the rest frame of the Planck function from Rayleigh-Jeans \citep{Scoville2016}:
\begin{equation}
    \Gamma_{RJ}(T_{dust},\nu_{obs},z)\,=\,\frac{h\nu_{obs}(1+z)/(k_b T_{dust})}{\text{e}^{(h\nu_{obs}(1+z)/(k_b T_{dust}))}- 1},
\end{equation}
\noindent where $h$ is the Planck's constant and $k_b$ is the Boltzmann constant.
As explained in \cite{Scoville2016}, the temperature we should use in this equation is not the temperature derived by the MBB fit for each individual galaxy but the mass-weighted temperature. We have therefore adopted a constant mass-weighted temperature value of 25\,K (see Apendix A.2 in \cite{Scoville2016}).

Using a fixed  mass-weighted temperature of 25\,K, we find a difference between the calculated gas mass (M$_{(gas,\ this\, work)}$) and that derived following \cite{Scoville2016}: M$_{(mol,\ Scoville)}$/M$_{(gas,\ this\, work)}$\,=\,1.1\,$\pm$\,0.6.

The gas mass is directly related to the depletion time ($\tau_{\text{dep}}$) by:
\begin{equation}
\tau_{\text{dep}}\,[yr^{-1}]\,=\,\frac{M_{\text{gas}}}{SFR}.
\end{equation}

\subsection{SFR}
In this section, we detail the computation of the SFR of our sample. Although the IR part totally dominates the SFR, we took advantage of the UV coverage of GOODS-ALMA to calculate the total SFR (SFR$_{tot}$\,=\,SFR$_{UV}$\,+\,SFR$_{IR}$).
\subsubsection{SFR$_{IR}$}\label{sec:SFR_IR}

The infrared luminosity of each galaxy has been converted to SFR using the Kennicutt relation \citep{Kennicutt1998} below: 

\begin{equation}
SFR_{IR}\ [M_{\odot}\,yr^{-1}]\,=\,1.72 \times 10^{-10} L_{IR},
\label{SFR_IR}
\end{equation}
\noindent with $L_{IR}$ in $L_{\odot}$, and 
\begin{equation}
L_{IR}\ [L_{\odot}]\,=\,4\pi d_L^2\int_{8\mu m}^{1000\mu m} F_{\nu}(\lambda) \times \frac{c}{\lambda^2}d\lambda, 
\label{LIR}
\end{equation}
\noindent where $d_L$ is the luminosity distance.

\begin{figure}
   \centering
   \includegraphics[width=1.\hsize]{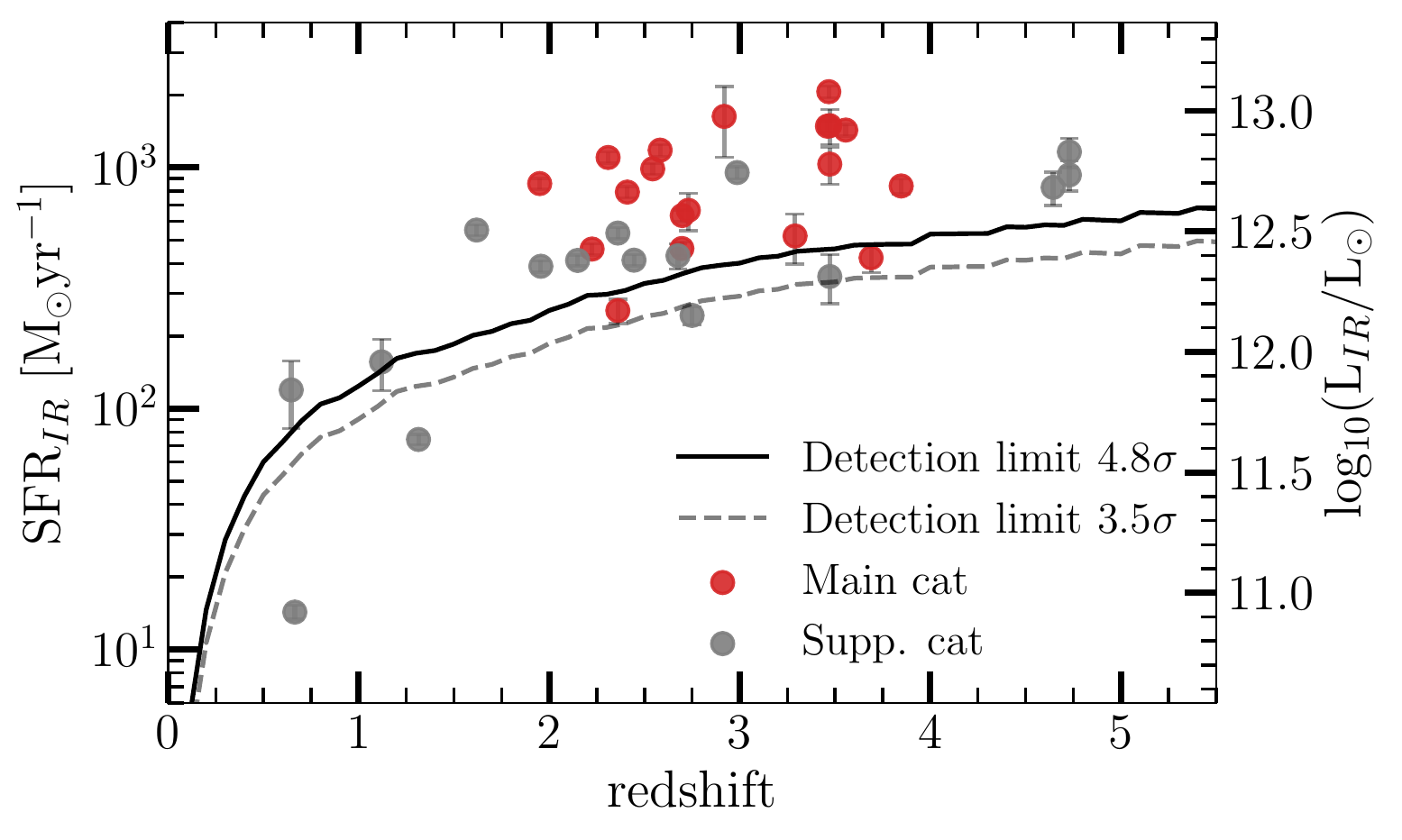}
      \caption{Infrared star formation rate as a function of redshift, for galaxies from the main (red dots) and supplementary (gray dots) catalogs respectively. The SFR has been computed from the IR luminosity following Eq.~\ref{SFR_IR}. The detection limits at 4.8\,$\sigma$ (solid black line) and 3.5\,$\sigma$ (dashed black line) have also been computed from the IR library of \cite{Schreiber2017}, with a dust temperature evolving with redshift taking into account the average value of the RMS at 0.182\,mJy.}
         \label{SFR_lim}
   \end{figure}

In Fig.~\ref{SFR_lim}, we illustrate the distribution of SFRs as a function of redshift for the ALMA-detected galaxies. We also represent the theoretical detection limit of the galaxies present in the survey at the limit of 4.8$\sigma$ (solid black line) used to create the main catalog, assuming a constant RMS (RMS\,=\,0.182\,mJy) over the whole map, as well as the 3.5$\sigma$ (dashed black line) limit used to build the supplementary catalog. However, as the RMS is not constant, and therefore may be lower at some points in the map, some galaxies (AGS21, for example) appear below this line.

We note that there is a galaxy (AGS36) that is clearly offset from the detection limit, with a SFR\,$\sim$\,20 M$_{\odot}$yr$^{-1}$. This galaxy is atypical, as it has the lowest redshift in our sample ($z$\,=\,0.66, the same redshift as AGS30) and it also hosts a powerful AGN with an X-ray luminosity\,=\,1.39 $\times$$10^{43}$erg\,sec$^{-1}$.

The SFR limit has been computed taking into account the main sequence SED from \cite{Schreiber2017}, with the temperature and the fraction of polycyclic aromatic hydrocarbon (PAH) emission evolving as a function of redshift. The IR luminosity was calculated by integrating the flux from the SED using Eq.~\ref{LIR}, and was then converted into SFR using Eq.~\ref{SFR_IR}.

This IR luminosity limit allows us to detect galaxies down to an IR luminosity of 10$^{12}$\,L$_\odot$ at redshift $z$\,=\,1.5, and down to 3$\times$10$^{12}$\,L$_\odot$ at redshift $z$\,=\,4. In other words, for a MS galaxy, this allows us to detect galaxies with a minimum stellar mass of 2.5\,$\times$\,10$^{11}$\,M$_\odot$, 1.8\,$\times$\,10$^{11}$\,M$_\odot$ and 1.5\,$\times$\,10$^{11}$\,M$_\odot$ for redshifts $z$\,=\,2, $z$\,=\,3 and $z$\,=\,4 respectively, using Eq.~9 of \cite{Schreiber2015}.

\begin{figure}
   \centering
\includegraphics[width=1.\hsize]{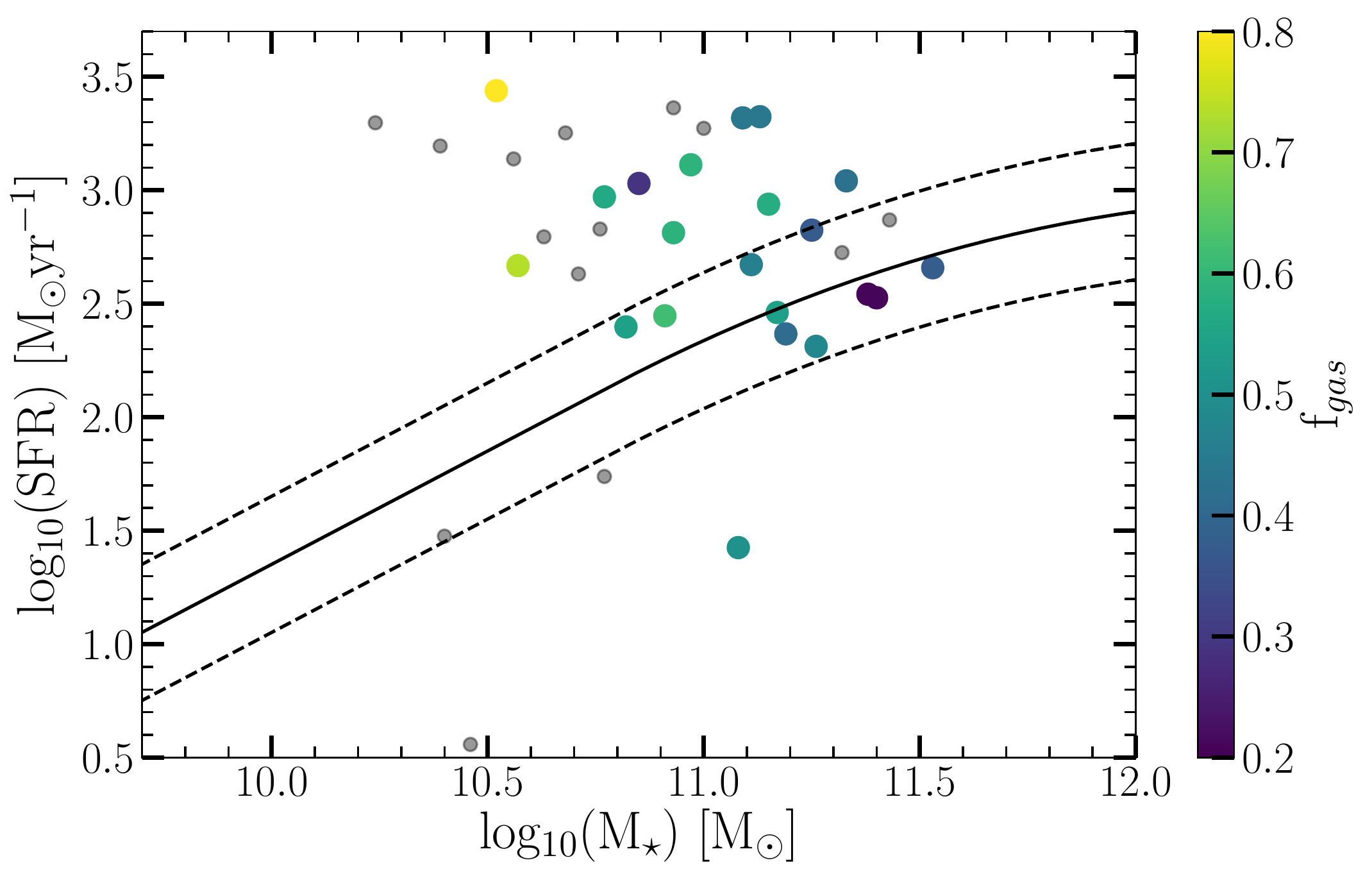}
      \caption{Location of our ALMA detected galaxies in the SFR-M$_\star$ plane. Galaxies with \textit{Herschel} counterparts are color-coded as a function of the f$_\text{gas}$. The other galaxies are represented by gray dots. We have rescaled all of the SFRs by multiplying by SFR$_{MS}$($z$)/SFR$_{MS}$($z$\,=\,2.7), in order to maintain their relative positions with respect to the main sequence. We indicated the MS using Eq.~9 from \cite{Schreiber2015}, with a dispersion of 0.3 dex (solid and dashed lines respectively).}
         \label{SFR_Mstar}
   \end{figure}

The majority of the galaxies detected in this ALMA survey are starbursts, or in the upper part of the MS (see Fig.~\ref{SFR_Mstar}). Among the galaxies for which we determined the SFR, 54\% of them have a R$_{SB}$(SFR/SFR$_{MS}$)\,$>$\,3 (see Table~\ref{derived_properties}).

Not surprisingly, the most IR luminous galaxies have been listed in the main catalog. However, we note the presence of a portion of galaxies from the supplementary catalog that are also among the most IR luminous galaxies. The size of the galaxies explains this behavior. The galaxies detected in the supplementary catalog generally have larger sizes than those in the main catalog (\citetalias{Franco2020}). Even though the peak flux is fainter on average, the integrated flux can reach values close to those of the main catalog.

The vast majority (86\%) of the galaxies analyzed in this study can be classified as ultraluminous infrared galaxies (ULIRGs) with 12\,$<$\,log$_{10}$(L$_{IR}$/L$_\odot$)\,$<$\,13. Only one galaxy has an infrared luminosity slightly above this threshold. All of the galaxies with log$_{10}$(L$_{IR}$/L$_\odot$)\,$<$\,12 are galaxies less distant than the average of the galaxies detected in this survey, with $z$\,$<$\,1.5.

\subsubsection{SFR$_{UV}$}

Massive galaxies are known to be heavily dust-obscured at $z$\,$>$\,2 \citep[e.g.,][]{Magnelli2009,Murphy2011}. While the SFR$_{IR}$ is derived from the dust emission, we also consider the unobscured contribution to the total SFR, observed through UV emission. For the most massive galaxies (M$_\star$\,$>$\,10$^{10.5} M_\odot$), the fraction of obscured to unobscured star formation (SFR$_{IR}$/SFR$_{IR+UV}$) is greater than 90\% \citep{Whitaker2017}.

We derive L$_{UV}$ from the observed magnitude as follows:
\begin{equation}
L_{UV}\ [L_{\odot}]\,=\,\frac{4\pi d_L^2\nu_{1600}10^{-0.4(48.6+m)}}{1+z},
\end{equation}
\noindent where $d_L$ is the luminosity distance and $m$ is the observed magnitude. 
The SFR$_{UV}$, uncorrected for dust attenuation, is in turn derived from the L$_{UV}$, following \citep{Daddi2004}:

\begin{equation}
SFR_{UV}\ [M_{\odot}\ yr^{-1}]\,=\,2.17 \times 10^{-10} \times L_{UV}.
\end{equation}

The total SFR (SFR$_{tot}$\,=\,SFR$_{UV}$\,+\,SFR$_{IR}$) is given in Table~\ref{derived_properties}. Unless otherwise noted, in the following, SFR refers to the total SFR. The median contribution from SFR$_{UV}$ to SFR$_{tot}$ is $\sim$1\%.

\begin{table*}\footnotesize
\caption{Derived properties of the GOODS-ALMA sources.}
\centering        
\begin{threeparttable}
\begin{tabular}{l c c c c c c c c c c}    
\hline       
 ID & $z$ & log$_{10}$(M$_\star$) & log$_{10}$(L$_{\text{IR}}$) & SFR  & R$_{SB}$ & log$_{10}$(M$_{\text{dust}}$)  & log$_{10}$(M$_{\text{gas}}$) & f$_{\text{gas}}$  &   T$_{\text{dust}}$ & S$_{1.1mm}$ \\
       & & M$_{\odot}$ &  L$_{\odot}$   &M$_{\odot}yr^{-1}$&                & M$_{\odot}$   & M$_{\odot}$  &                   & K   & mJy\\
\hline  
\hline
   AGS1 		& 2.309 &11.15 &  12.81 $\pm$   0.02 &   1103$_{-    60}^{+    54}$ &      4.8$_{-  0.3}^{+  0.2}$ &    9.2 $\pm$    0.1 &   11.3 $\pm$    0.1&   0.57$_{- 0.06}^{+ 0.07}$ &  38.3 $\pm$    0.9  & 1.90 $\pm$            0.20\\
  AGS2$\dag$ 	& 2.918 &10.68 &  12.98 $\pm$   0.14 &   1642$_{-   534}^{+   530}$ &     14.0$_{-  4.5}^{+  4.3}$ &    8.8 $\pm$    0.1 &   11.0 $\pm$    0.1&   0.67$_{- 0.24}^{+ 0.30}$ &  44.8 $\pm$    4.6  & 1.99 $\pm$            0.22\\
  AGS3 			& 2.582 &11.33 &  12.84 $\pm$   0.02 &   1187$_{-    72}^{+    60}$ &      3.3$_{-  0.2}^{+  0.2}$ &    9.1 $\pm$    0.1 &   11.2 $\pm$    0.1&   0.42$_{- 0.05}^{+ 0.05}$ &  41.0 $\pm$    0.9  & 1.84 $\pm$            0.21\\
  AGS4 			& 3.556 &11.09 &  12.93 $\pm$   0.02 &   1435$_{-    83}^{+    76}$ &      3.8$_{-  0.5}^{+  1.9}$ &    8.9 $\pm$    0.1 &   11.0 $\pm$    0.1&   0.44$_{- 0.04}^{+ 0.05}$ &  47.0 $\pm$    1.8  & 1.72 $\pm$            0.20\\
  AGS5 			& 3.46  &11.13 &  12.94 $\pm$   0.02 &   1487$_{-    82}^{+    78}$ &      3.8$_{-  0.2}^{+  0.2}$ &    8.9 $\pm$    0.1 &   11.0 $\pm$    0.1&   0.44$_{- 0.06}^{+ 0.06}$ &  43.9 $\pm$    1.8  & 1.56 $\pm$            0.19\\
  AGS6      	& 2.698 &10.93 &  12.57 $\pm$   0.02 &    651$_{-    52}^{+    16}$ &      3.5$_{-  0.3}^{+  0.1}$ &    9.0 $\pm$    0.1 &   11.1 $\pm$    0.1&   0.59$_{- 0.08}^{+ 0.08}$ &  39.5 $\pm$    0.8  & 1.27 $\pm$            0.18\\
  AGS7$\dag$  	& 3.29  &11.43 &  12.48 $\pm$   0.10 &    522$_{-   126}^{+   120}$ &      0.8$_{-  0.2}^{+  0.2}$ &    8.7 $\pm$    0.1 &   10.7 $\pm$    0.1&   0.17$_{- 0.04}^{+ 0.04}$ &  37.4 $\pm$    2.4  & 1.15 $\pm$            0.17\\
  AGS8 			& 1.95  &11.53 &  12.70 $\pm$   0.02 &    861$_{-    47}^{+    42}$ &      3.1$_{-  0.2}^{+  0.2}$ &    9.3 $\pm$    0.1 &   11.3 $\pm$    0.1&   0.38$_{- 0.03}^{+ 0.03}$ &  32.4 $\pm$    2.1  & 1.43 $\pm$            0.22\\
  AGS9 			& 3.847 &10.97 &  12.69 $\pm$   0.03 &    843$_{-    61}^{+    55}$ &      2.7$_{-  0.2}^{+  0.2}$ &    9.0 $\pm$    0.1 &   11.1 $\pm$    0.1&   0.59$_{- 0.14}^{+ 0.16}$ &  38.8 $\pm$    2.2  & 1.25 $\pm$            0.21\\
 AGS10 			& 2.41  &11.25 &  12.66 $\pm$   0.02 &    803$_{-    53}^{+    30}$ &      2.8$_{-  0.2}^{+  0.1}$ &    9.0 $\pm$    0.1 &   11.0 $\pm$    0.1&   0.37$_{- 0.05}^{+ 0.05}$ &  39.0 $\pm$    1.9  & 0.88 $\pm$            0.15\\
 AGS11$\dag$ 	& 3.472 &10.24 &  12.94 $\pm$   0.07 &   1492$_{-   252}^{+   252}$ &     28.7$_{- 23.6}^{+  0.1}$ &    8.5 $\pm$    0.1 &   10.8 $\pm$    0.1&   0.80$_{- 0.16}^{+ 0.19}$ &  50.1 $\pm$    2.5  & 1.34 $\pm$            0.25\\
 AGS12 			& 2.543 &10.77 &  12.76 $\pm$   0.02 &    998$_{-    63}^{+    40}$ &      8.1$_{-  0.5}^{+  0.3}$ &    8.7 $\pm$    0.1 &   10.9 $\pm$    0.1&   0.56$_{- 0.06}^{+ 0.07}$ &  50.6 $\pm$    1.3  & 0.93 $\pm$            0.18\\
 AGS13 			& 2.225 &11.40 &  12.43 $\pm$   0.02 &    476$_{-    41}^{+     7}$ &      1.6$_{-  0.1}^{+  0.0}$ &    8.8 $\pm$    0.1 &   10.8 $\pm$    0.1&   0.21$_{- 0.03}^{+ 0.03}$ &  39.7 $\pm$    1.4  & 0.78 $\pm$            0.15\\
 AGS15$\dag$ 	& 3.472 &10.56 &  12.78 $\pm$   0.08 &   1034$_{-   179}^{+   180}$ &      9.5$_{-  1.7}^{+ 11.2}$ &    8.9 $\pm$    0.1 &   11.1 $\pm$    0.1&   0.78$_{- 0.16}^{+ 0.19}$ &  38.9 $\pm$    2.4  & 1.21$^\star$$\pm$    0.11\\
 AGS17 			& 3.467 &10.52 &  13.08 $\pm$   0.02 &   2070$_{-   117}^{+   112}$ &     20.9$_{- 12.6}^{+  3.1}$ &    9.0 $\pm$    0.1 &   11.2 $\pm$    0.1&   0.84$_{- 0.09}^{+ 0.09}$ &  49.6 $\pm$    1.7  & 2.30$^\star$$\pm$    0.44\\
 AGS18 			& 2.696 &11.11 &  12.43 $\pm$   0.03 &    471$_{-    39}^{+    19}$ &      1.8$_{-  0.1}^{+  0.1}$ &    8.9 $\pm$    0.1 &   11.0 $\pm$    0.1&   0.46$_{- 0.07}^{+ 0.08}$ &  38.1 $\pm$    1.4  & 1.70$^\star$$\pm$    0.30\\
 AGS20$\dag$ 	& 2.73  &10.76 &  12.59 $\pm$   0.08 &    666$_{-   120}^{+   118}$ &      5.1$_{-  0.9}^{+  0.9}$ &    8.6 $\pm$    0.1 &   10.8 $\pm$    0.1&   0.52$_{- 0.10}^{+ 0.11}$ &  40.6 $\pm$    2.4  & 1.11 $\pm$            0.24\\
 AGS21$\dag$ 	& 3.689 &10.63 &  12.39 $\pm$   0.06 &    437$_{-    71}^{+    42}$ &      3.2$_{-  0.5}^{+  0.3}$ &    8.3 $\pm$    0.1 &   10.5 $\pm$    0.1&   0.41$_{- 0.06}^{+ 0.06}$ &  43.7 $\pm$    2.0  & 0.64 $\pm$            0.11\\
 AGS23 			& 2.36  &11.26 &  12.17 $\pm$   0.05 &    256$_{-    31}^{+    29}$ &      0.9$_{-  0.1}^{+  0.1}$ &    9.2 $\pm$    0.1 &   11.2 $\pm$    0.1&   0.48$_{- 0.15}^{+ 0.17}$ &  29.1 $\pm$    0.6  & 0.98 $\pm$            0.21\\
 AGS24$\dag$    & 3.472 &11.32 &  12.31 $\pm$   0.11 &    353$_{-    82}^{+    80}$ &      0.6$_{-  0.1}^{+  0.5}$ &    8.5 $\pm$    0.1 &   10.6 $\pm$    0.1&   0.16$_{- 0.02}^{+ 0.02}$ &  37.3 $\pm$    2.5  & 0.88 $\pm$            0.22\\
 AGS25$\dag$ 	& 4.64  &10.39 &  12.68 $\pm$   0.07 &    832$_{-   135}^{+   126}$ &      8.0$_{-  1.3}^{+  1.3}$ &    8.2 $\pm$    0.1 &   10.5 $\pm$    0.1&   0.55$_{- 0.19}^{+ 0.22}$ &  51.5 $\pm$    2.6  & 0.81 $\pm$            0.19\\
 AGS26 			& 1.619 &10.91 &  12.51 $\pm$   0.02 &    553$_{-    32}^{+    26}$ &      6.1$_{-  0.4}^{+  0.3}$ &    9.0 $\pm$    0.1 &   11.1 $\pm$    0.1&   0.62$_{- 0.06}^{+ 0.06}$ &  38.2 $\pm$    2.5  & 0.97 $\pm$            0.15\\
 AGS27$\dag$ 	& 4.73  &10.93 &  12.83 $\pm$   0.06 &   1180$_{-   183}^{+   144}$ &      3.2$_{-  0.5}^{+  0.4}$ &    8.5 $\pm$    0.1 &   10.6 $\pm$    0.1&   0.33$_{- 0.03}^{+ 0.03}$ &  48.0 $\pm$    2.6  & 1.43 $\pm$            0.28\\
 AGS28 			& 2.15  &11.17 &  12.38 $\pm$   0.02 &    413$_{-    22}^{+    19}$ &      2.0$_{-  0.1}^{+  0.1}$ &    9.2 $\pm$    0.1 &   11.2 $\pm$    0.1&   0.54$_{- 0.16}^{+ 0.19}$ &  33.3 $\pm$    0.5  & 1.56 $\pm$            0.21\\
 AGS29$\dag$ 	& 1.117 &10.77 &  11.96 $\pm$   0.10 &    161$_{-    41}^{+    32}$ &      3.6$_{-  0.9}^{+  0.7}$ &    8.7 $\pm$    0.1 &   10.9 $\pm$    0.1&   0.55$_{- 0.07}^{+ 0.08}$ &  31.5 $\pm$    2.4  & 0.61 $\pm$            0.18\\
 AGS30$\dag$ 	& 0.646 &10.40 &  11.84 $\pm$   0.14 &    119$_{-    37}^{+    36}$ &      8.5$_{-  2.7}^{+  2.6}$ &    8.5 $\pm$    0.1 &   10.8 $\pm$    0.1&   0.71$_{- 0.18}^{+ 0.23}$ &  32.0 $\pm$    2.4  & 0.67 $\pm$            0.17\\
 AGS31 			& 2.45  &11.38 &  12.38 $\pm$   0.02 &    415$_{-    24}^{+    19}$ &      1.2$_{-  0.1}^{+  0.1}$ &    8.8 $\pm$    0.1 &   10.8 $\pm$    0.1&   0.21$_{- 0.03}^{+ 0.03}$ &  38.8 $\pm$    2.9  & 0.72 $\pm$            0.19\\
 AGS32$\dag$ 	& 4.73  &11.00 &  12.73 $\pm$   0.06 &    941$_{-   145}^{+   121}$ &      2.2$_{-  0.3}^{+  0.3}$ &    8.5 $\pm$    0.1 &   10.7 $\pm$    0.1&   0.31$_{- 0.04}^{+ 0.04}$ &  44.9 $\pm$    2.5  & 1.23 $\pm$            0.16\\
 AGS33$\dag$ 	& 2.68  &10.71 &  12.40 $\pm$   0.05 &    432$_{-    54}^{+    52}$ &      3.8$_{-  0.5}^{+  0.4}$ &    9.0 $\pm$    0.1 &   11.1 $\pm$    0.1&   0.73$_{- 0.15}^{+ 0.17}$ &  ....               & 1.77 $\pm$            0.27\\
 AGS34      	& 2.75  &10.82 &  12.15 $\pm$   0.04 &    244$_{-    21}^{+    19}$ &      1.6$_{-  0.1}^{+  0.1}$ &    8.7 $\pm$    0.2 &   10.9 $\pm$    0.2&   0.54$_{- 0.27}^{+ 0.35}$ &  ....               & 0.55 $\pm$            0.15\\
 AGS35 			& 2.99  &10.85 &  12.74 $\pm$   0.02 &    955$_{-    57}^{+    53}$ &      5.4$_{-  0.3}^{+  0.3}$ &    8.3 $\pm$    0.1 &   10.5 $\pm$    0.1&   0.29$_{- 0.03}^{+ 0.03}$ &  60.3 $\pm$    3.6  & 1.16 $\pm$            0.21\\
 AGS36$\dag$ 	& 0.665 &10.46 &  10.92 $\pm$   0.03 &     14$_{-     1}^{+     1}$ &      0.9$_{-  0.1}^{+  0.0}$ &    9.4 $\pm$    0.1 &   11.7 $\pm$    0.1&   0.94$_{- 0.10}^{+ 0.11}$ &  ....               & 0.74 $\pm$            0.21\\
 AGS37 			& 1.956 &11.19 &  12.36 $\pm$   0.02 &    390$_{-    21}^{+    19}$ &      2.1$_{-  0.1}^{+  0.1}$ &    9.0 $\pm$    0.1 &   11.0 $\pm$    0.1&   0.41$_{- 0.03}^{+ 0.03}$ &  38.0 $\pm$    1.6  & 1.10 $\pm$            0.16\\
 AGS38 			& 1.314 &11.08 &  11.64 $\pm$   0.02 &     79$_{-     8}^{+     1}$ &      0.9$_{-  0.1}^{+  0.1}$ &    9.0 $\pm$    0.3 &   11.1 $\pm$    0.3&   0.50$_{- 0.37}^{+ 0.53}$ &  ....               & 1.00 $\pm$            0.16\\
 AGS39 			& 2.36  &10.57 &  12.49 $\pm$   0.02 &    537$_{-    30}^{+    24}$ &      7.4$_{-  0.4}^{+  0.3}$ &    8.8 $\pm$    0.1 &   11.0 $\pm$    0.1&   0.73$_{- 0.11}^{+ 0.13}$ &  40.6 $\pm$    1.7  & 0.80 $\pm$            0.23\\ 
\hline   
\end{tabular}
\begin{tablenotes}
\item Columns: (1) Source name; (2) Redshifts (spectroscopic redshifts are shown to three decimal places); (3) Stellar Mass; (4) L$_{\text{IR}}$ derived from SED fitting; (5) SFR\,=\,SFR$_{IR}$\,+\,SFR$_{UV}$ ; (6) R$_{SB}$\,=\,SFR/SFR$_{MS}$, where SFR$_{MS}$ is the average SFR of MS galaxies following \cite{Schreiber2015}; (7) Dust Mass. For galaxies for which we used the dust spectral energy distribution library presented in \cite{Schreiber2017} (labeled by a $\dag$ in this table), the dust mass is multiplied by a factor of 2, to be consistent with the dust mass derived by the \cite{Draine2014} model (see \cite{Schreiber2017} for details); (8) Gas mass derived from Eq.~\ref{eq:gdr}; (9) Gas fraction defined by f$_{\text{gas}}$\,=\,M$_{\text{gas}}$/(M$_{\text{gas}}$\,+\,M$_\star$); (10) Dust temperature derived from a MBB model assuming $\beta$=1.5. $\dag$ indicates galaxies without a $Herschel$ counterpart and whose L$_{IR}$ is determined only by the ALMA contribution. For these galaxies, we show the mass of gas as an indication but we do not use it in the rest of this paper; (11) Flux density at 1.1\,mm. $^{\star}$ indicates changes in the flux density since \citetalias{Franco2018}. A summary of the fluxes (peak and integrated) as well as the sizes measured in the Main and Supplementary catalogs are given in Table~\ref{blobcat_uvmodelfit}.
\end{tablenotes}
\label{derived_properties}   
\end{threeparttable}
\end{table*}

\subsection{AGN}

Of the 1008 sources detected in X-ray during the 7\,Ms exposure survey of the \textit{Chandra} Deep Field--South presented in \cite{Luo2017}, 397 lie in the GOODS-ALMA field. We adopted a cross-matching radius of 0.6", after applying the offset corrections presented in  \citetalias{Franco2020}. We found that 13/23 (6/20) of our main (supplementary) catalog galaxies had matches with the \cite{Luo2017} catalog. However, the detection in X-rays is not definitive proof that a galaxy hosts an AGN. We corrected the \cite{Luo2017} cataloged X-ray luminosities when redshift deviations were observed, using the following formula:
\begin{equation}
L_X\,=\,4\pi d_L^2(1+z)^{\Gamma\,-\,2}f_X,
\end{equation}
\noindent and assuming a fixed $\Gamma$\,=\,2. In the following paragraphs, a galaxy will be considered as hosting an AGN if the galaxy has an X-ray luminosity  L$_{X,int}$\,$>$\,10$^{43}$\,erg\,s$^{-1}$ (luminous X-ray sources).

\section{The slow downfall of star formation in $z$\,=\,2-3 massive galaxies}
\label{sec:slow_death}
\subsection{A large fraction of galaxies in our sample with low gas fractions}

\begin{figure*}
   \centering
   \begin{minipage}[t]{1.0\textwidth}
\resizebox{\hsize}{!} {
\includegraphics[width=\hsize]{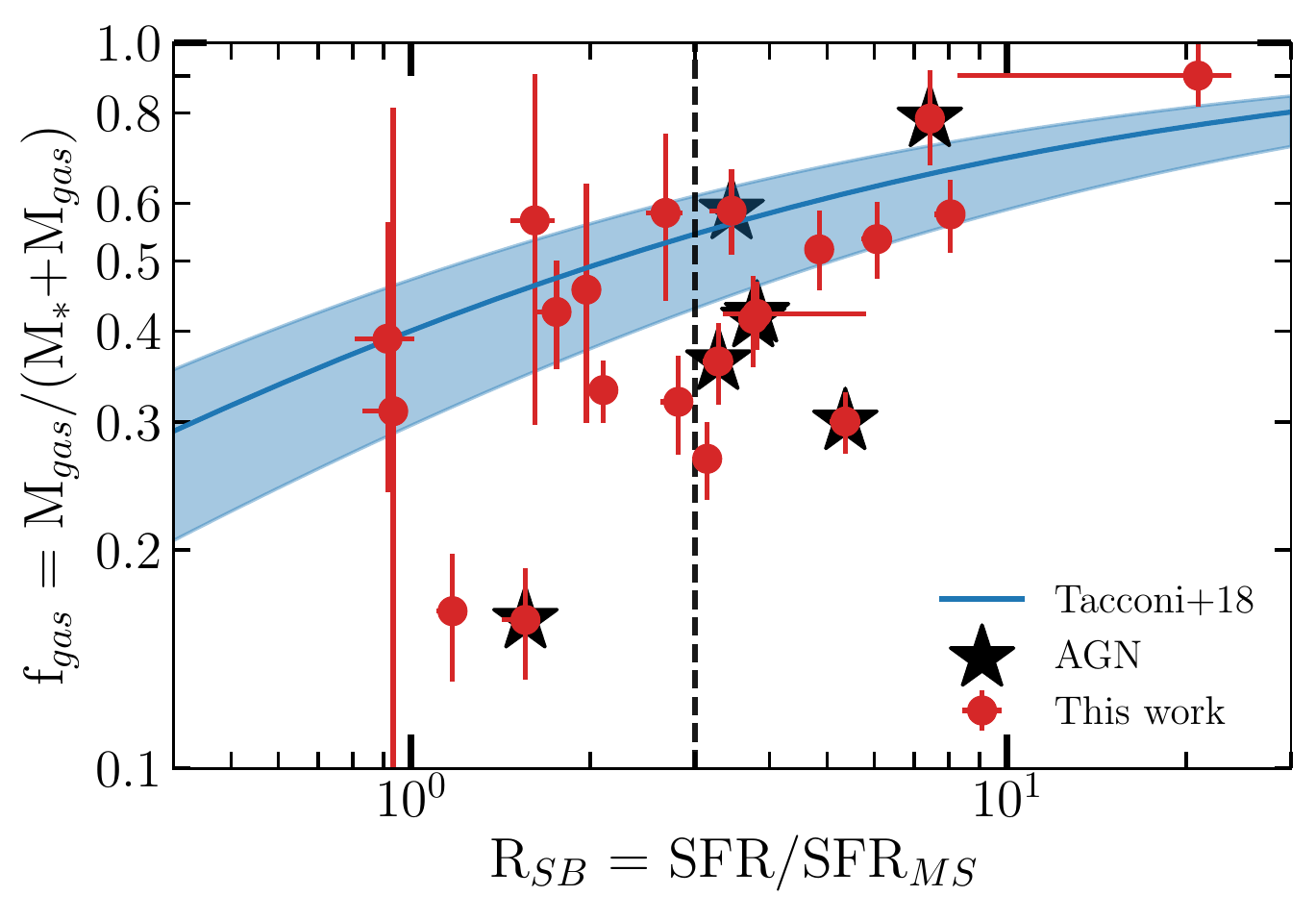}\\
\includegraphics[width=\hsize]{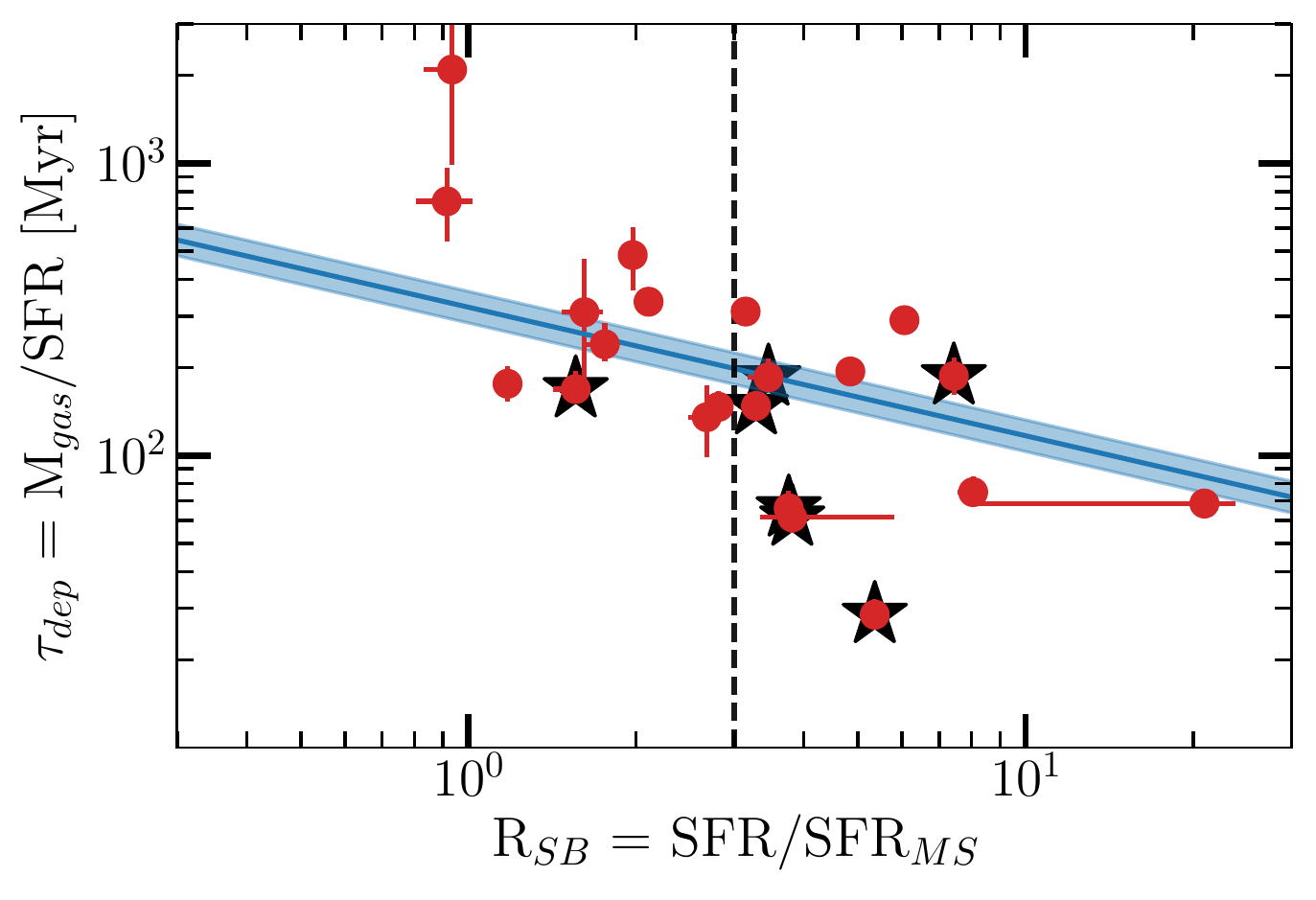}
}
\end{minipage}
      \caption{Evolution of the molecular gas fraction (f$_{\text{gas}}$) and the gas depletion timescale ($\tau_{dep}$) as a function of the distance to the main sequence of star-forming galaxies (R$_{SB}$\,=\,SFR/SFR$_{MS}$) for the main and supplementary catalog of galaxies detected by ALMA in the GOODS-ALMA field. The solid blue line shows the relation obtained by \cite{Tacconi2018} for the median redshift and stellar mass of our sample ($z_{med}$\,=2.7, M$_{\star, med}$\,=8.5\,$\times$\,10$^{10}$\,M$_\odot$). The uncertainty on the mean trend is obtained by Monte-Carlo simulations. In order to compare the gas fractions of all of the galaxies in our sample, we have rescaled our gas fractions according to the median redshift and the stellar mass of our sample.}
         \label{gas_fraction_RSB}
   \end{figure*}

In this survey, we have detected particularly massive galaxies, the majority of which are beyond cosmic noon at $z$\,$\sim$1-2. The study of the gas mass reservoirs is essential to understand how the galaxies will evolve with redshift and whether these galaxies could be the progenitors of passive galaxies at $z$\,$\sim$\,2. To obtain the most robust results possible, we have considered in the following section only galaxies with a \textit{Herschel} counterpart. The galaxies without a \textit{Herschel} counterpart are marked with $\dag$ in Table~\ref{derived_properties}. In Fig.~\ref{gas_fraction_RSB} (left panel), we compare the gas fraction of our galaxies as a function of their deviation from the MS, with the relationship presented in \cite{Tacconi2018}:
\begin{equation}
    M_{\text{gas}}/M_\star\,=\,\left[ 0.66^{+0.22}_{-23} \right] \times R_{SB}^{0.53}.
\end{equation}

In the same way, we compare the depletion time with the relationship presented in \cite{Tacconi2018}:

\begin{equation}
    \tau_{dep}\ [Myr]\,=\,\left[ 322^{+43}_{-38} \right] \times R_{SB}^{-0.44}.
\end{equation}

We have rescaled this relationship to correspond to the median redshift ($z_{med}$\,=\,2.7) and the median stellar mass of our sample (M$_{\star, med}$\,=\,8.5\,$\times$ 10$^{10}$M$_\odot$) by multiplying our results by $\tau_{dep}$($z$, M$_{\star}$)/$\tau_{dep}$($z_{med}$, M$_{\star,med}$). To be able to directly compare the gas fraction of our galaxies to the relationship of \cite{Tacconi2018}, we have also scaled our gas fractions according to the median redshift and stellar mass of our sample. The gas fractions, before rescaling, are presented in Table~\ref{derived_properties}.

The depletion times span a large range, between 30 and 1600\,Myr. The galaxies studied here show a dependence between depletion time and distance to the main sequence (R$_\text{SB})$, although very scattered (see Fig.~\ref{gas_fraction_RSB}). 

About half of the GOODS-ALMA galaxies follow the $f_{\text{gas}}$-$R_{SB}$ relation from (\citealt{Tacconi2018}, Eq.~20). 
However, we find a surprisingly large fraction (40\%) of galaxies lying well below this relation, that is, with excessively short depletion times (see Fig.~\ref{gas_fraction_RSB}). 
Galaxies lying below this relationship do not have a preferential redshift, we were detecting galaxies with redshifts close to the median redshift of our sample. On the other hand, the galaxies under this relation preferentially have stellar masses higher than the average of our sample with all but one exhibiting a stellar mass higher than 10$^{11}$M$_\odot$.  Moreover galaxies with a low gas fraction can either be on the main sequence or be starbursts.

The galaxies with the shortest depletion times are also those with the lowest gas fraction. This is because despite exhibiting lower gas masses, these galaxies keep forming stars with a high SFR.

We note that the majority of the ALMA galaxies experiencing a strong AGN episode with L$_X$\,$>$\,10$^{43}$\,erg\,sec$^{-1}$ lie below the $\tau_{dep}$-R$_{SB}$ and f$_{\text{gas}}$-R$_{SB}$ relations (stars in Fig.~\ref{gas_fraction_RSB}). This suggests that the low gas content and associated short depletion time of the galaxies may be due to the AGN feedback, heating the surrounding extragalactic medium and preventing further infall of gas. In other words, about half of the galaxies at these flux densities and redshifts appear to suffer from starvation and constitute excellent candidate progenitors of $z$\,$\simeq$\,2 massive and compact elliptical galaxies. To further investigate this possibility, we show in Sect.~\ref{sec:size} that the ALMA sizes, i.e., where the stars are formed, are consistent with the compact cores of $z$\,=\,2 elliptical galaxies.

 However, there is a trend between R$_{SB}$ and the stellar mass of the galaxies, in that the less massive galaxies in our sample have a larger average R$_{SB}$. This is partly due to a selection effect. A deeper survey would be needed to investigate the population of galaxies on the main sequence with intermediate stellar masses. We also investigated the evolution of the depletion time as a function of the stellar mass but we found no correlation. This means that the star-formation efficiency (SFE\,=\,SFR/M$_{gas}$\,=\,1/$\tau_{dep}$) does not change according to the stellar mass of the galaxy. 

The gas fractions cover a significant range of values, between f$_{gas}$\,=\,0.21 and 0.84, with a median of f$_{gas}$\,=\,0.52 (mean\,=\,0.52).
These values are consistent with other studies, such as \cite{Wiklind2014} for a sample with comparable stellar masses and redshifts. We do, however remark that for the two common galaxies between this work and \cite{Wiklind2014}, there is a significant difference in the calculated gas fractions. These two common galaxies are outliers from the rest of the \cite{Wiklind2014} sample as they have gas fractions close to unity, and in fact, correspond to two HST-dark galaxies that were previously falsely attributed with optical counterparts.

We note that a significant number of the outliers with low gas fractions are classified as AGN. The presence of an AGN can influence the measurement of the stellar mass of the galaxy and artificially lower the calculated gas fraction of the galaxies. This result is consistent with the findings of \cite{Perna2018} who found systematically low gas fractions in obscured AGN at $z$\,$>$\,$1$ and suggests that AGN feedback could lead to the expulsion of gas.
One of these galaxies has a low gas fraction (21\%) and does not show any sign of an AGN. This galaxy is a particularly striking example of interacting galaxies, with strong tidal tails. This galaxy does not have a high star formation rate, it lies on the MS, but it does display a starburst-like behavior since it exhibits a short gas depletion time. This galaxy could, therefore, be a member of the population of galaxies described in \cite{Elbaz2018}, a starburst galaxy hidden in the main sequence.

   \begin{figure*}
   \centering
   \begin{minipage}[t]{1.0\textwidth}
\resizebox{\hsize}{!} {
\includegraphics[width=0.49\hsize]{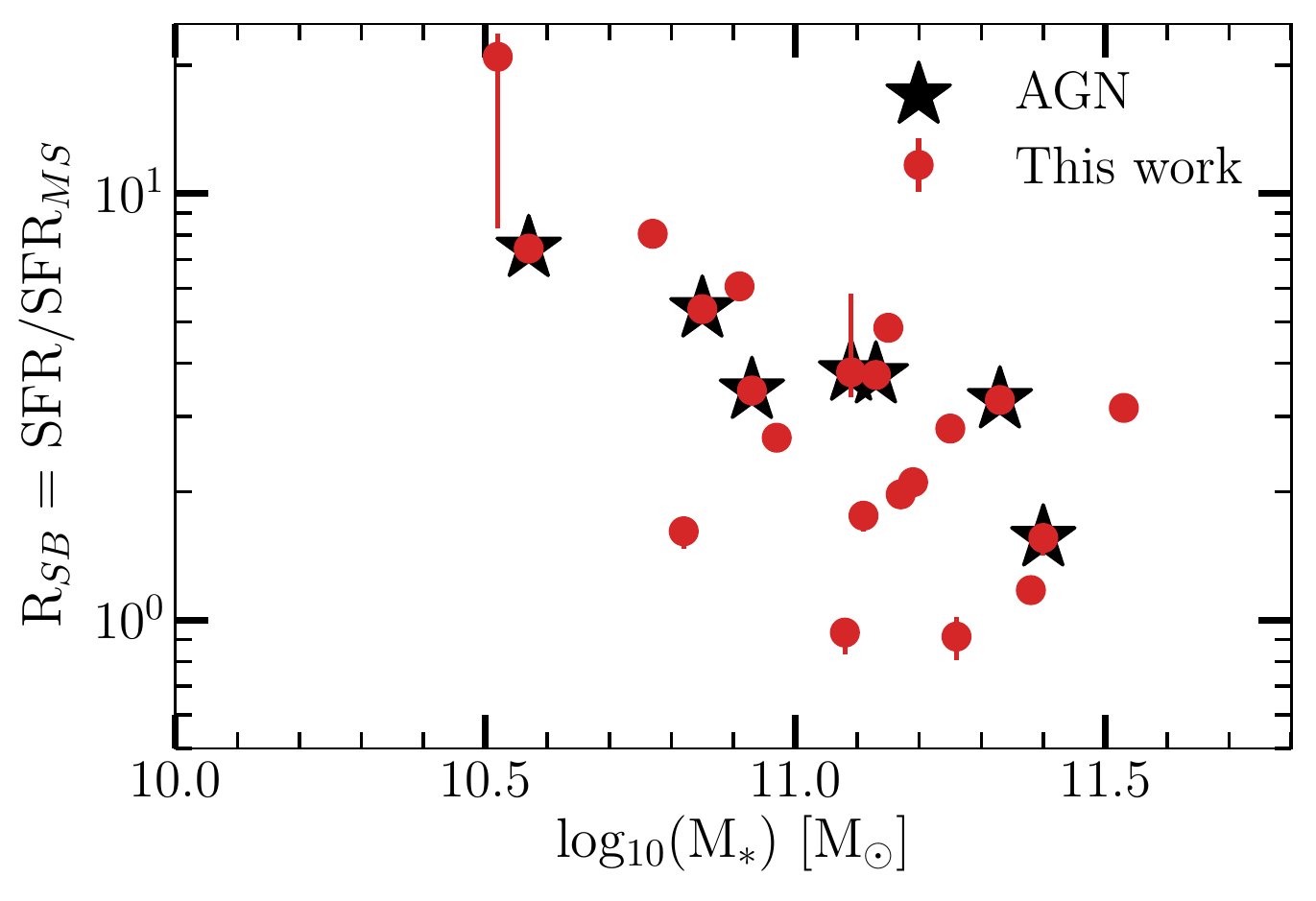}\\
\includegraphics[width=0.51\hsize]{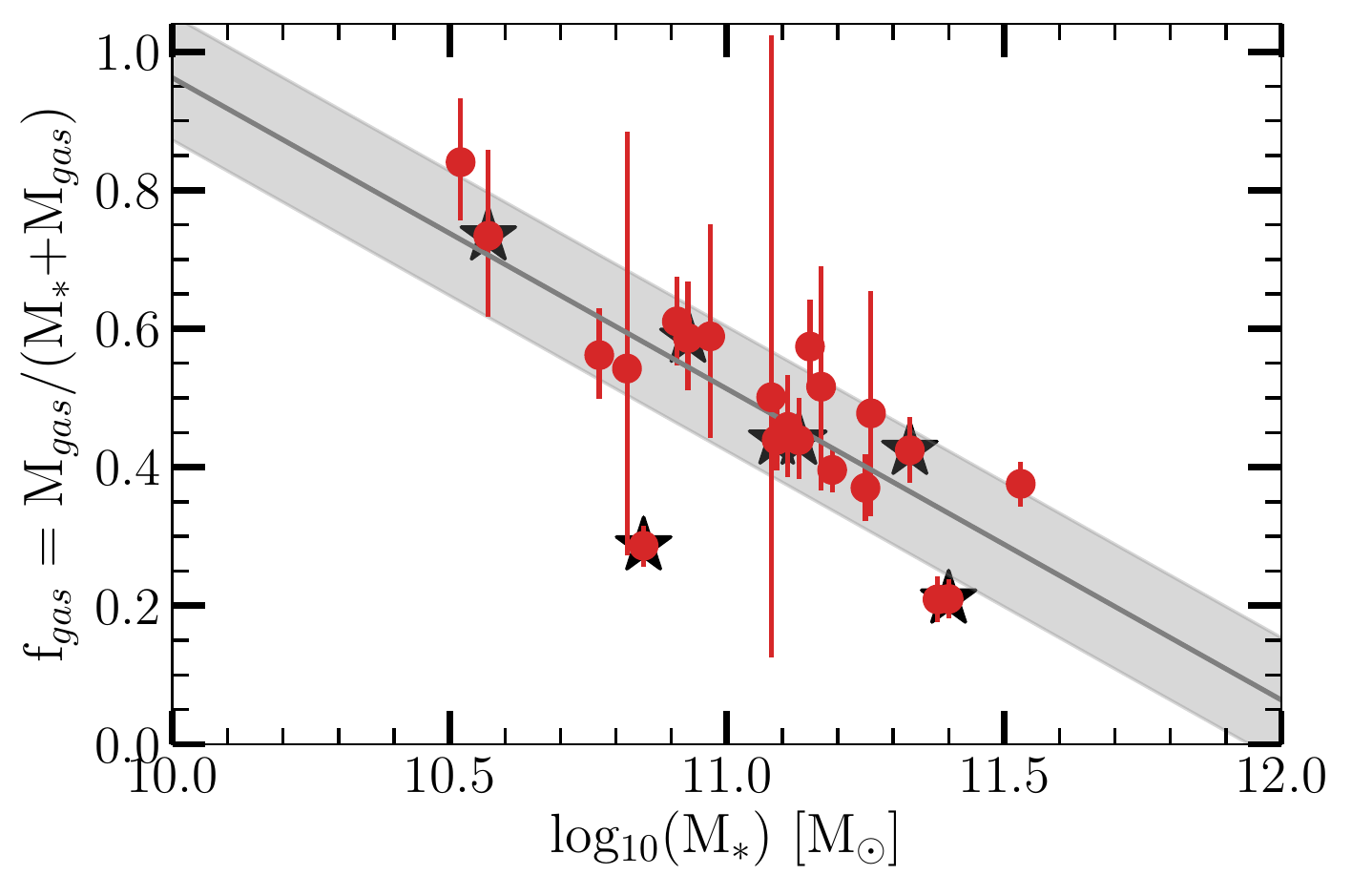}
}
\end{minipage}
      \caption{Evolution of the distance to the main sequence of star-forming galaxies (R$_{SB}$\,=\,SFR/SFR$_{MS}$, left panel) and the molecular gas fraction (f$_{gas}$, right panel) as a function of the stellar mass. The best fit, given by Eq.~\ref{gas_fraction_moi}, is shown by the gray shaded region}.
         \label{gas_fraction_Mass}
   \end{figure*}

We find a negative correlation between the stellar mass and the gas fraction (see Fig.~\ref{gas_fraction_Mass}, right panel). The following equation characterizes this relationship: 
\begin{equation}
    f_{gas}\,=\,(-0.45\pm0.09) \times \log_{10}(M_\star)\,+\,5.44\pm0.95.
    \label{gas_fraction_moi}
\end{equation}
A similar relationship has been found in other studies \cite[e.g.,][]{Popping2012,Magdis2012,Sargent2014,Schinnerer2016}. Galaxies hosting an AGN do not seem to occupy a particular position in Fig.~\ref{gas_fraction_Mass}. We also indicate in the left panel the distance to the main sequence as a function of the stellar mass. We can also see a clear negative correlation between the stellar mass and R$_{SB}$.
On the other hand, it is not possible to say whether selection effects are driving this trend. To be detected, a galaxy of low mass must have a larger R$_{SB}$ than a massive galaxy. On the other hand, we do not find massive galaxies (M$_\star$\,$>$\,10$^{11}$M$_\odot$ with R$_{SB}$\,$>$\,5). We remain cautious in our interpretation of this result. As the size of our survey is modest, selection bias may be significant.

We found no correlation between the depletion time and the stellar mass. This means that the star-formation efficiency (SFE\,=\,M$_{gas}$/SFR\,=\,1/$\tau_{dep}$) does not change according to the mass of the galaxy. In our sample, the efficiency for the galaxies to transform their gas into stars is independent of the stellar mass of the galaxy. The Pearson's correlation coefficient between these two quantities is 0.05, indicating the absence of a correlation.
Galaxies with the lowest gas fractions also appear to be the most massive, suggesting that we are witnessing a slow downfall of the galaxies with the most massive galaxies dying first to become elliptical galaxies, in a similar way to what has been shown in \cite{Schreiber2016}, but at higher redshifts. This is consistent with the idea of ``downsizing'' where the most massive are also the ones that form their stars the earliest and fastest \cite[e.g.,][]{Cowie1996, Guzman1997, Brinchmann2000, Neistein2006, Fontanot2009}.


\subsection{Toward a reduction in the size of galaxies}
\subsubsection{Size}
\label{sec:size}
Several studies have reported the observation of massive star-forming galaxies, compact at a rest-frame wavelength of 5000\,$\AA$ or in the $H$-band (e.g., blue nuggets; \citealt{Barro2013,Dekel2014}). It has been proposed that these galaxies are the progenitors of massive, compact and passive galaxies at $z$\,=\,2 \cite[e.g.,][]{Barro2013, Williams2014, Toft2014, Van_der_Wel2014, Barro2016, Kocevski2017}.

We have, thanks to the GOODS-ALMA survey, selected a sample of massive star-forming galaxies. These galaxies are among the most massive ones within the UVJ active -- in other words, star-forming -- galaxies (\citealt{Williams2009}, using the same definition as in \citetalias{Franco2018}) listed in the ZFOURGE catalog (see Fig.~10 in \citetalias{Franco2020}). For example, with ALMA we have detected the most massive ZFOURGE galaxy in the redshift range 1\,$<$\,$z$\,$<$\,2, the most massive galaxy at 2\,$<$\,$z$\,$<$\,3, the second most massive galaxy at 3\,$<$\,$z$\,$<$\,4.  These galaxies cannot continue to form stars for long periods. If this were the case, they would become much more massive than the most massive galaxies we have observed at $z$\,$\sim$\,1, or in the local universe.

The galaxies in the present paper have not been selected to be compact at a rest-frame wavelength of 5000\,$\AA$. They are flux-selected. Due to the low dispersion of the main sequence, this selection can be seen as a stellar mass selection. We aim to study here whether galaxies that have not been selected to be compact at 5000\,$\AA$ can also be the progenitors of compact galaxies at $z$\,$\simeq$\,2. To do this, we have compared the 5000\,$\AA$ sizes of the galaxies detected by ALMA with the 5000\,$\AA$ sizes of the UVJ active galaxies located in the area defined by the GOODS-ALMA survey and detected during the CANDELS program.

\begin{figure*}
   \centering
   \begin{minipage}[t]{1.\textwidth}
\resizebox{\hsize}{!} {
\includegraphics[width=4cm]{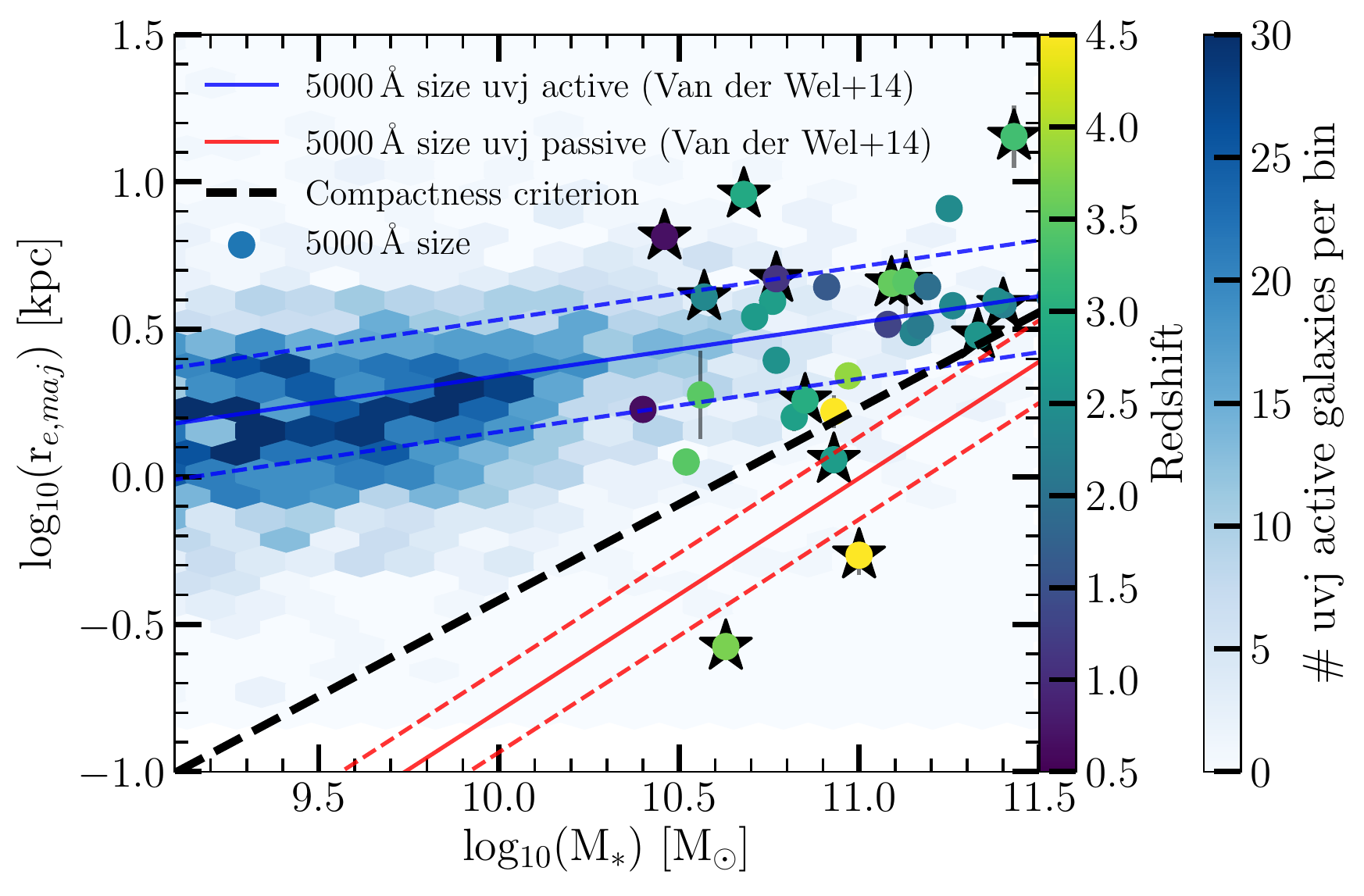}\\
\includegraphics[width=3.85cm]{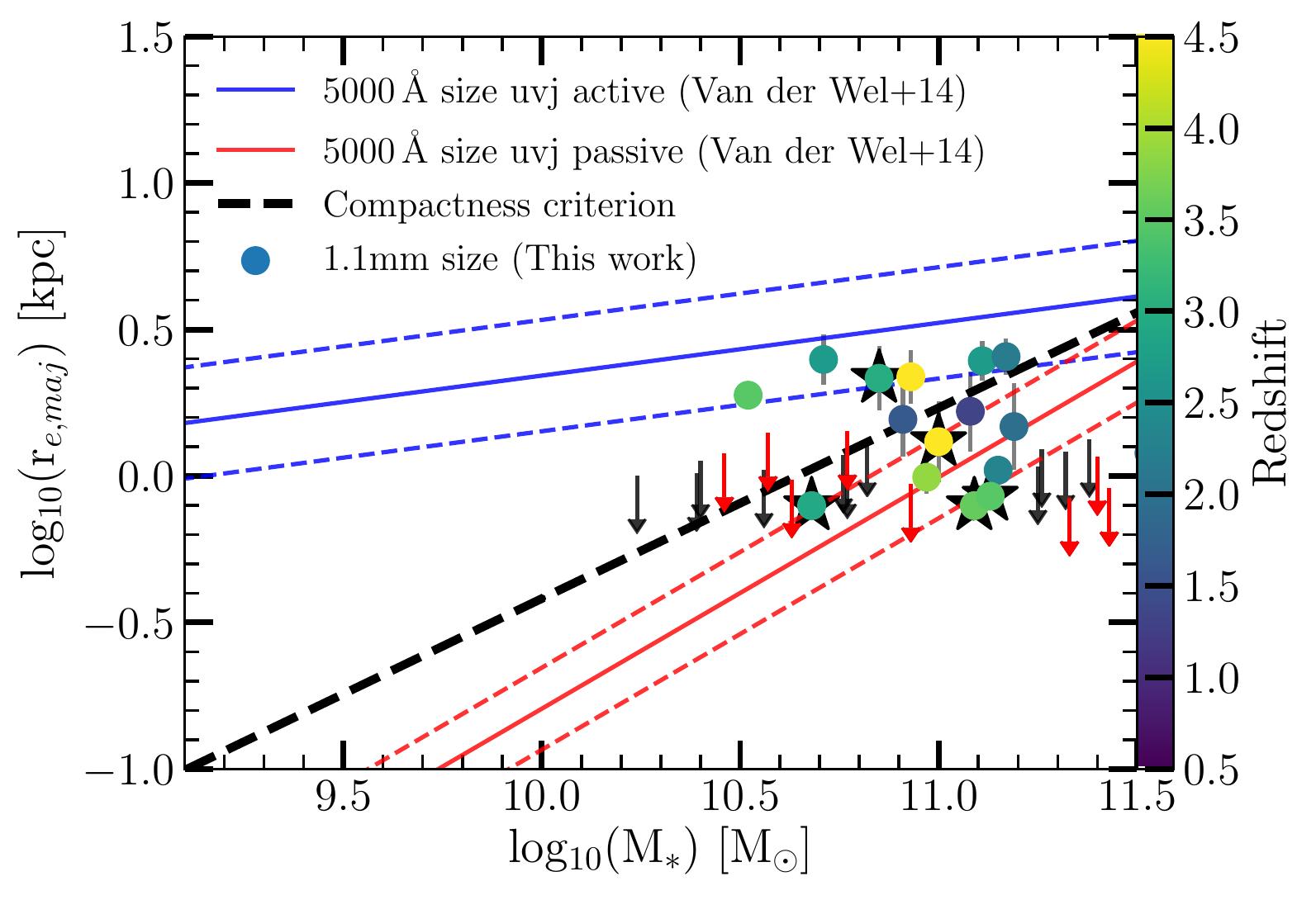}
}
\end{minipage}
\caption{Left panel: 5000\,$\AA$ size-mass plane for the galaxies located in the GOODS--South field for which sizes have been measured in \cite{Van_der_Wel2014}. The density of the UVJ active galaxies (with 2\,$<$\,$z$\,$<$\,4) in the GOODS-ALMA field is represented by the blue hexagons. The blue and red lines represent the trends of active and passive galaxies respectively, while the dashed lines give the scatter on these relations \citep{Van_der_Wel2014}. The ALMA-detected galaxies are shown with dots color-coded by redshift. The stars represent galaxies with L$_{X,int}$\,$>$\,10$^{43}$\,erg\,s$^{-1}$. Right panel: $ALMA$ size-mass plane for the ALMA detected galaxies. For comparison, the trends for active and passive galaxies are also shown. We indicate the compactness criterion described in Eq.~\ref{barro_criterion} to visualize which galaxies are compact at 5000\,$\AA$. In this Figure, ALMA sizes have been divided by a factor of $\sqrt{0.65}$, which corresponds to the median of the b/a ratio, to reflect size differences with HST. Galaxies with L$_{X,int}$\,$>$\,10$^{43}$\,erg\,s$^{-1}$ and for which only an upper limit in ALMA size has been determined are displayed with a red arrow.}
\label{size_mass_ALMA}
\end{figure*}

The majority of the galaxies studied in this paper have a redshift between $z$\,=\,2 and 4. We report in Fig.~\ref{size_mass_ALMA}-left panel, the 5000\,$\AA$ sizes of all galaxies within 2\,$<$\,$z$\,$<$\,4 located in the area defined by the GOODS-ALMA survey, as a function of stellar mass, in blue. We also show the 5000\,$\AA$ sizes of the ALMA-detected galaxies color-coded with redshift. Galaxy sizes and S\'ersic indices are obtained from \cite{Van_der_Wel2014}. These values have been computed by fitting a single-component S\'ersic profile using \texttt{GALFIT} \citep{Peng2010} at both 1.25 and 1.6 $\mu$m. Following \cite{Van_der_Wel2014}, the effective radius at 5000\,$\AA$ has been estimated as:
\begin{equation}\label{Eq::5000AA}
    r_e =
    \begin{cases}
      r_{\mathrm{e, F125W}}\left(\frac{1+z}{2.5}\right)^{-0.35+0.12 z-0.25 \log_{10}  \left(\frac{M_{\star}}{1.7\times10^{10} M_{\odot}}\right)}, & \text{if}\ z<1.5 \\
      r_{\mathrm{e, F160W}}\left(\frac{1+z}{3.2}\right)^{-0.35+0.12 z-0.25 \log_{10}    \left(\frac{M_{\star}}{1.7\times10^{10} M_{\odot}}\right)}, & \text{if}\ z>1.5
    \end{cases}
\end{equation}
\noindent where $r_{\mathrm{e, F125W}}$ and $r_{\mathrm{e, F160W}}$ are the effective radius though the F125W and F160W filter respectively.
We also show the trends for the UVJ active and UVJ passive galaxies with blue and red lines respectively. These two relations were parametrized by \mbox{\cite{Van_der_Wel2014}} following:
\begin{equation}
r_e =A\left[(\text{M}_{\star} / 5 \times 10^{10} ) / 1.7 \right]^{\alpha},
\end{equation}
\noindent where $r_e$ is the effective radius, in other words, the semi-major axis of the ellipse that contains half of the total flux of the best-fitting S\'ersic model, in kpc. We use the following parameters: log$_{10}(A)$\,=\,-0.06\,$\pm$\,0.03, $\alpha$\,=\,0.79\,$\pm$\,0.07, and the scatter in (r$_e$) in logarithmic units $\sigma$log$_{10}(r_e)$\,=\,0.14\,$\pm$\,0.03 for early-type galaxies and log$_{10}(A)$\,=\,0.51\,$\pm$\,0.01, $\alpha$\,=\,0.18\,$\pm$\,0.02, and $\sigma$log$_{10}(r_e)$\,=\,0.19\,$\pm$\,0.01 for late-type galaxies \citep{Van_der_Wel2014}. 

We see that there is a significant difference in size between active and quiescent galaxies. The size of star-forming galaxies is on average larger than passive galaxies. \cite{Mosleh2011} noted, for example, that  UV-bright galaxies with 10$^{10}$\,$<$\,M$_\star$/M$_\odot$\,$<$\,10$^{11}$ and 0.5\,$<$\,$z$\,$<$\,3.5 are larger than quiescent galaxies in the same mass and redshift range by 0.45\,$\pm$\,0.09 dex.

For the vast majority of the ALMA detected galaxies, their optical rest-frame sizes are comparable to the 5000\,$\AA$ sizes of the UVJ active galaxies (blue hexagons) at 2\,$<$\,$z$\,$<$\,4 selected in the same field of view. We also over-plot, in Fig.~\ref{size_mass_ALMA}, the compactness criterion given in \cite{Barro2013} and modified by \cite{Barro2016}:
\begin{equation}
\Sigma_{1.5}=\frac{\text{M}_\star}{r_{e}^{1.5}} \ge 10^{10.4}\,\text{M}_{\odot} \mathrm{kpc}^{-1.5}.
\label{barro_criterion}
\end{equation}
This relation initially defined for $H$-band sizes was then rescaled according to Eq.~\ref{Eq::5000AA} with a redshift equal to the median redshift of our sample of galaxies to make it correspond to a size at 5000\,$\AA$. Only three GOODS-ALMA galaxies are compact following the compactness criterion of Eq.~\ref{barro_criterion}. These galaxies lie on the trend for quiescent galaxies. We note that those galaxies that do not follow the trend of star-forming galaxies systematically host an AGN. If these galaxies suddenly stopped forming stars, they would already be located on the right trend in the mass-size diagram to be compact massive galaxies. With the data available to us, it is not possible to distinguish whether the compaction of the galaxy has triggered the AGN or, on the contrary, it is the presence of the AGN that has caused its compaction. We note that for galaxies with higher stellar mass (M$_\star$\,$>$\,10$^{11}$M$_\odot$), the star-forming and quiescent size-mass relations converge making it more difficult to investigate these trends.

We also show the ALMA 1.1\,mm sizes in comparison to the 5000\,$\AA$ sizes in Fig.~\ref{size_mass_ALMA}, right panel. The ALMA sizes for the main and supplementary catalogs are given in \citetalias{Franco2020} and in \ref{blobcat_uvmodelfit}. The size distribution differs slightly between the two samples. We showed in \citetalias{Franco2018} that we were biased toward compact sources with our detection limit of $4.8\sigma$. By lowering this detection threshold in the supplementary catalog, which was made possible as a result of basing our detections on IRAC and VLA detections, we are now detecting galaxies with larger ALMA sizes.
For the 26 galaxies for which we have both HST 5000\,$\AA$ sizes and could measure a 1.1\,mm size with ALMA, we find that ALMA sizes are generally smaller, with a median $r_{e,\text{HST}}$/$r_{e,\text{ALMA}}$\,=\,2.3. This ratio is significantly higher than the ratio of 1.4 found by \cite{Fujimoto2017} at 870\,$\mu$m, for a sample of 1034 ALMA sources.

Considering that dust emission is a good indicator of dust-obscured star formation, this result indicates that compact dust-obscured star formation (at least more compact than optical emission) is taking place in the core of the galaxies studied here. This study confirms the comparison of optical and millimeter sizes performed at $z$\,$\simeq$\,1.3 by \cite{Puglisi2019}, and extends it to higher redshifts, at and before the epoch of the peak of cosmic star formation.

For these galaxies to be the progenitors of compact elliptical galaxies at $z$\,$\simeq$\,2, they need to become more compact than their 5000\,$\AA$ size. The observed strong star formation activity concentrated in a small region of the galaxy can morphologically transform a galaxy into a more compact object. Assuming that there is no addition of gas, the majority of these galaxies have gas reservoirs equal to or close to their stellar mass. If this gas is transformed into stars in the compact emission region detected by ALMA, these galaxies will become compact and gradually migrate into the location of the mass-size diagram reserved for passive galaxies.

The ALMA galaxies presented here exhibit a present S\'ersic index in the $H$-band of $<$$n$$_\text{AGS}$$>$\,=\,1.63. 
We have seen that the amount of star formation associated to the compact 1.1\,mm emission is large enough to bring the half-light radius of the ALMA galaxies on top of the one expected for passive compact galaxies at $z$\,$\sim$\,2, hence the question that remains to be answered is whether this evolution will also be accompanied with an increase of the S\'ersic index that will bring them closer to the one observed for passive compact galaxies, i.e., increasing from $n$\,=\,1.6 to $n$\,=\,2.6. To answer this question, we would need to know with enough accuracy what is the actual S\'ersic index of the ALMA sources in the 1.1\,mm band. Unfortunately our resolution and depth are not sufficient to derive a S\'ersic index for the dust emission, hence we cannot answer the question without supplementary information. We note however, that at least some of the ALMA sources may present S\'ersic indices similar to those measured by \cite{Hodge2016}, \cite{Elbaz2018} and \cite{Rujopakarn2019} who measured S\'ersic indices close to $n$\,$\sim$\,1 for galaxies with similar behavior in term of stellar mass, SFR and redshift. A simple model of the impact of the newly formed stars following such an index to the final stellar distribution of the galaxies suggests that they would remain below $n$\,=\,2.6. Hence we conclude that despite the fact that the ALMA galaxies will inevitably have compact final half-light radii, only a fraction of them will end up showing the high S\'ersic index of the compact ellipticals observed at $z$\,$\sim$\,2. We note that this index itself presents a distribution, hence we cannot reject the possibility that most of the present ALMA sources represent reliable progenitors of compact ellipticals at $z$\,$\sim$\,2. This galaxy sample is building compact bulges. This is consistent with a scenario of rapid bulge growth in galaxies and that these bulges can already be in place at $z$\,$\sim$\,2 \citep{Tacchella2015}.

\subsubsection{Morphology}
We here aim to look at the mechanisms that may have driven the gas in the center of the ALMA galaxies. This may be violent disk instabilities \citep{Dekel2014}, or other dissipative processes, including mergers \citep{Wellons2015}. To investigate the role of mergers in the compaction process, we now investigate the morphology of the ALMA-detected galaxies.

Increasing numbers of observations have demonstrated that elliptical galaxies at $z$\,=\,2 are particularly compact \cite[e.g.,][]{Trujillo2006, vanDokkum2008, Conselice2014, Van_der_Wel2014}. Major merger events can give rise to elliptical galaxies \cite[e.g.,][]{Dekel2006,Hopkins2006}, but can also influence the compactness of the star formation in galaxies \cite[e.g.,][]{Wuyts2010, Ceverino2015}. Due to their large stellar masses, which has generated and retained a large amount of metals, and hence dust, against outflows \citep[e.g.,][]{Dekel1986, Dekel2003, Tremonti2004}, the galaxies detected in this study are extremely dust-obscured. In addition to this, their redshift makes them particularly faint in UV and optical filters. Some of them are $Y$-dropout (e.g., AGS5, AGS18), $V$-dropouts (e.g., AGS9, AGS10) or visible only in the $Ks$-band (AGS4, AGS11, etc.). The morphology of these galaxies is therefore difficult to obtain.  We cross-matched our sample with the catalog of \cite{Huertas-Company2015} that estimates the probability of being a spheroid, disk or irregular using the Convolutional Neural Network technique. In addition to the 6 HST-dark galaxies, which, by definition, cannot be categorized, nine other galaxies have $H$-band fluxes too faint to be classified (F160W\,$>$\,24.5 AB mag). This leaves only 20 of our galaxies that are present in this catalog. We use the simplified classification proposed in \cite{Huertas-Company2015a}:

\begin{itemize}
\item pure bulges: f$_{sph}$\,$>$\,2/3 AND f$_{disk}$\,$<$\,2/3 AND f$_{irr}$\,$<$\,1/10;
\item pure disks: f$_{sph}$\,$<$\,2/3 AND f$_{disk}$\,$>$\,2/3 AND f$_{irr}$\,$<$\,1/10;
\item disk+sph: f$_{sph}$\,$>$\,2/3 AND f$_{disk}$\,$>$\,2/3 AND f$_{irr}$\,$<$\,1/10;
\item irregular disks: f$_{disk}$\,$>$\,2/3 AND f$_{sph}$\,$<$\,2/3 AND f$_{irr}$\,$>$\,1/10;
\item irregulars/mergers: f$_{disk}$\,$<$\,2/3 AND f$_{sph}$\,$<$\,2/3 AND f$_{irr}$\,$>$\,1/10.
\end{itemize}

As a result, 61\% (11/18) of our galaxies are classified as irregulars/mergers (two galaxies do not fit into any of the categories presented above). If we also take into account irregular disks, 78\% (14/18) have an irregular morphology. Several galaxies show clear morphological characteristics of mergers, for example with large tidal tails. The galaxy AGS31, which exhibits large tidal tails, is an excellent illustration of this (see Appendix A in \citetalias{Franco2020}). For other galaxies, the interaction with another galaxy is more weak or uncertain. 

We compared these proportions against a control sample. We have for each of the 18 galaxies with estimated morphologies from the \cite{Huertas-Company2015} catalog, a galaxy closest to it in terms of redshift and stellar mass. This control sample exhibits significantly different morphological proportions. Only 6\% (1/18) of these galaxies can be classified as irregulars/mergers, 22\% (4/18) if we take into account irregular disks. The galaxy population detected by ALMA, therefore, tends to be on average biased toward irregular galaxies. By more precisely considering the morphological classification, we obtain for the sample galaxies detected by ALMA an average f$_{sph}$\,=\,0.16, f$_{disk}$\,=\,0.50, f$_{irr}$\,=\,0.34, while for the control sample, an average of f$_{sph}$\,=\,0.40, f$_{disk}$\,=\,0.53, f$_{irr}$\,=\,0.07. While the disk fraction is relatively constant between these two samples, we are witnessing an inversion of the fraction between the irregulars and the spheroids. This result is consistent with a scenario in which a gas-rich major merger could funnel the gas into the center \cite[e.g.,][]{Barnes1991, Hopkins2006b, Dekel2014, Zolotov2015,Tacchella2016}. This would explain the high fraction of irregular galaxies compared to the control sample.

We are therefore in the presence of a heterogeneous population of both secularly evolving disk and merger-type galaxies. The number of galaxies classified as irregulars/mergers is slightly higher with that found by models \citep{Hayward2011, Hayward2013}, which predict that for a population of SMGs with S$_{1.1mm}$\,$>$\,0.5\,mJy, star-forming galaxy-pairs account for $\sim$30-50 percent of the galaxies.

\subsection{IR surface brightness as a prior for the remaining lifetime of a galaxy}

\begin{figure}
   \centering
   \includegraphics[width=1.\hsize]{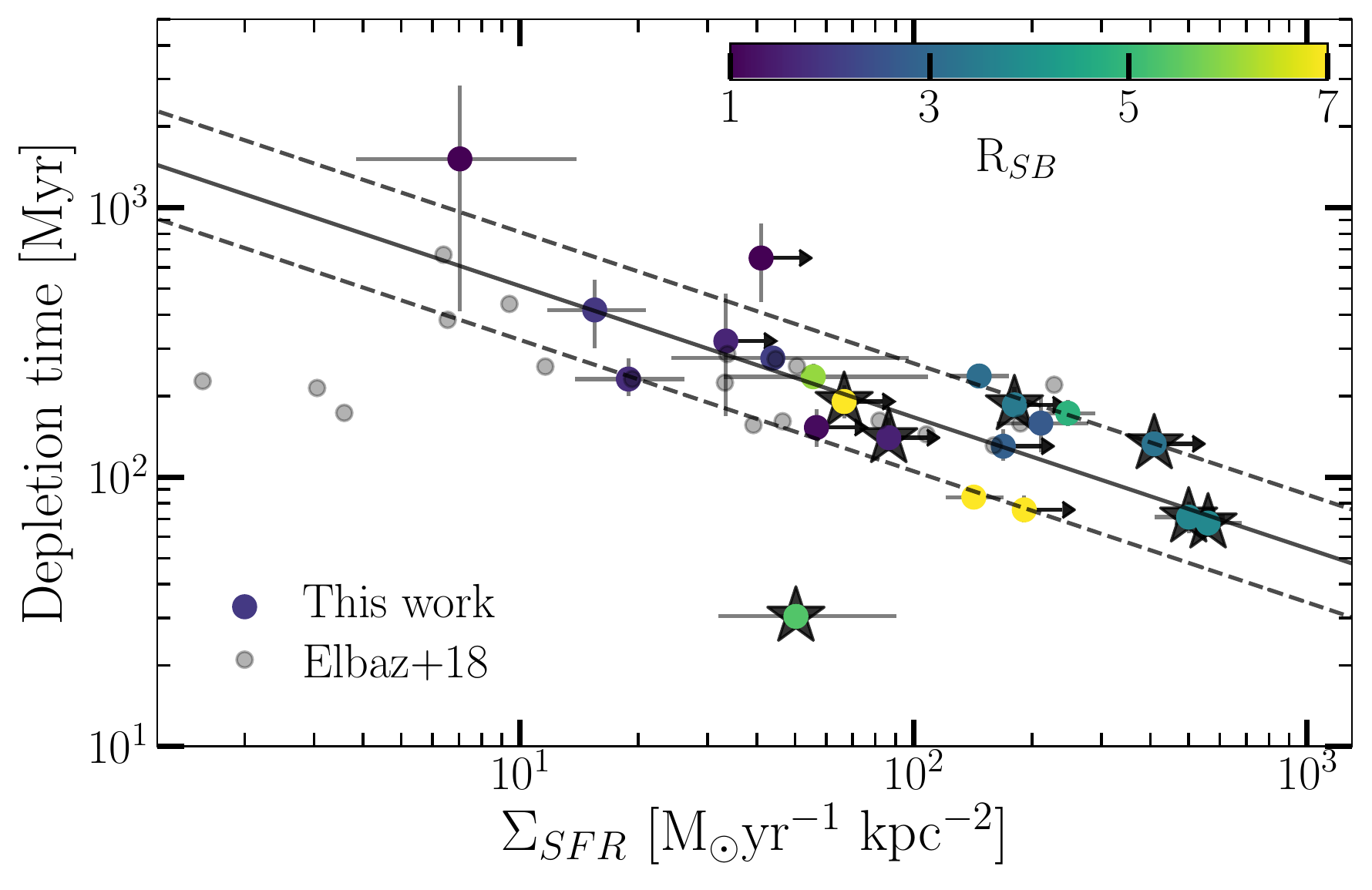}
      \caption{Depletion time as a function of the $\Sigma_{SFR}$, color-coded according to the distance to the main sequence. The solid and dashed lines are the fit to the sliding median and its 68\% scatter respectively. The stars represent galaxies with L$_{X,int}$\,$>$\,10$^{43}$\,erg\,s$^{-1}$. For comparison, the results of \cite{Elbaz2018} are shown by gray dots.}
         \label{Depletion_time}
   \end{figure}

The role of compact star formation in enhancing the efficiency of star formation is illustrated in Fig.~\ref{Depletion_time}. Galaxies forming stars with the largest star-formation surface density, $\Sigma_{SFR}$, experience the strongest star formation episodes with the shortest depletion times (see Table~\ref{Table:size}).

The SFR surface density ($\Sigma_{SFR}$) can be defined as:
\begin{equation}
\Sigma_{SFR}\,=\,\text{SFR}/(2\pi R_{1.1mm}^2 ),
\end{equation}
\noindent where $R_{1.1mm}$ is the half light radius (see Sect.~\ref{sec:size} for a description of the determination of the millimeter size).

We have found a strong negative correlation between $\Sigma_{SFR}$ and depletion time (see Fig.~\ref{Depletion_time}). A similar trend was found in \cite{Elbaz2018}. This correlation can be characterized by the following equation:

\begin{equation}
\tau_{dep} [Myr]\,=\,10^{(3.20\pm0.25)} \times \Sigma_{SFR}^{(-0.49\pm0.12)}.
\end{equation}

\begin{table}
\centering       
\caption{Size, depletion time, and star-formation surface density.}
\begin{threeparttable}
\begin{tabular}{r c c c}     
\hline       
 ID &   FWHM  &  $\tau_{dep}$ & $\Sigma_{SFR}$  \\
      &  arcsec    & Myr & M$_{\odot}$yr$^{-1}$kpc$^{-2}$ \\
\hline  
\hline
  AGS1          &   0.21 $\pm$   0.02         &   172$_{-   18}^{+   19}$           &   246$_{-   35}^{+   42}$ \\
  AGS3          &$<$0.17                      &   132$_{-   14}^{+   16}$           &$>$408                     \\
  AGS4          &   0.18 $\pm$   0.02         &    67$_{-    7}^{+    7}$           &   560$_{-   98}^{+  117}$ \\
  AGS5          &   0.19 $\pm$   0.02         &    71$_{-    9}^{+    9}$           &   500$_{-   93}^{+  120}$ \\
  AGS6          &$<$0.19                      &   185$_{-   19}^{+   29}$           &$>$180                     \\
  AGS8          &   0.23 $\pm$   0.02         &   237$_{-   22}^{+   24}$           &   146$_{-   23}^{+   28}$ \\
  AGS9          &   0.23 $\pm$   0.03         &   158$_{-   36}^{+   38}$           &   210$_{-   46}^{+   66}$ \\
 AGS10          &$<$0.21                      &   130$_{-   15}^{+   20}$           &$>$169                     \\
 AGS12          &$<$0.23                      &    75$_{-    7}^{+   10}$           &$>$191                     \\
 AGS13          &$<$0.23                      &   139$_{-   15}^{+   26}$           &$>$ 86                     \\
 AGS17          &   0.41 $\pm$   0.03         &    84$_{-    7}^{+    8}$           &   142$_{-   22}^{+   27}$ \\
 AGS18          &   0.50 $\pm$   0.08         &   231$_{-   30}^{+   44}$           &    18$_{-    5}^{+    7}$ \\
 AGS23          &$<$0.24                      &   650$_{-  203}^{+  222}$           &$>$ 41                     \\
 AGS26          &   0.30 $\pm$   0.09         &   235$_{-   23}^{+   26}$           &    55$_{-   22}^{+   53}$ \\
 AGS28          &   0.50 $\pm$   0.07         &   416$_{-  117}^{+  120}$           &    15$_{-    3}^{+    5}$ \\
 AGS29          &$<$0.28                      &   447$_{-   85}^{+  173}$           &$>$ 19                     \\
 AGS31          &$<$0.27                      &   152$_{-   23}^{+   25}$           &$>$ 56                     \\
 AGS34          &$<$0.27                      &   319$_{-  149}^{+  162}$           &$>$ 33                     \\
 AGS35          &   0.45 $\pm$   0.12         &    30$_{-    3}^{+    3}$           &    50$_{-   18}^{+   38}$ \\
 AGS37          &   0.28 $\pm$   0.10         &   276$_{-   24}^{+   27}$           &    43$_{-   19}^{+   57}$ \\
 AGS38          &   0.32 $\pm$   0.10         &  1516$_{- 1103}^{+ 1256}$           &     7$_{-    3}^{+    7}$ \\
 AGS39          &$<$0.28                      &   190$_{-   26}^{+   29}$           &$>$ 66                     \\
\hline
\end{tabular}
\end{threeparttable}
\begin{tablenotes}
\item Columns:  (1) ALMA ID for galaxies with $Herschel$ measurements; (2) FWHM measured from uvmodelfit in \texttt{CASA}; (3) Depletion time ($\tau_{dep}$\,=\,M$_\text{gas}$/SFR), in Myr; (4) SFR surface density ($\Sigma_{SFR}$\,=\,\text{SFR}/(2$\pi$ R$_{1.1mm}^2$)), in M$_{\odot}$yr$^{-1}$kpc$^{-2}$.
\end{tablenotes}
\label{Table:size}   
\end{table}

\section{Conclusions}

We have taken advantage of the excellent multiwavelength supporting data in the GOODS--South field and the largest contiguous ALMA survey to derive the physical properties of 35 ALMA flux-selected galaxies. This sample of galaxies comes both from a purely blind search (galaxies with a peak flux $>\,$4.8\,$\sigma$, see \citetalias{Franco2018}) and from an extension of this catalog that we have built down to the 3.5\,$\sigma$ limit using IRAC and VLA to probe fainter millimeter galaxies (\citetalias{Franco2020}). 
The number of galaxies comprising our sample is comparable to the expected number of galaxies  reported  in  the  number  counts  literature  at  similar  flux  and  wavelengths  to  those  of  our  study  \cite[e.g.,][]{Hatsukade2013,Oteo2015,Aravena2016,Umehata2017,Fujimoto2017,Dunlop2017,Franco2018,Hatsukade2018, Gonzalez-Lopez2020, Popping2020}  indicating that our sample of galaxies is almost complete.

These galaxies are massive (M$_{\star,med}$\,=\,8.5\,$\times$\,10$^{10}$\,M$_\odot$) and therefore rare, so in order to be able to detect and analyze them, a sufficiently large survey, such as GOODS-ALMA was needed. It is possible now, for the first time with this survey, covering $\sim$69\, arcmin$^2$. The analysis of the SEDs of these galaxies has made it possible to derive some of the physical properties of these galaxies. We are confronted with a heterogeneous population of galaxies. However, we highlight that about 40\% of our galaxy sample exhibits a particularly small gas fraction. We remark that the most massive galaxies in our sample are also the galaxies with the lowest gas fractions. With their high star formation rates (the galaxies are mostly starbursts, or on the upper part of the main sequence) and without a gas refill mechanism, they will consume their gas reservoirs in a typical time of 100~Myr.

We also studied the sizes of these galaxies. The advantage of conducting a survey is that it does not impose a priori criteria for selecting the galaxies studied. The ALMA detected galaxies have observed $H$-band or 5000\,$\AA$ sizes comparable to the majority of galaxies with the same stellar masses and redshifts, whereas their dust emission regions, i.e., the regions tracing the obscured part of the star formation, are relatively compact and have sizes comparable to passive galaxies at $z$\,$\sim$\,2.

We have investigated the link between depletion time and star formation surface density. We confirm the result showing a tight correlation between these two quantities. The denser the galaxy star-forming region is, the shorter the gas depletion time is. Mechanisms leading to a compaction of the obscured star-forming regions are to be confirmed, but a compact region massively forming stars at the center of a galaxy can lead to a rapid morphological transition from a spiral to a compact elliptical galaxy such as those observed at $z$\,$\sim$\,2, despite the fact that the ALMA selected galaxies are not yet compact at 5000\,$\AA$ or in the $H$-band (they are not yet blue nuggets). 

All of these different pieces of evidence indicate that our ALMA-detected galaxies are the ideal progenitors of passive galaxies at $z$\,$\sim$\,2 and natural exhaustion of their gas reservoirs  (slow downfall) is sufficient for this transition to happen quickly without needing to invoke a (``fast'') quenching mechanism. The large fraction of AGN among galaxies with the shortest depletion times and gas fractions suggest however that they may act by a starvation mechanism in preventing any further growth.

\section{Acknowledgements}

  We thank the anonymous referee for the insightful comments and suggestions that improved the clarity and quality of this work. M.F. acknowledges support from the UK Science and Technology Facilities Council (STFC) (grant number ST/R000905/1). B.M. acknowledges support from the Collaborative Research Centre 956, sub-project A1, funded by the Deutsche Forschungsgemeinschaft (DFG) -- project ID 184018867. L.Z. acknowledges the support from the National Key R\&D Program of China (No. 2017YFA0402704, No. 2018YFA0404502),  the National Natural Science Foundation of China (NSFC grants 11825302, 11733002 and 11773013) and China Scholarship Council (CSC). R.D. gratefully acknowledges support from the Chilean Centro de Excelencia en Astrof\'isica y Tecnolog\'ias Afines (CATA) BASAL grant AFB-170002". GEM acknowledges support from  the Villum Fonden research grant 13160 “Gas to stars, stars to dust: tracing star formation across cosmic time”, the Cosmic Dawn Center of Excellence funded by the Danish National Research Foundation and the ERC Consolidator Grant funding scheme (project ConTExt, grant number No. 648179). MP is supported by the ERC-StG 'ClustersXCosmo', grant agreement 71676. DMA acknowledges support from the Science and Technology Facilities Council (ST/P000541/1; ST/T000244/1). This work was supported by the Programme National Cosmology et Galaxies (PNCG) of CNRS/INSU with INP and IN2P3, co-funded by CEA and CNES. This paper makes use of the following ALMA data: ADS/JAO.ALMA\#2015.1.00543.S. ALMA is a partnership of ESO (representing its member states), NSF (USA) and NINS (Japan), together with NRC (Canada), MOST and ASIAA (Taiwan), and KASI (Republic of Korea), in cooperation with the Republic of Chile. The Joint ALMA Observatory is operated by ESO, AUI/NRAO and NAOJ. 
  
\bibliographystyle{aa}
\bibliography{biblio}
\onecolumn
\begin{appendix}

\section{SEDs}

\begin{figure*}[h!]
\centering
\begin{minipage}[t]{.96\textwidth}
\resizebox{\hsize}{!} { 
\includegraphics[width=3cm,clip]{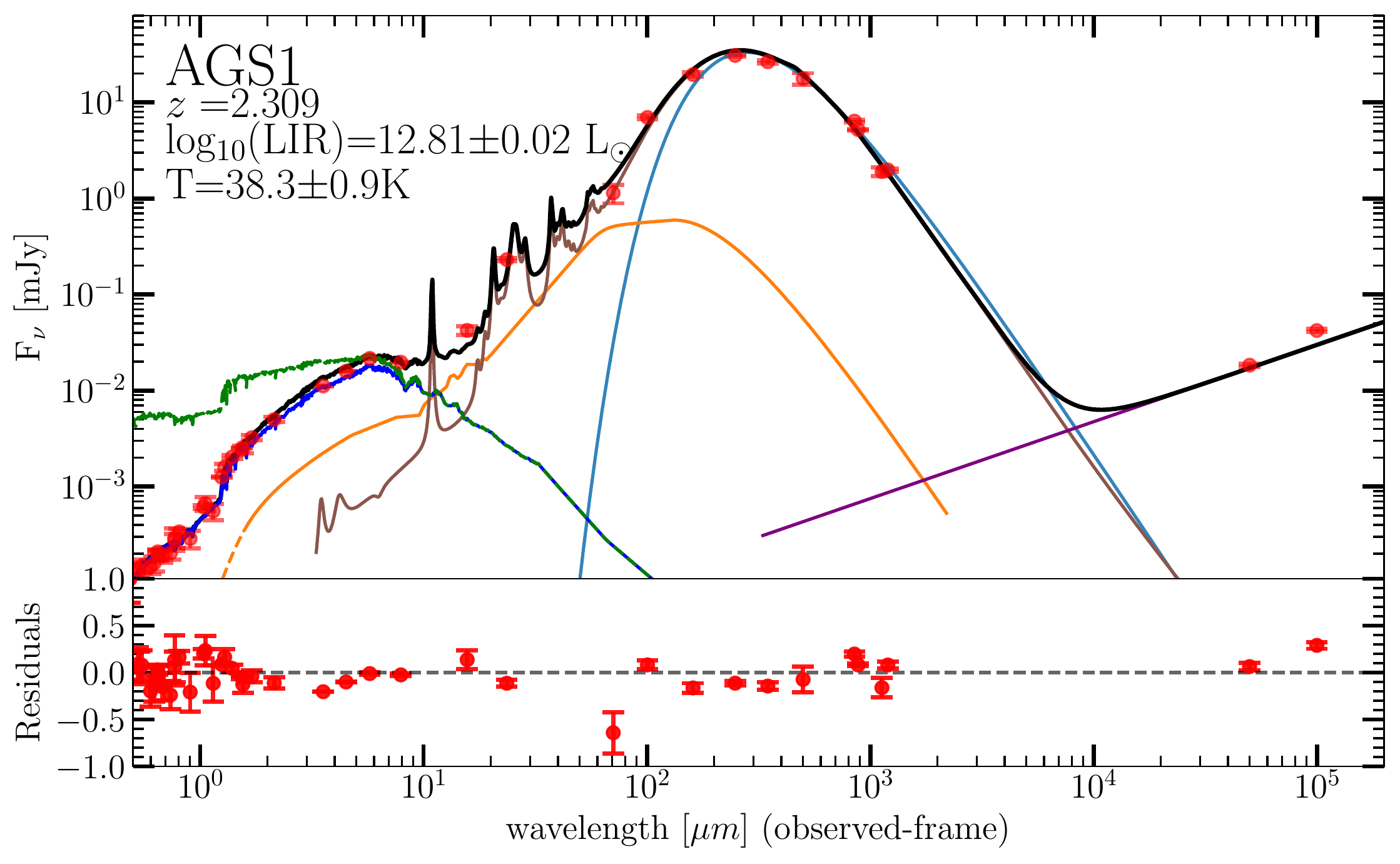} 
\includegraphics[width=3cm,clip]{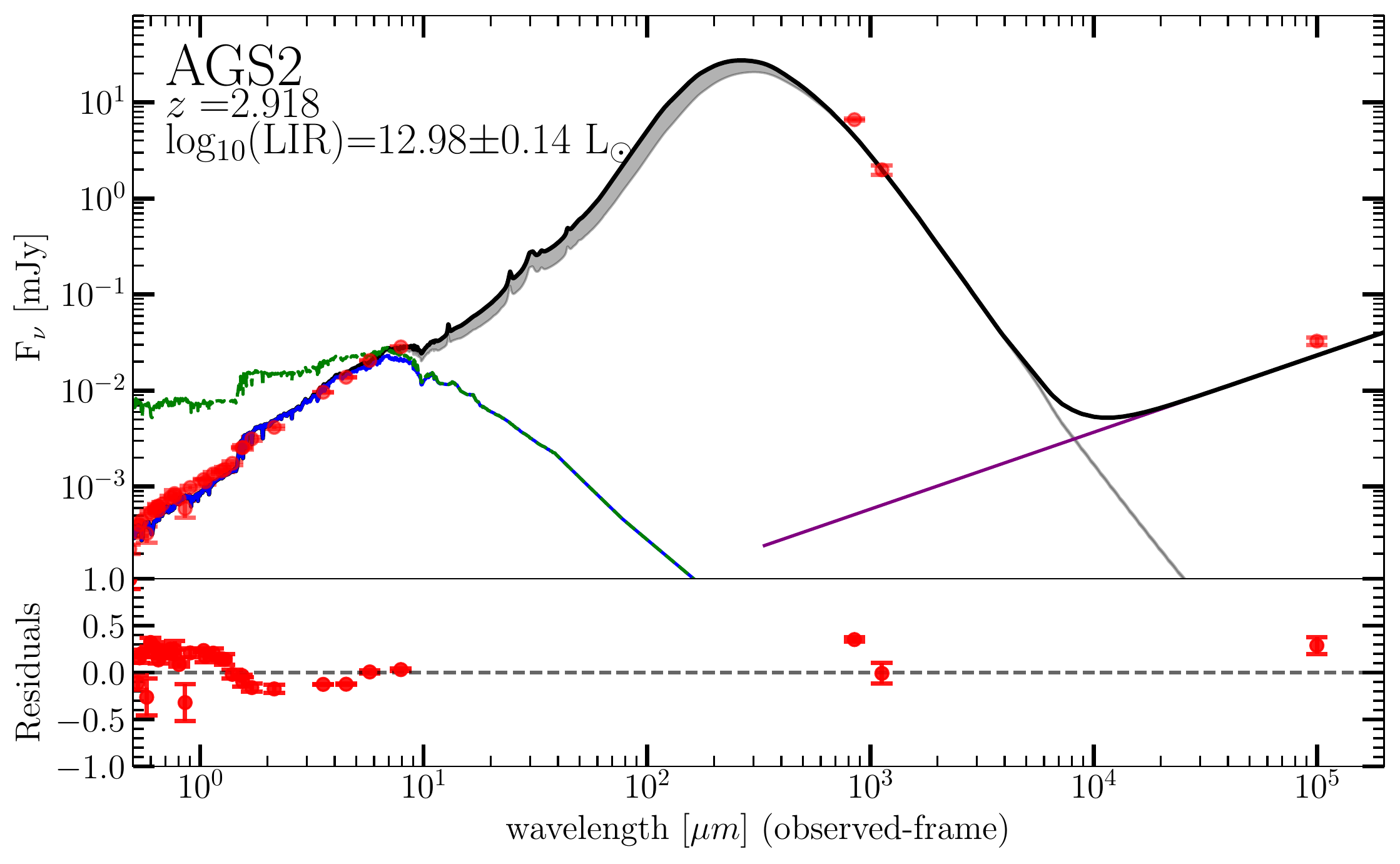} 
\includegraphics[width=3cm,clip]{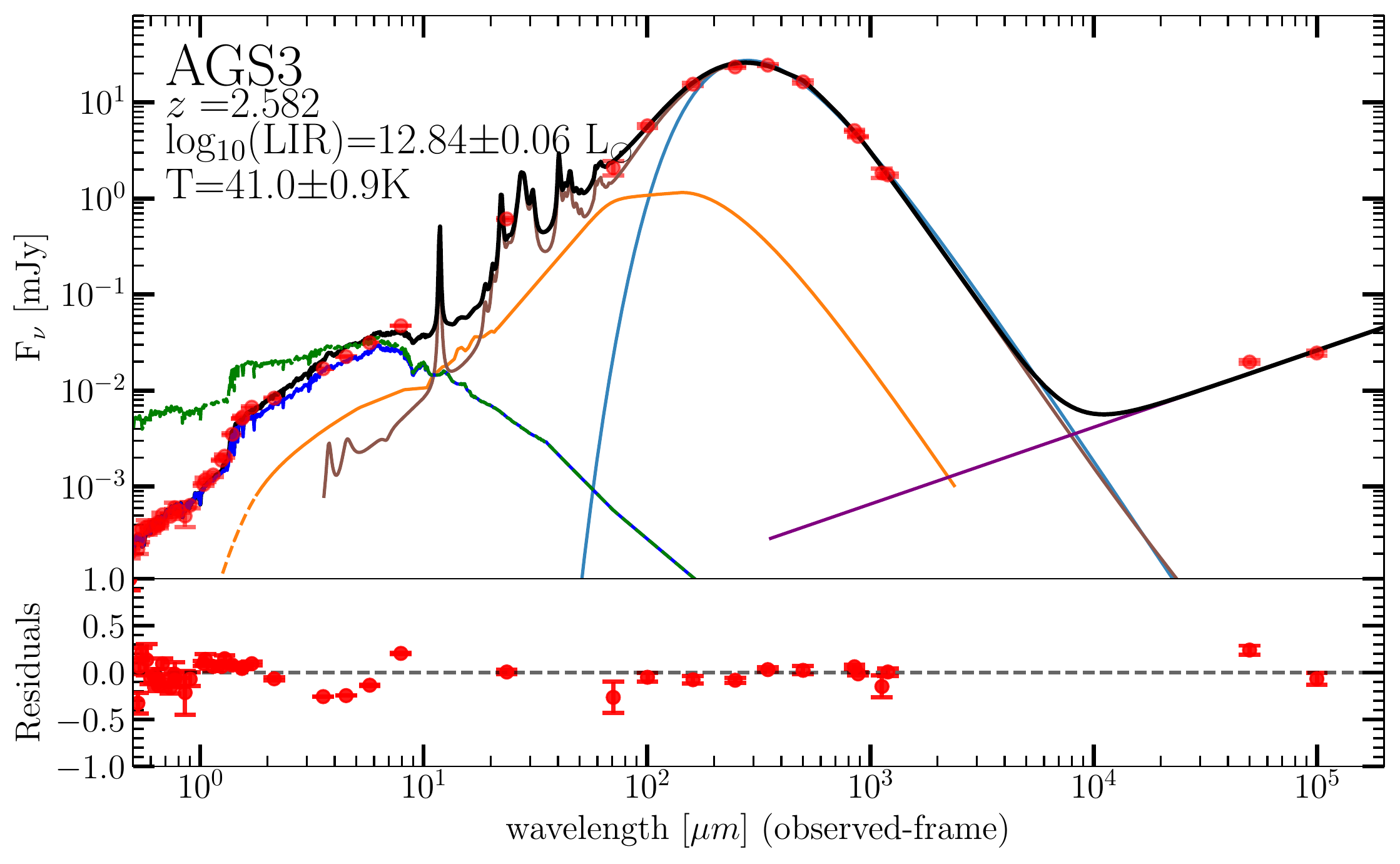} 
}
\end{minipage}
\begin{minipage}[t]{.96\textwidth}
\resizebox{\hsize}{!} { 
\includegraphics[width=3cm,clip]{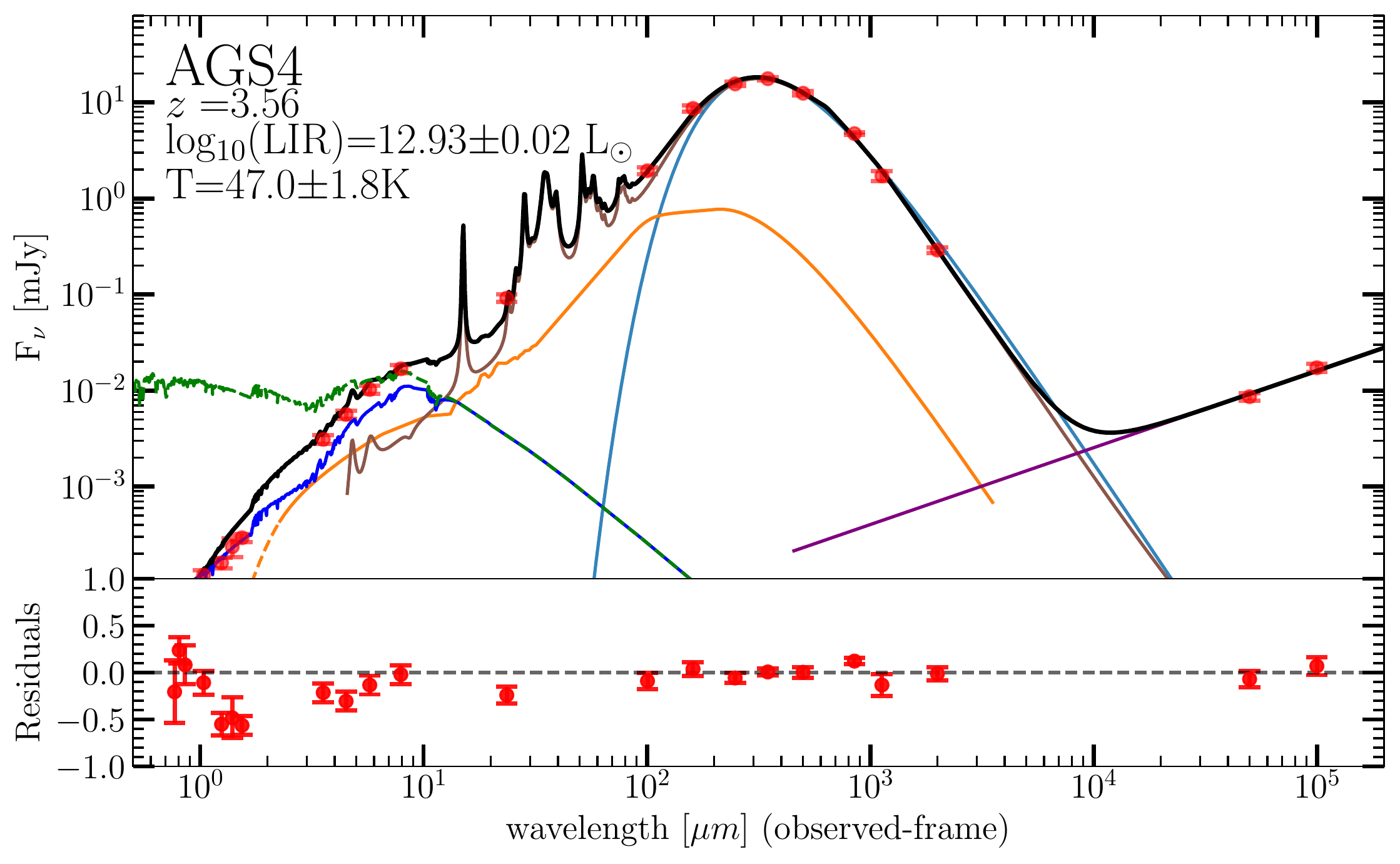} 
\includegraphics[width=3cm,clip]{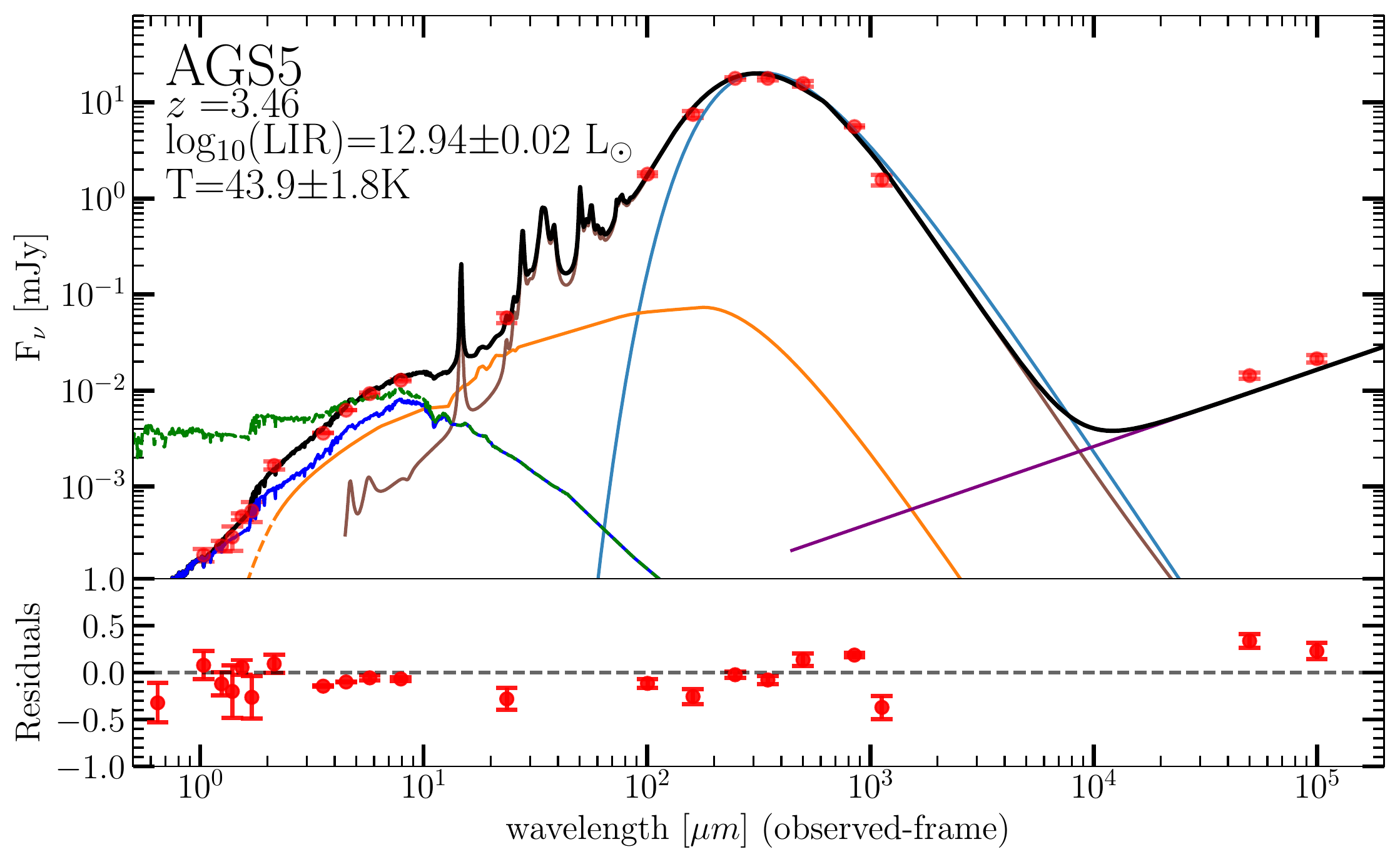} 
\includegraphics[width=3cm,clip]{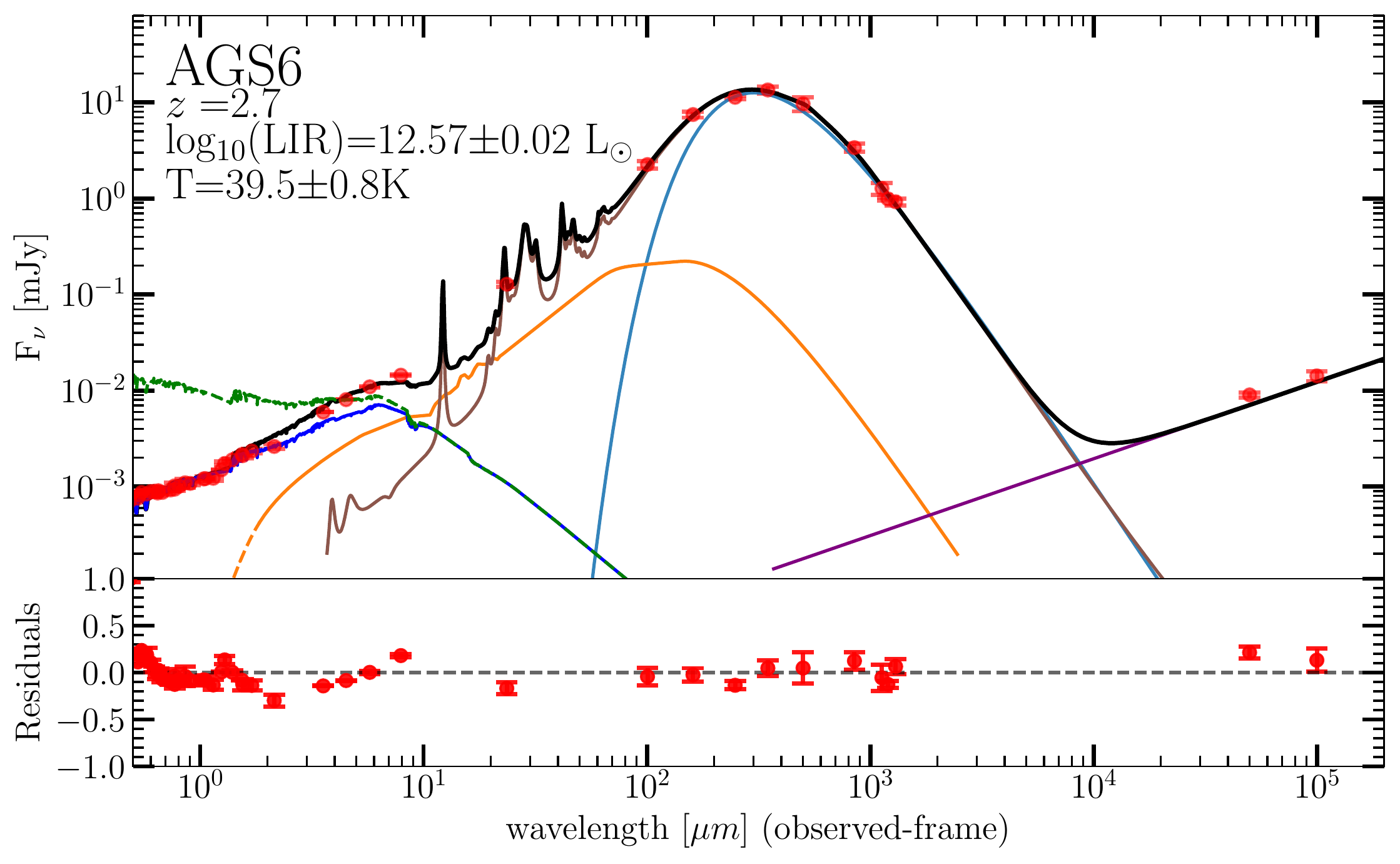} 
}
\end{minipage}
\begin{minipage}[t]{.96\textwidth}
\resizebox{\hsize}{!} { 
\includegraphics[width=3cm,clip]{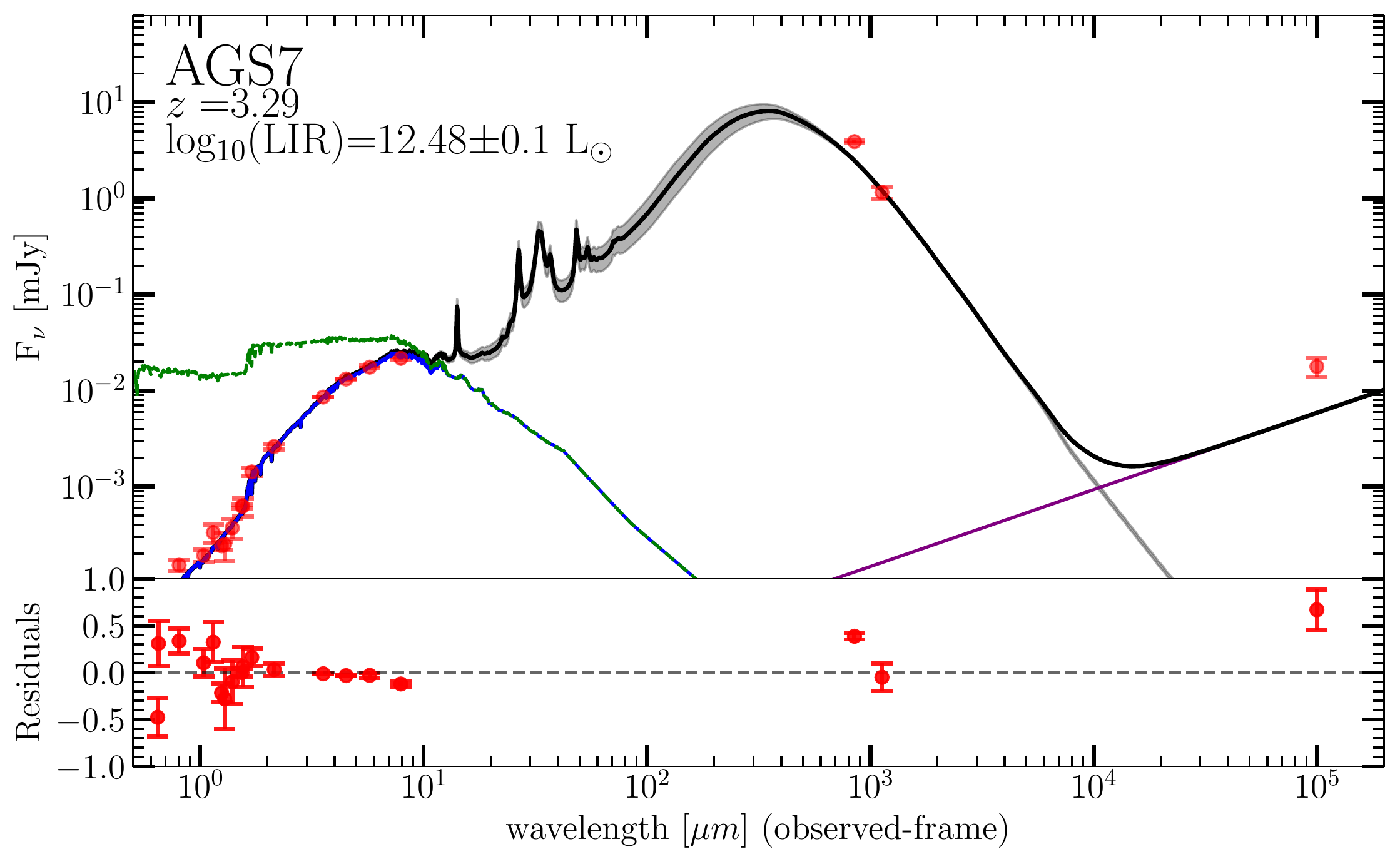} 
\includegraphics[width=3cm,clip]{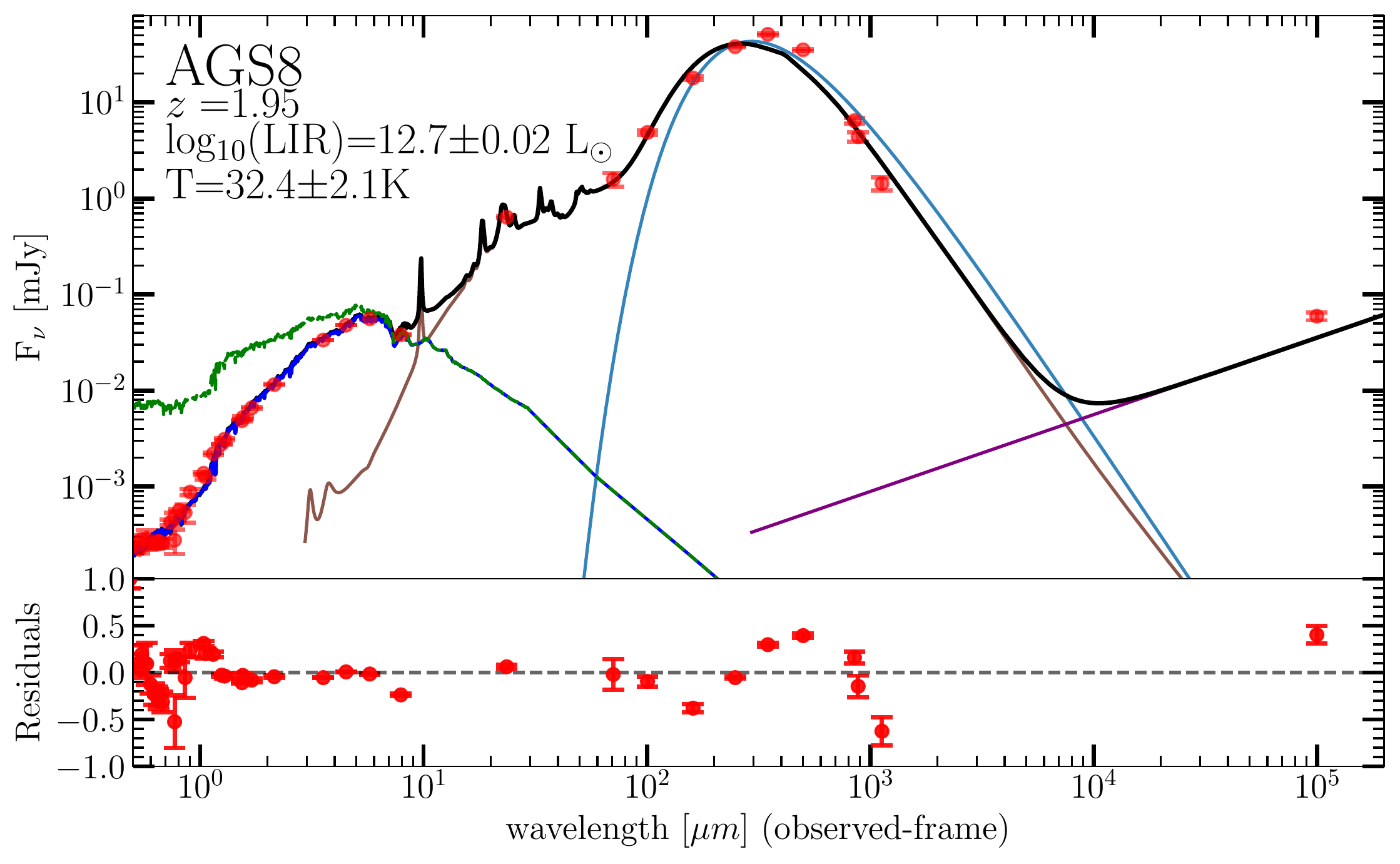} 
\includegraphics[width=3cm,clip]{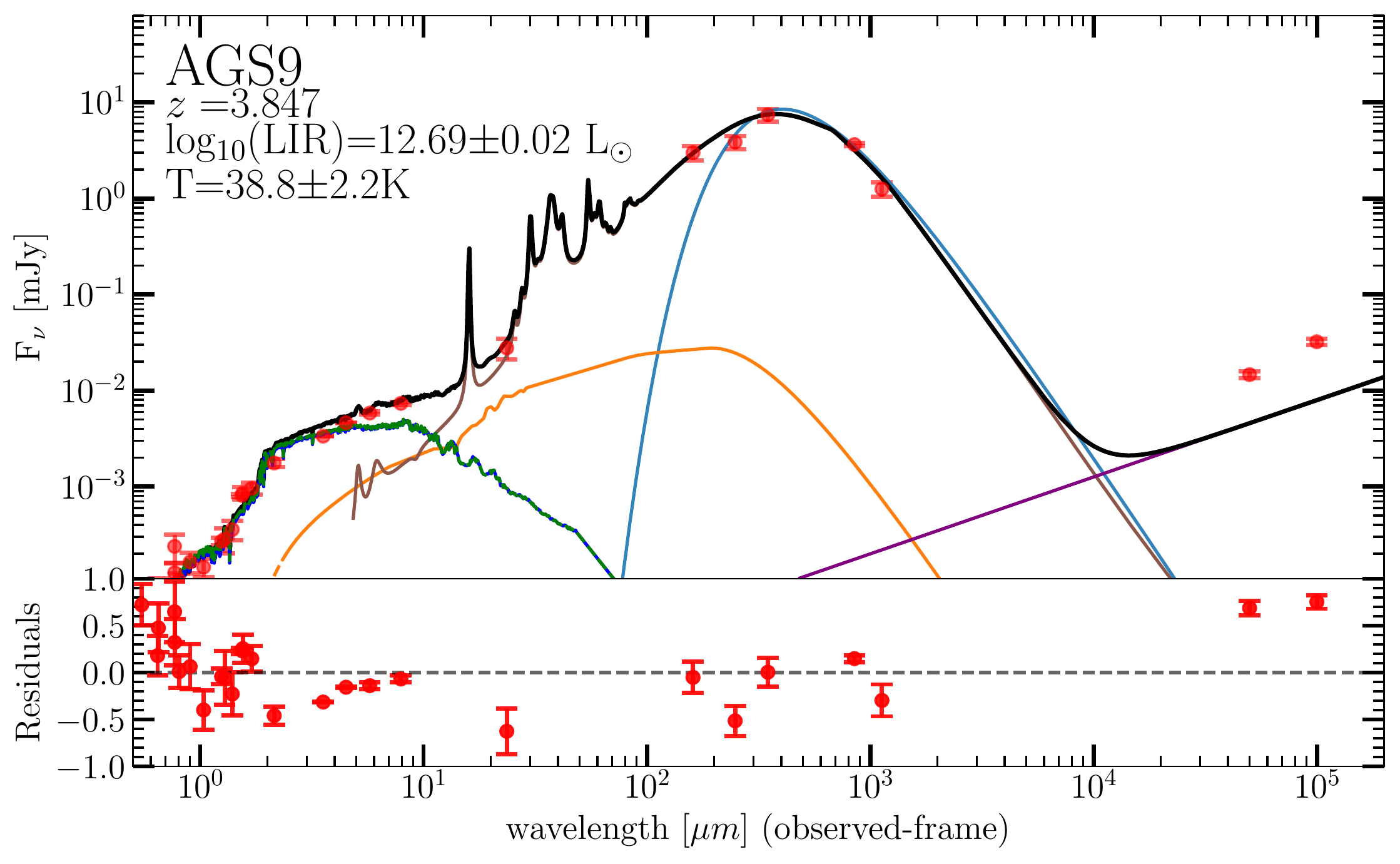} 
}
\end{minipage}
\begin{minipage}[t]{.96\textwidth}
\resizebox{\hsize}{!} { 
\includegraphics[width=3cm,clip]{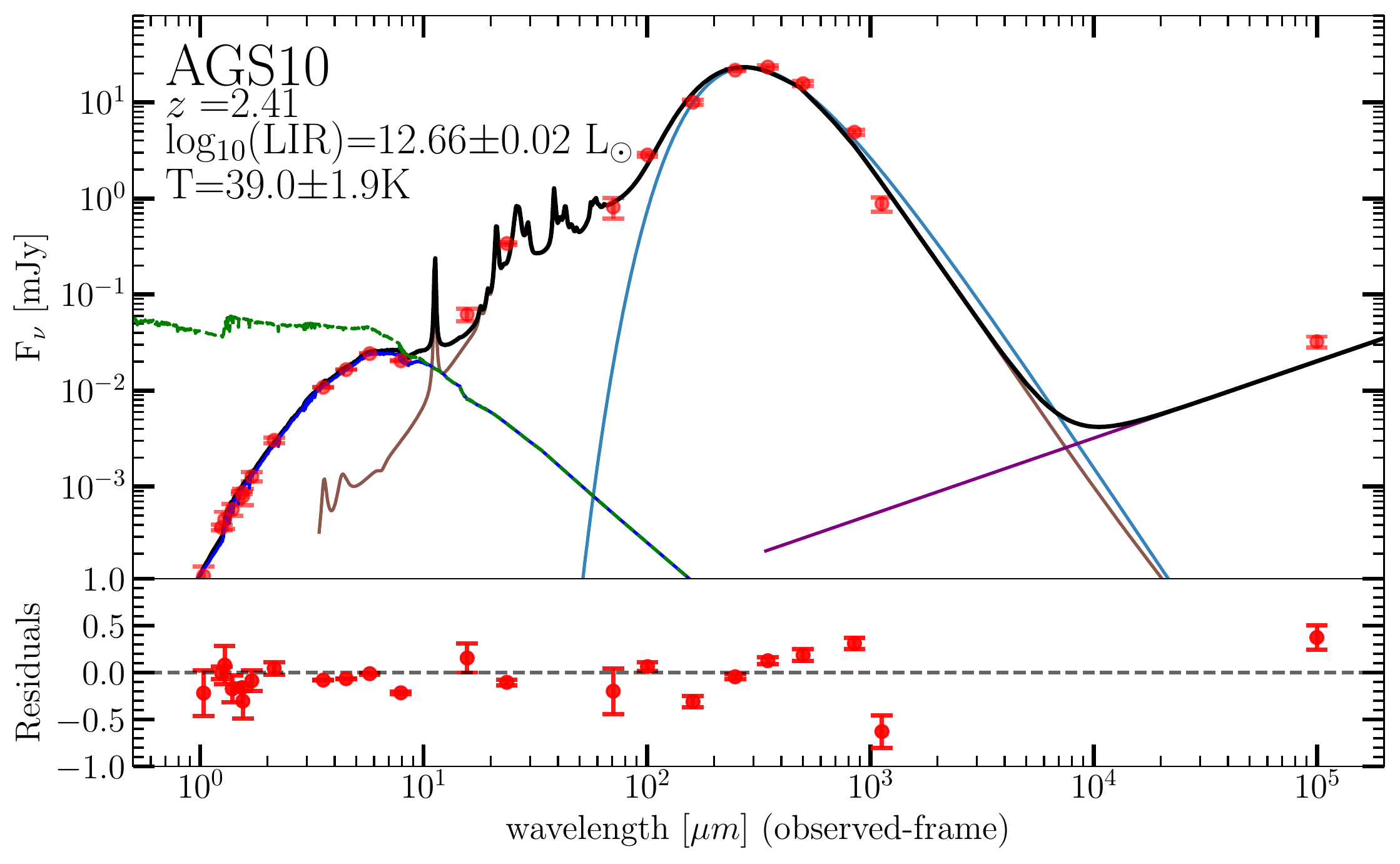} 
\includegraphics[width=3cm,clip]{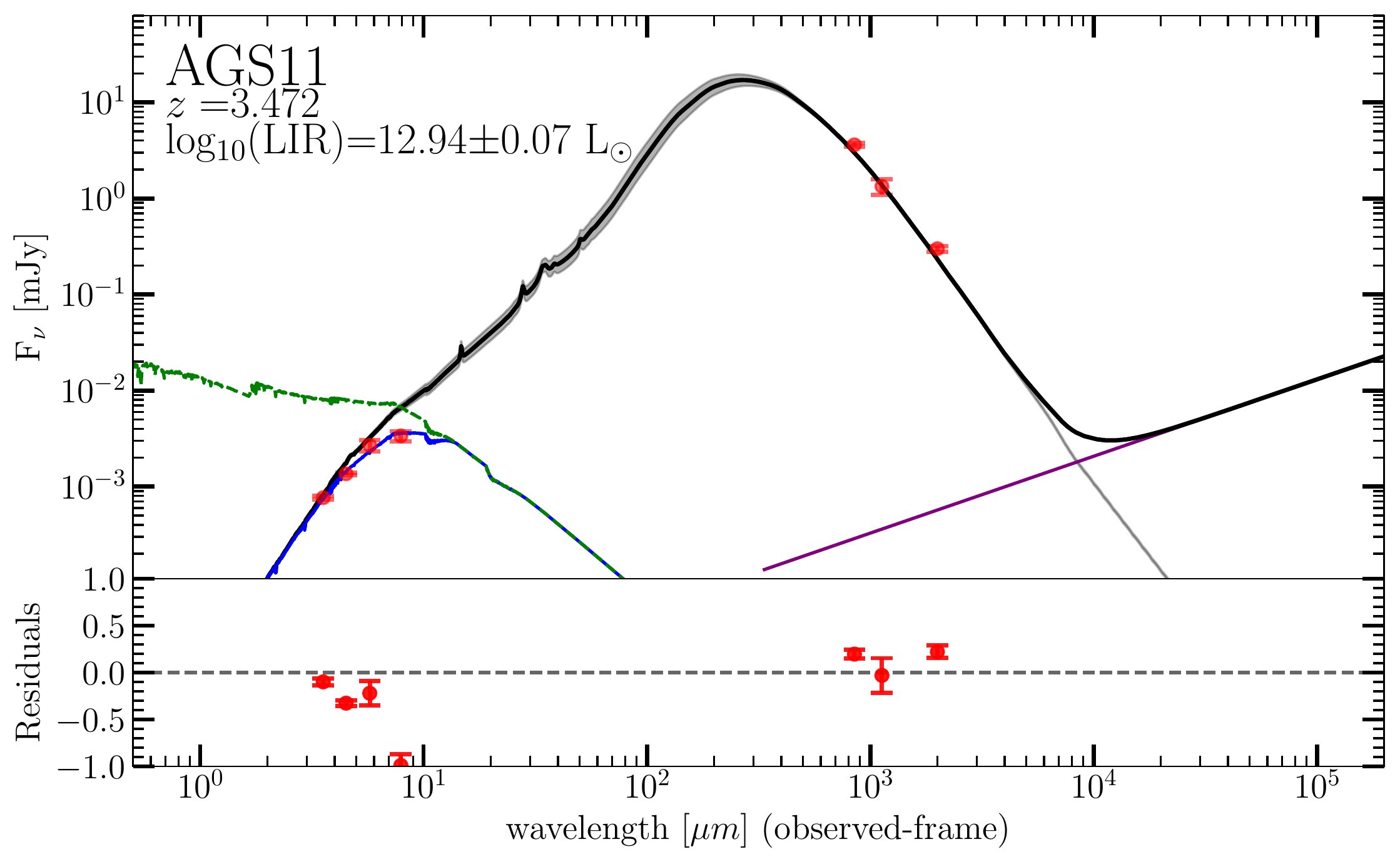} 
\includegraphics[width=3cm,clip]{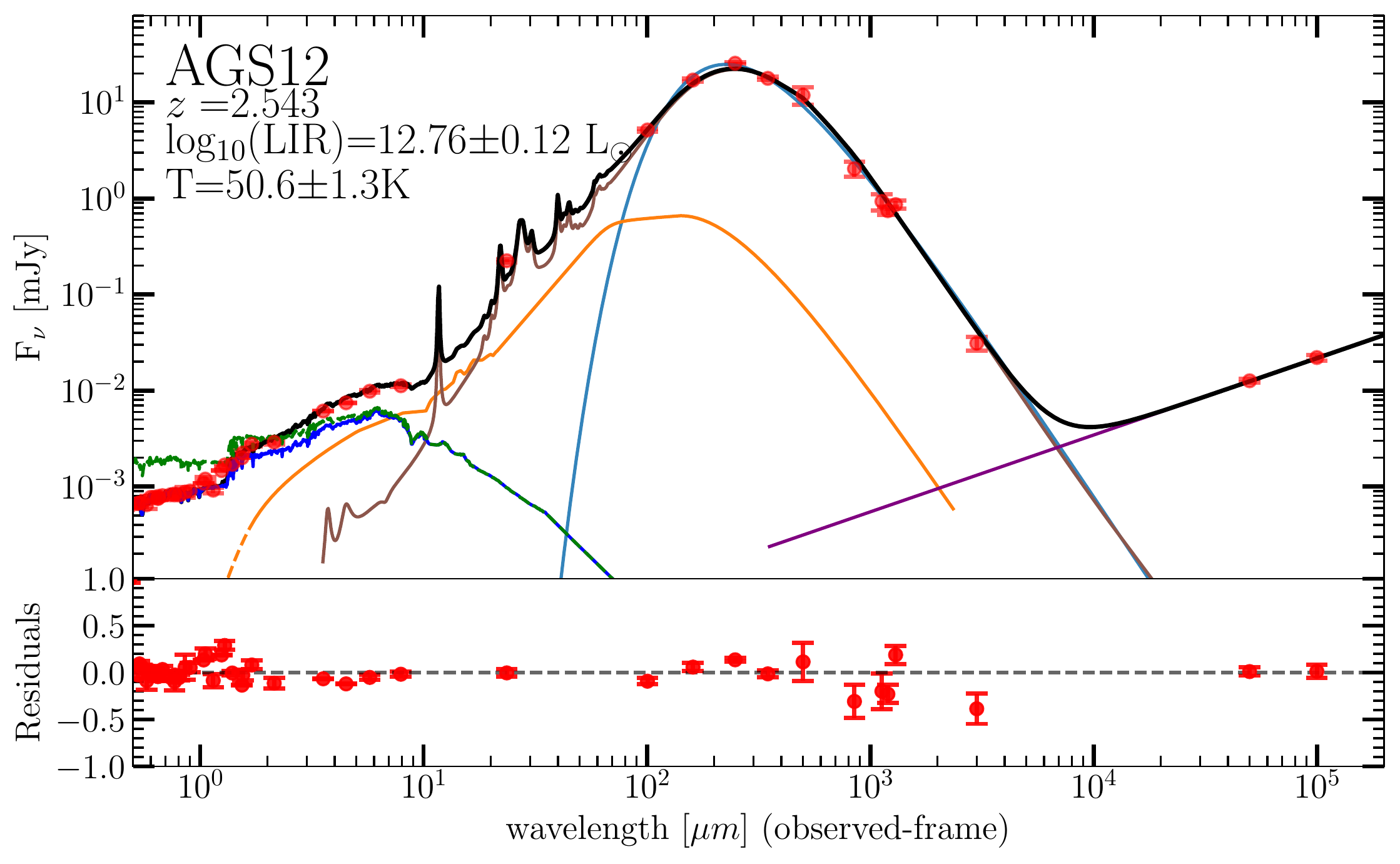} 
}
\end{minipage}
\begin{minipage}[t]{.96\textwidth}
\resizebox{\hsize}{!} { 
\includegraphics[width=3cm,clip]{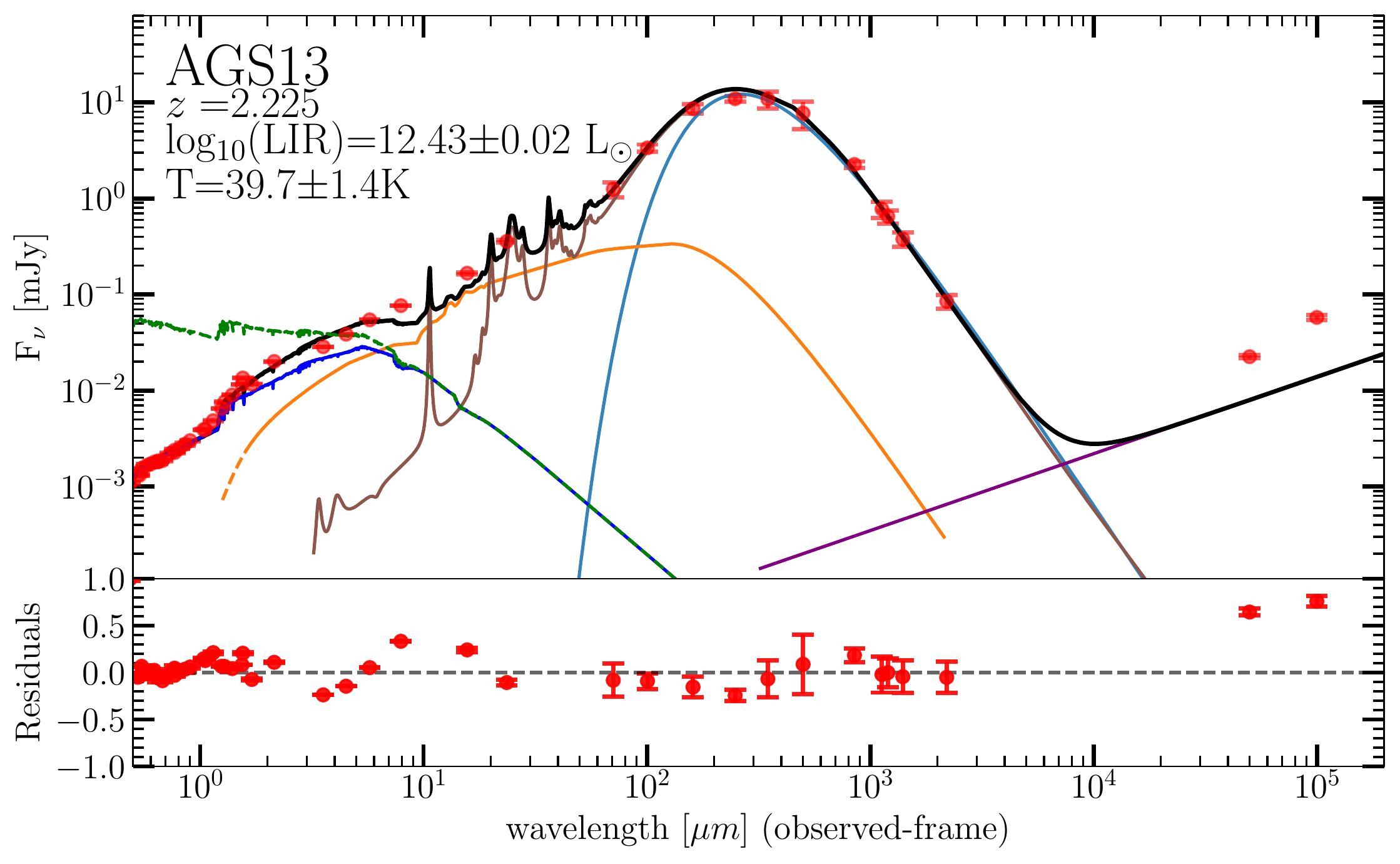} 
\includegraphics[width=3cm,clip]{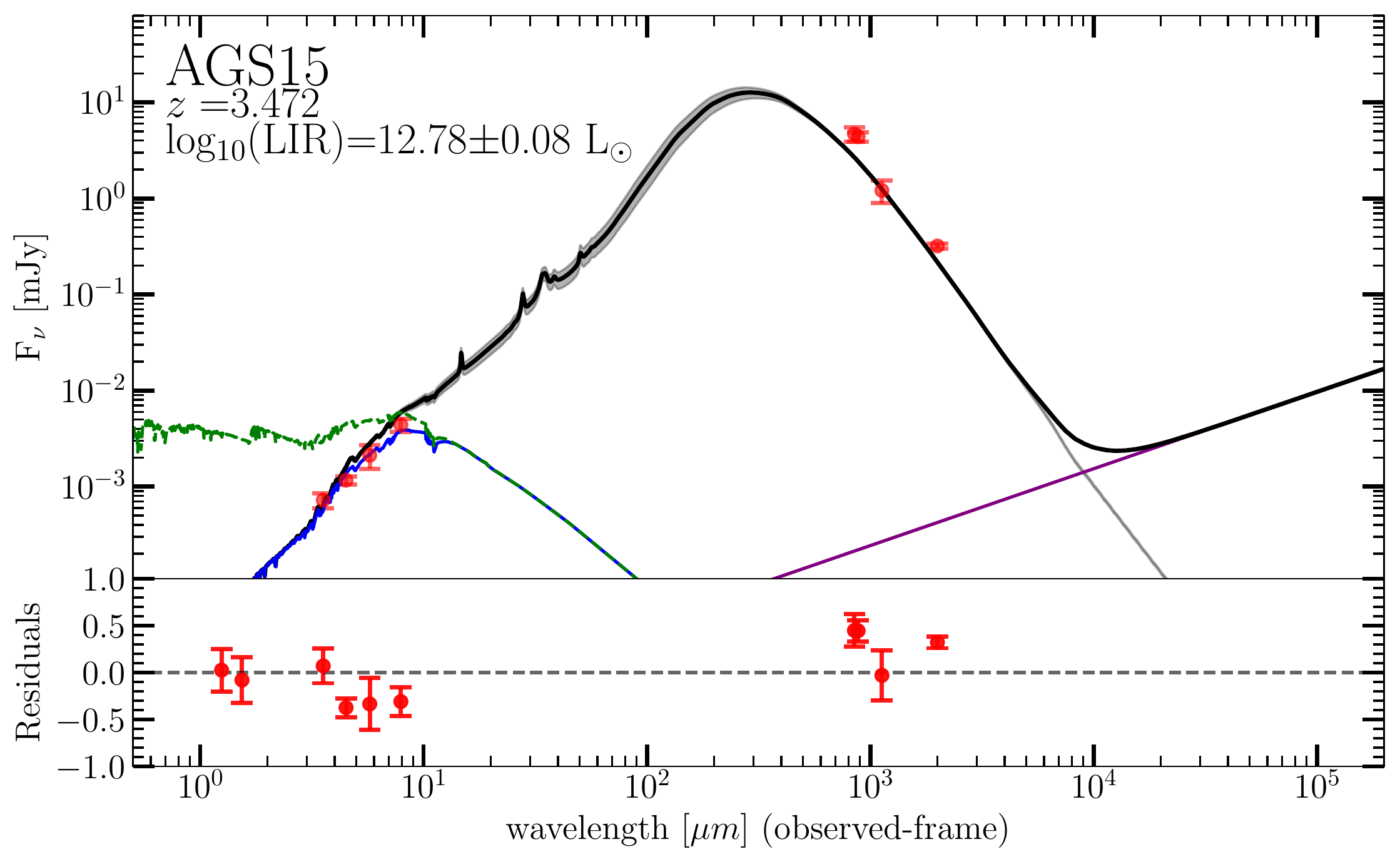} 
\includegraphics[width=3cm,clip]{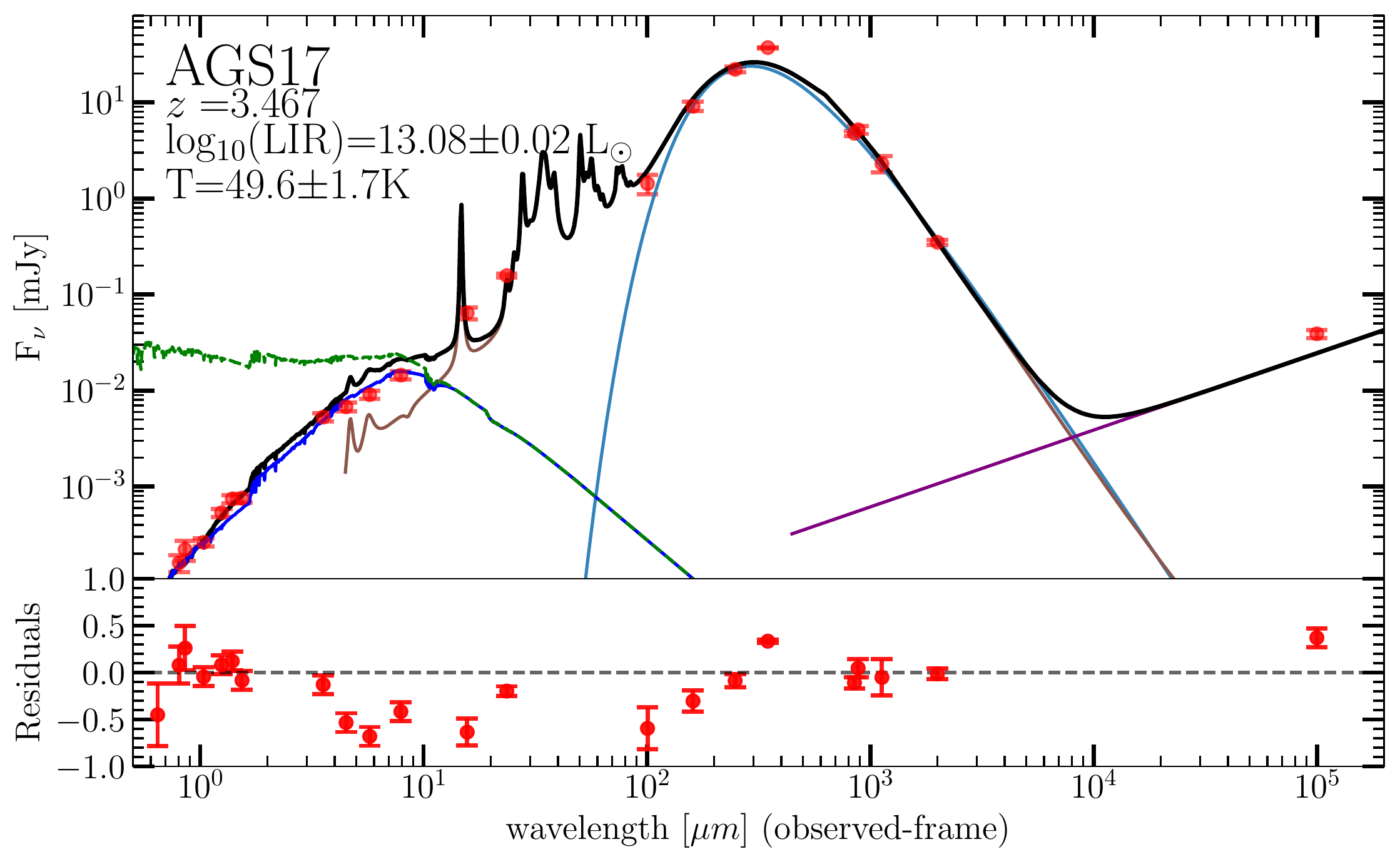} 
}
\end{minipage}
\begin{minipage}[t]{.96\textwidth}
\resizebox{\hsize}{!} { 
\includegraphics[width=3cm,clip]{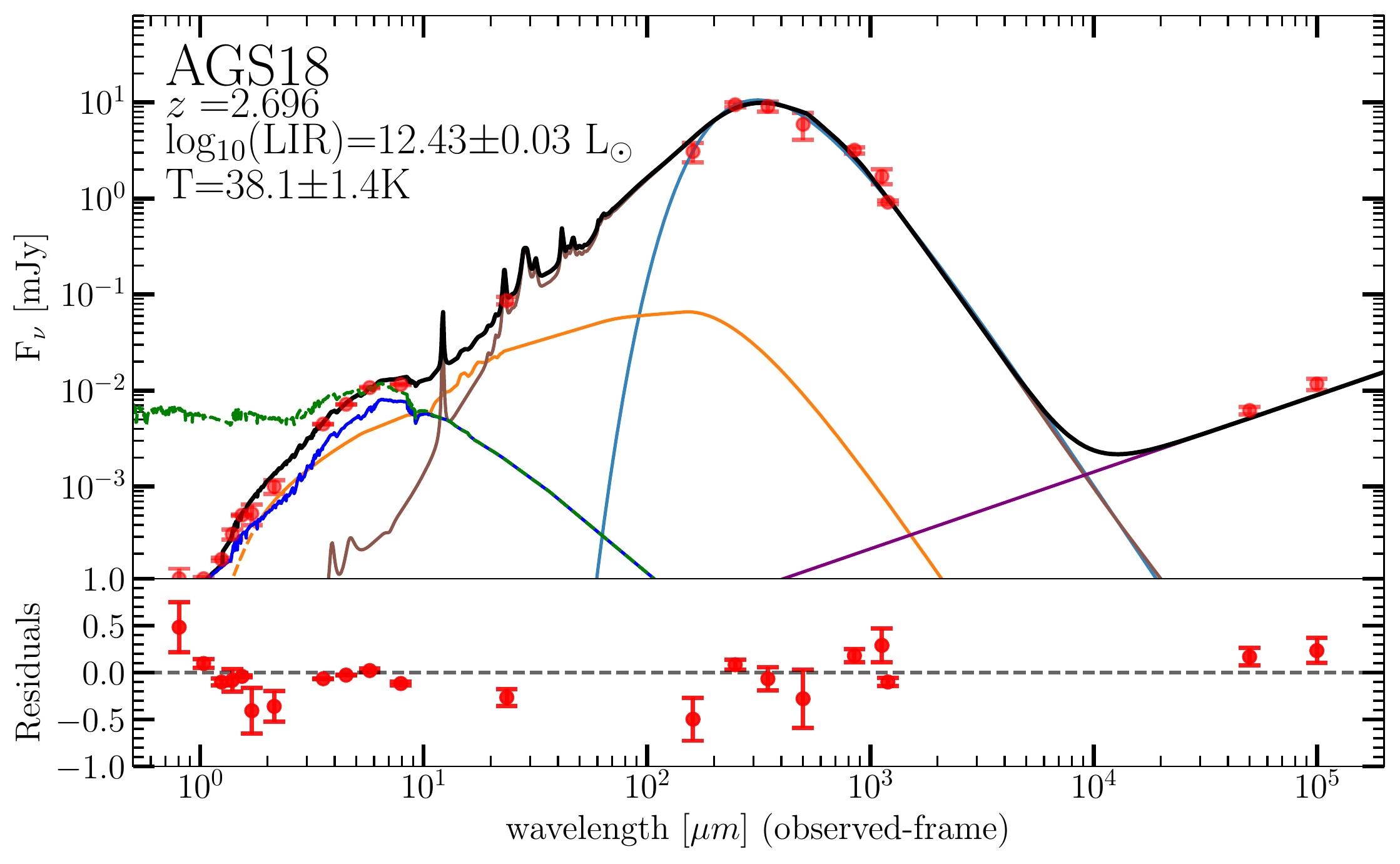} 
\includegraphics[width=3cm,clip]{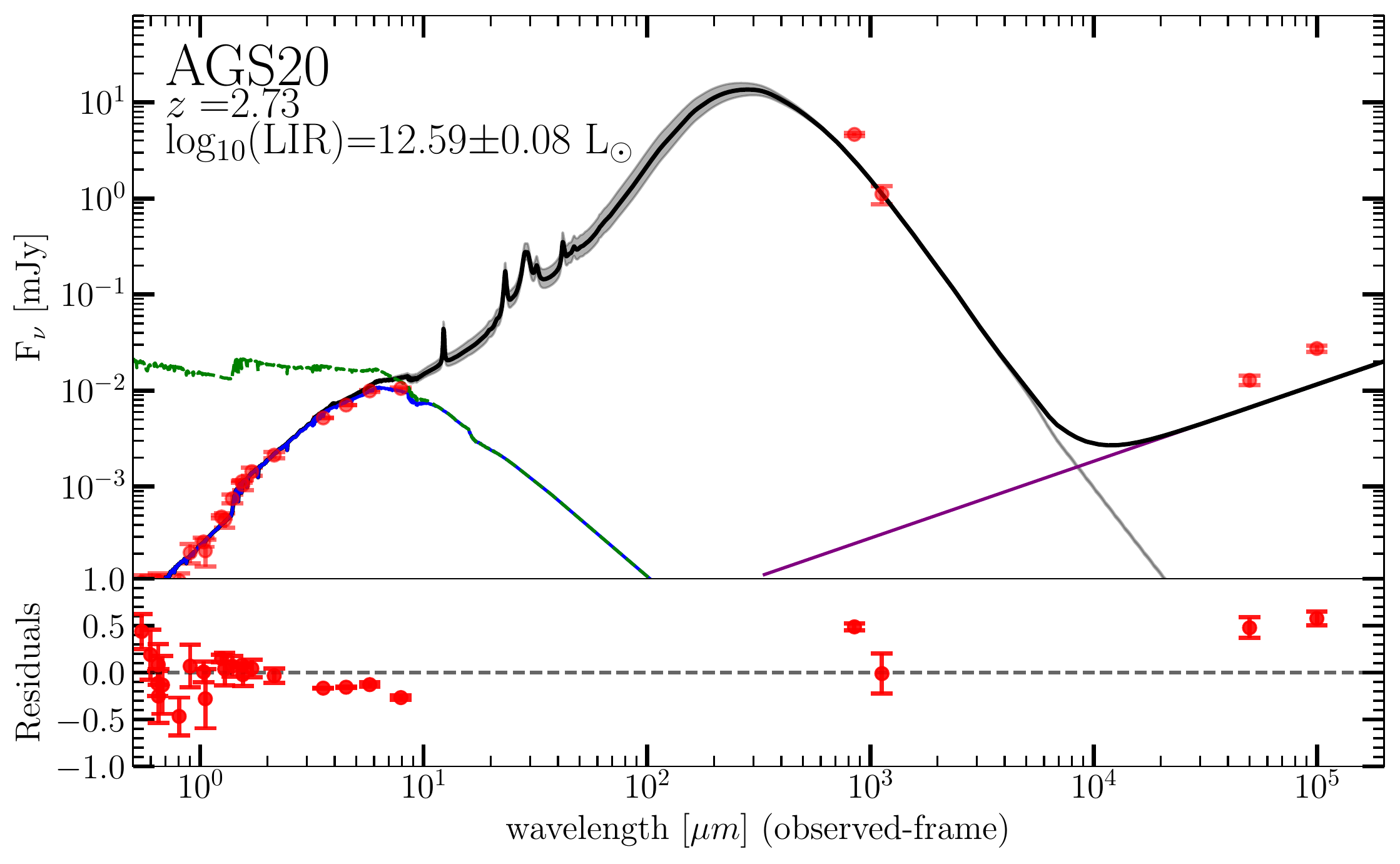} 
\includegraphics[width=3cm,clip]{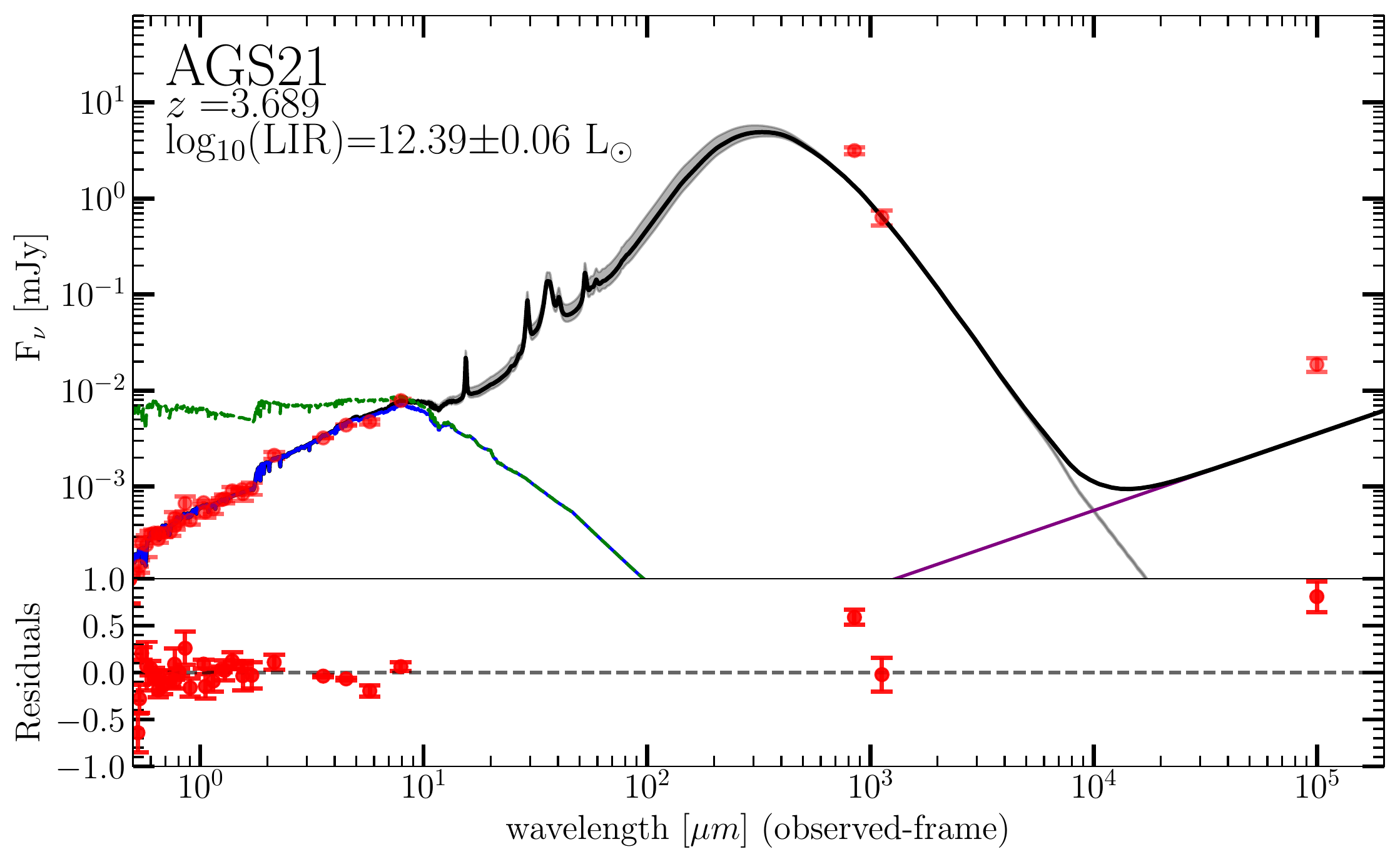} 
}
\end{minipage}
      \caption{Optical to radio Spectral Energy Distributions for the 35 galaxies detected in the GOODS-ALMA survey. If the studied galaxy has also been detected with \textit{Hershel}, we fit the SED using the \texttt{CIGALE} code, otherwise we use the dust spectral energy distribution library presented in \cite{Schreiber2017}. The solid black line represents the best fit, which can be decomposed into the IR dust contribution (brown line), a stellar component uncorrected for dust attenuation (dark blue line), synchrotron emission (purple line) and the AGN contribution (orange line). In addition, we show the best fit of a modified black body, with $\beta$\,=\,1.5 (light blue line). The corrected UV emission is also shown in green. The bottom panel shows the residuals: (observation - model)/observation.}
         \label{SEDs}
\end{figure*}

\begin{figure*}[h!]
\centering
\begin{minipage}[t]{.96\textwidth}
\resizebox{\hsize}{!} { 
\includegraphics[width=3cm,clip]{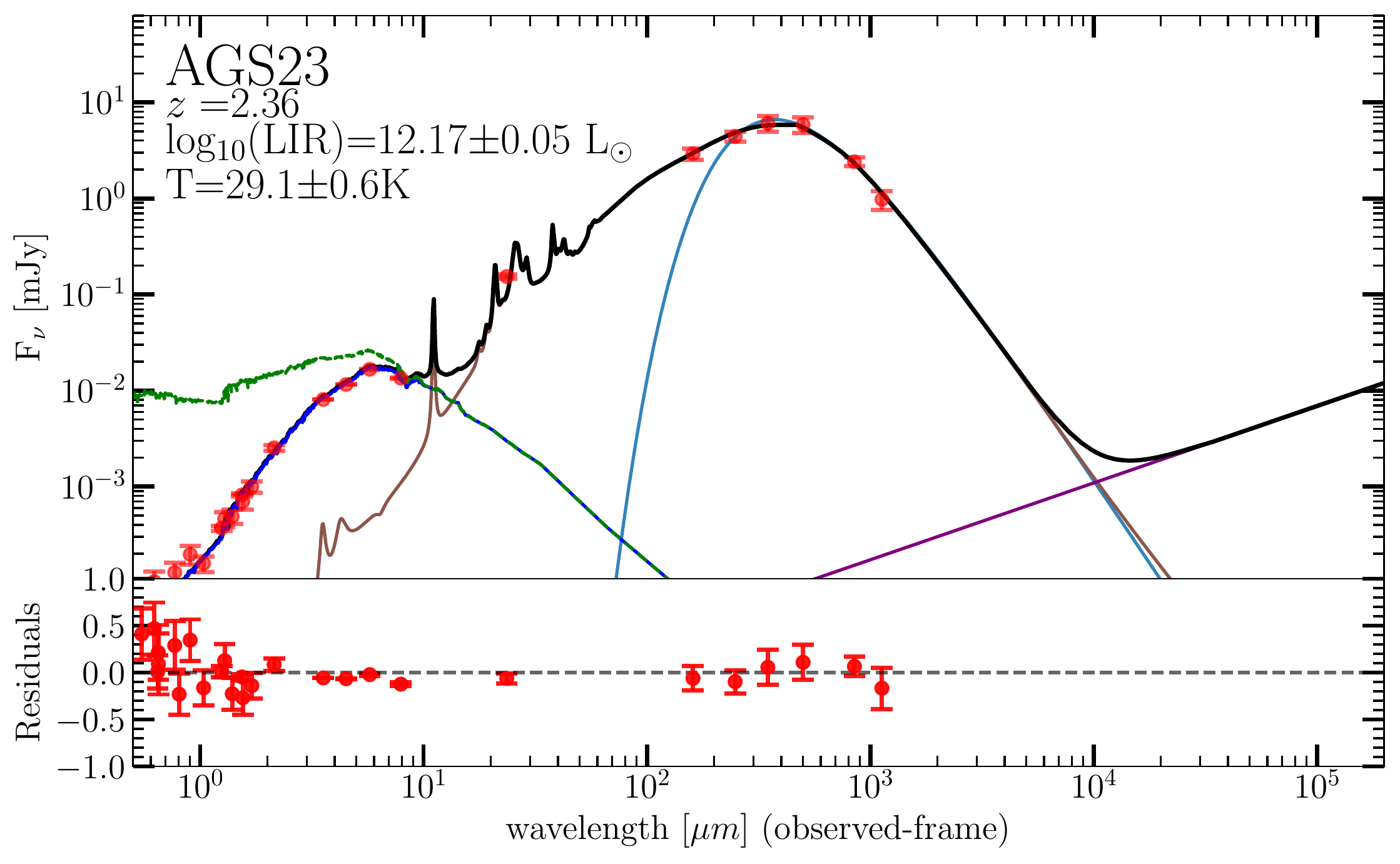} 
\includegraphics[width=3cm,clip]{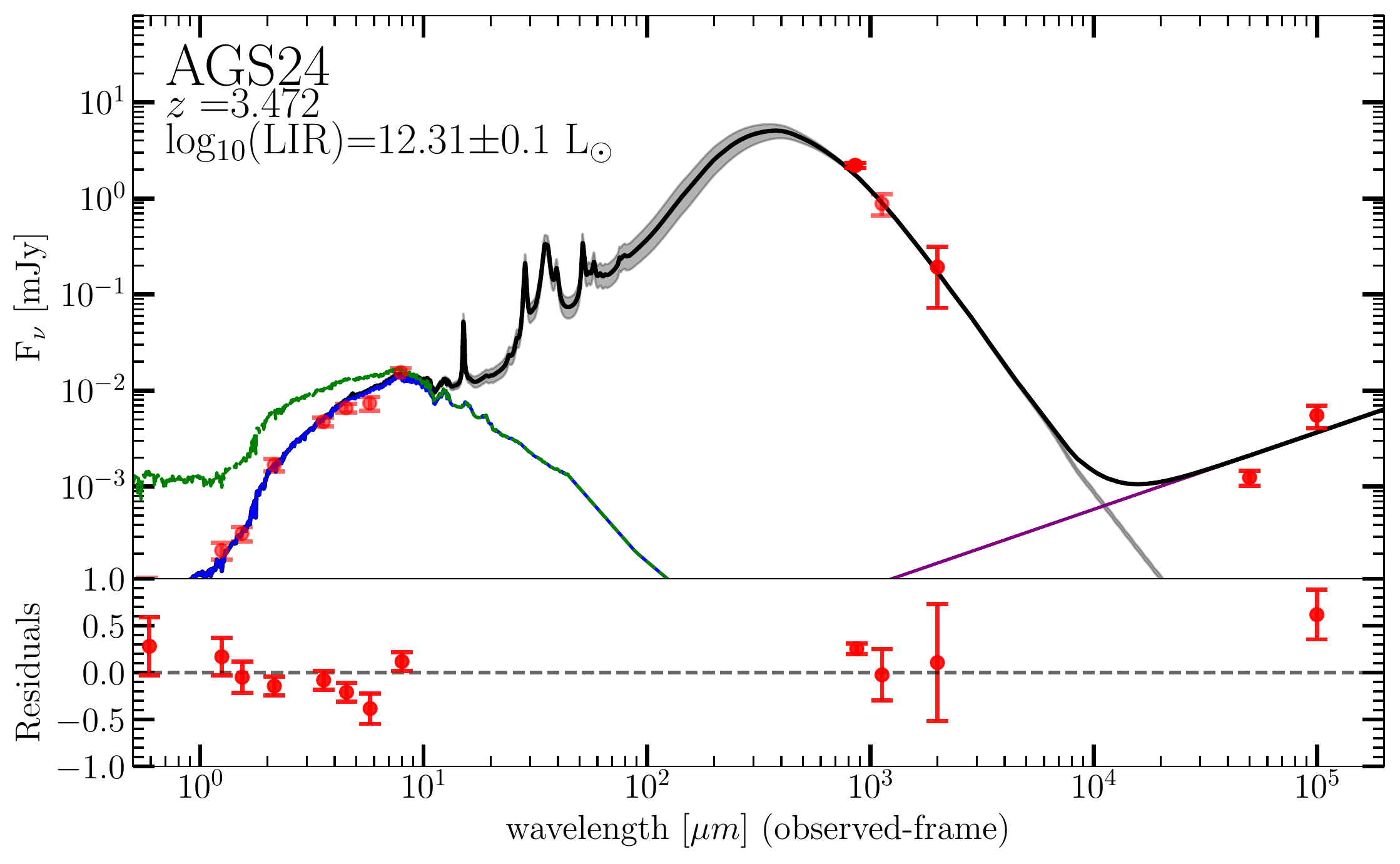} 
\includegraphics[width=3cm,clip]{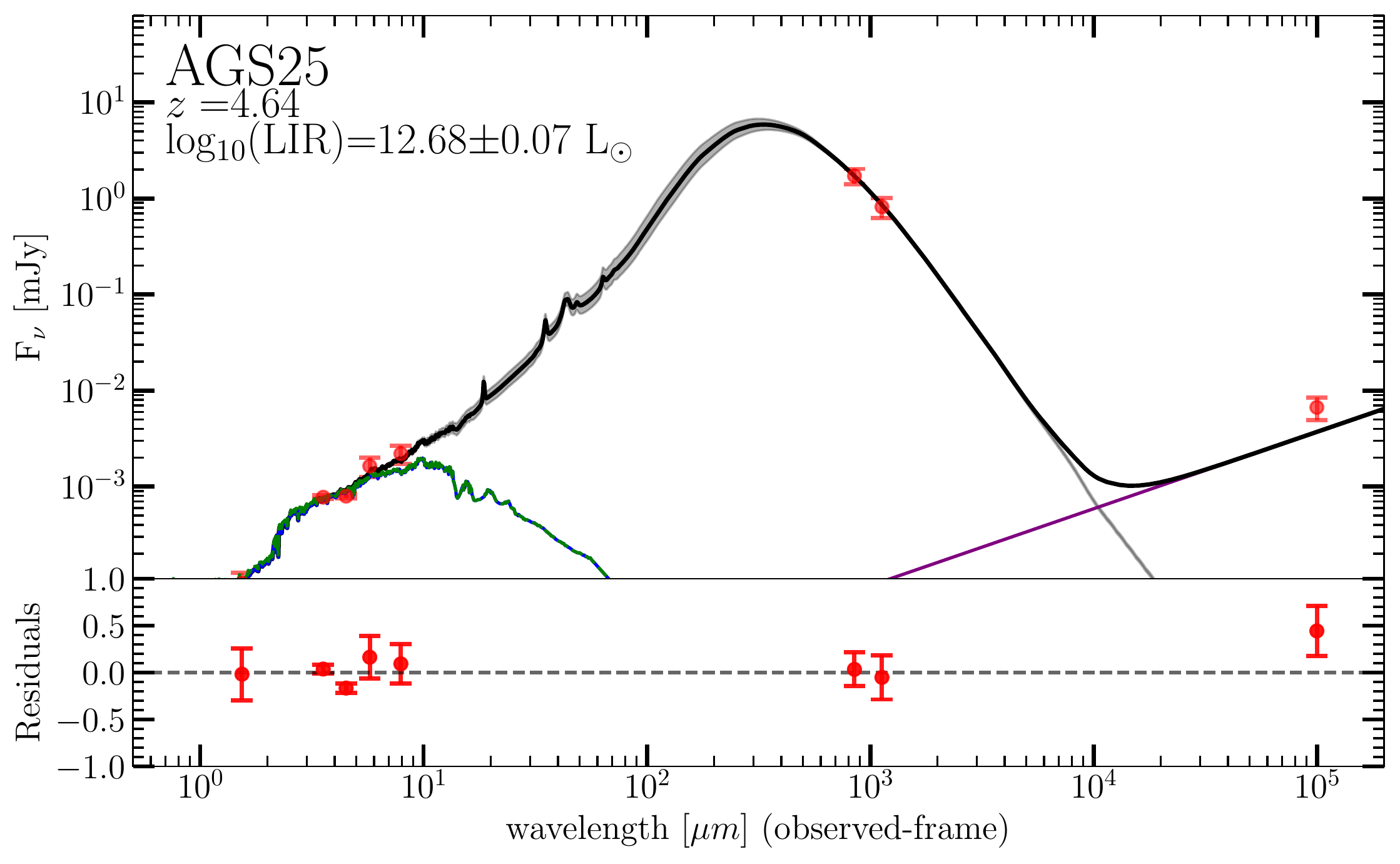} 
}
\end{minipage}
\begin{minipage}[t]{.96\textwidth}
\resizebox{\hsize}{!} { 
\includegraphics[width=3cm,clip]{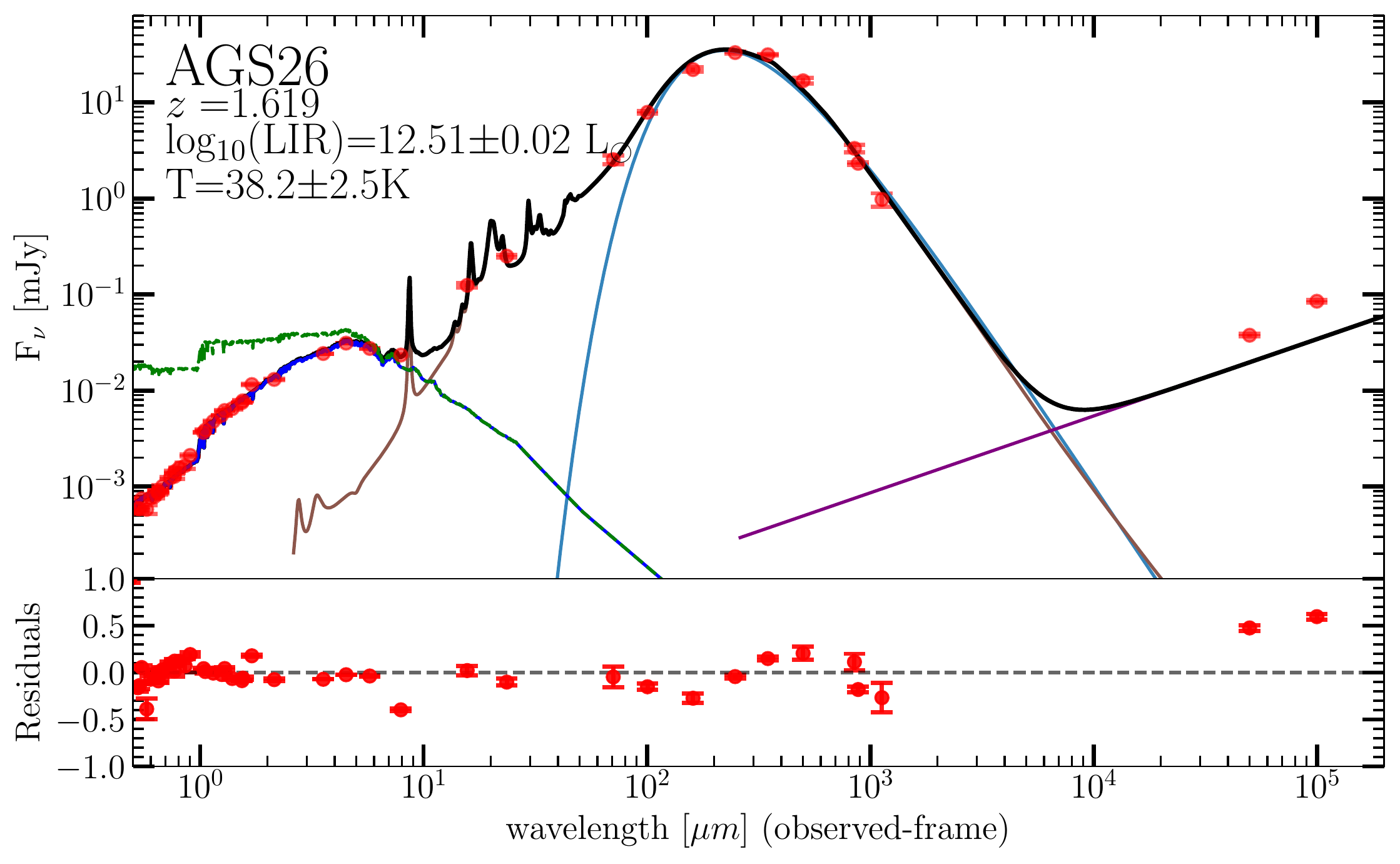} 
\includegraphics[width=3cm,clip]{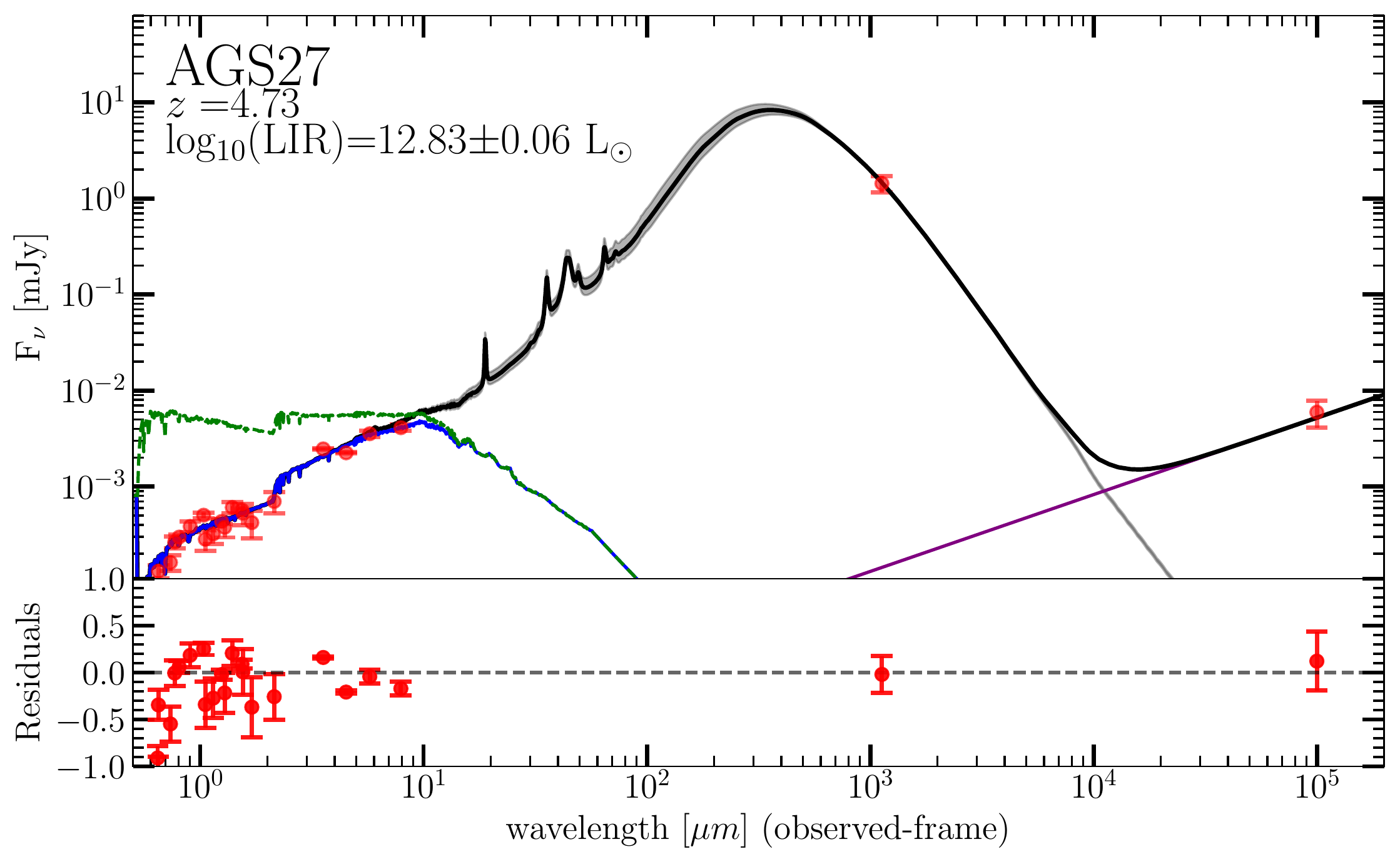} 
\includegraphics[width=3cm,clip]{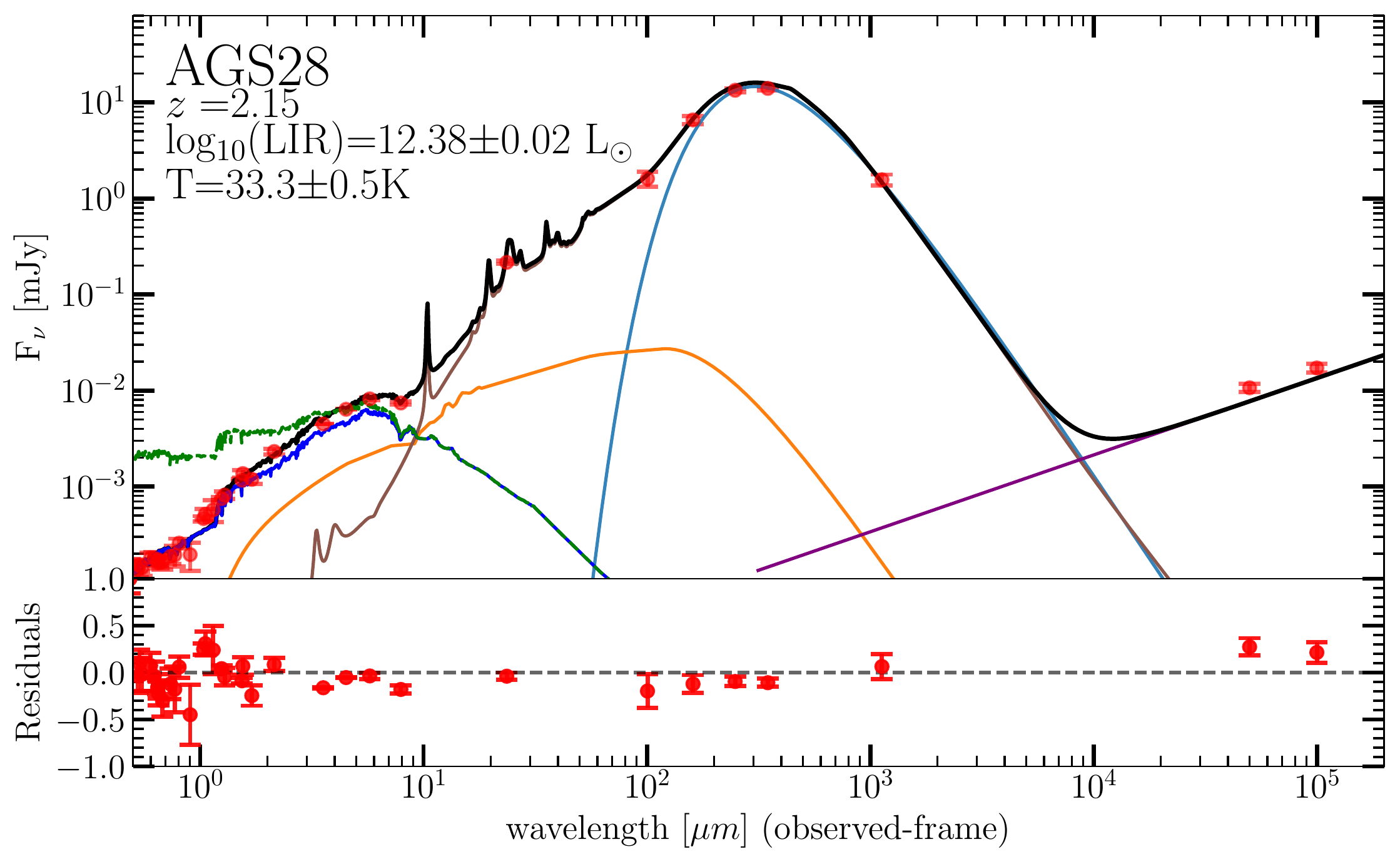} 
}
\end{minipage}
\begin{minipage}[t]{.96\textwidth}
\resizebox{\hsize}{!} { 
\includegraphics[width=3cm,clip]{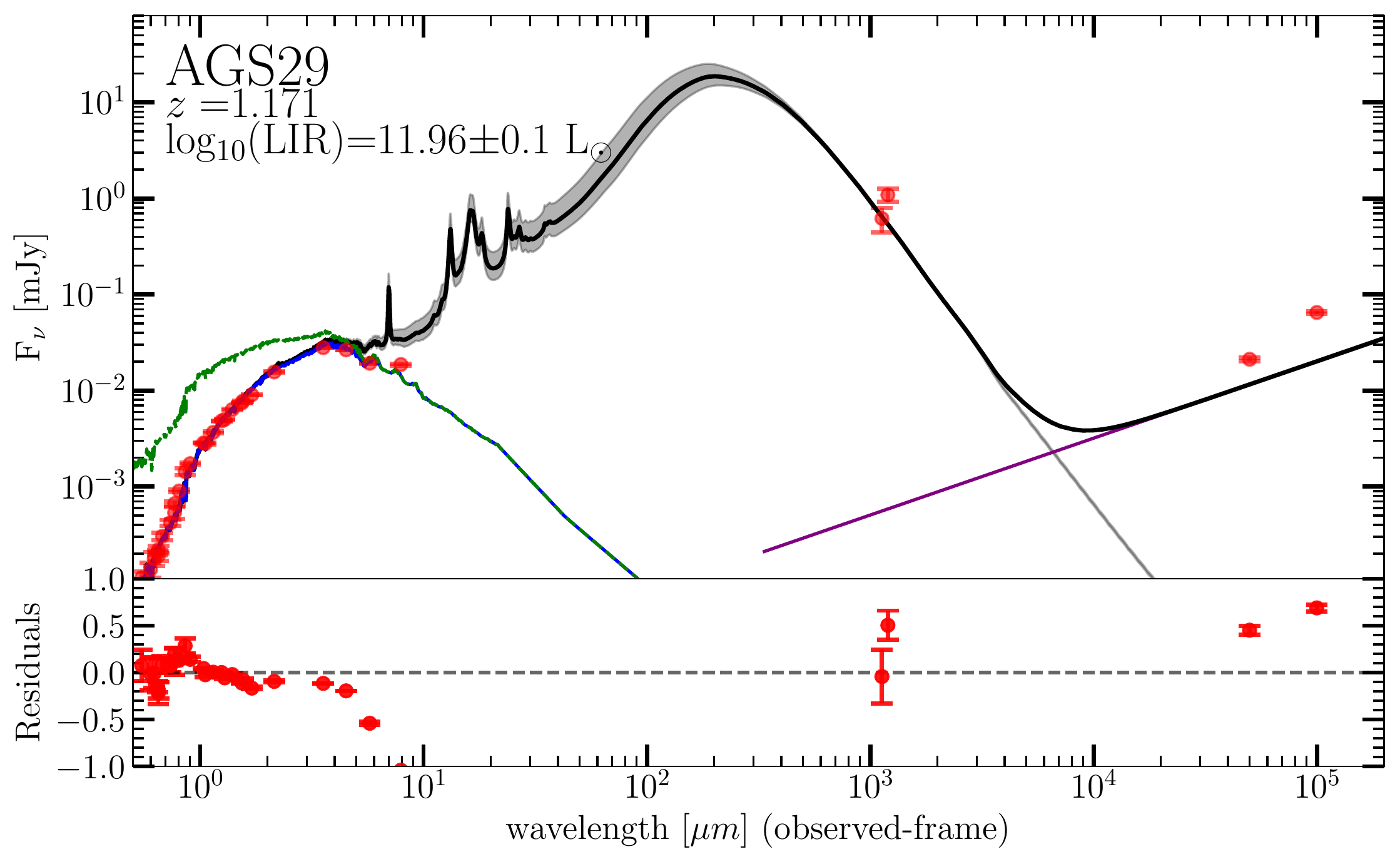} 
\includegraphics[width=3cm,clip]{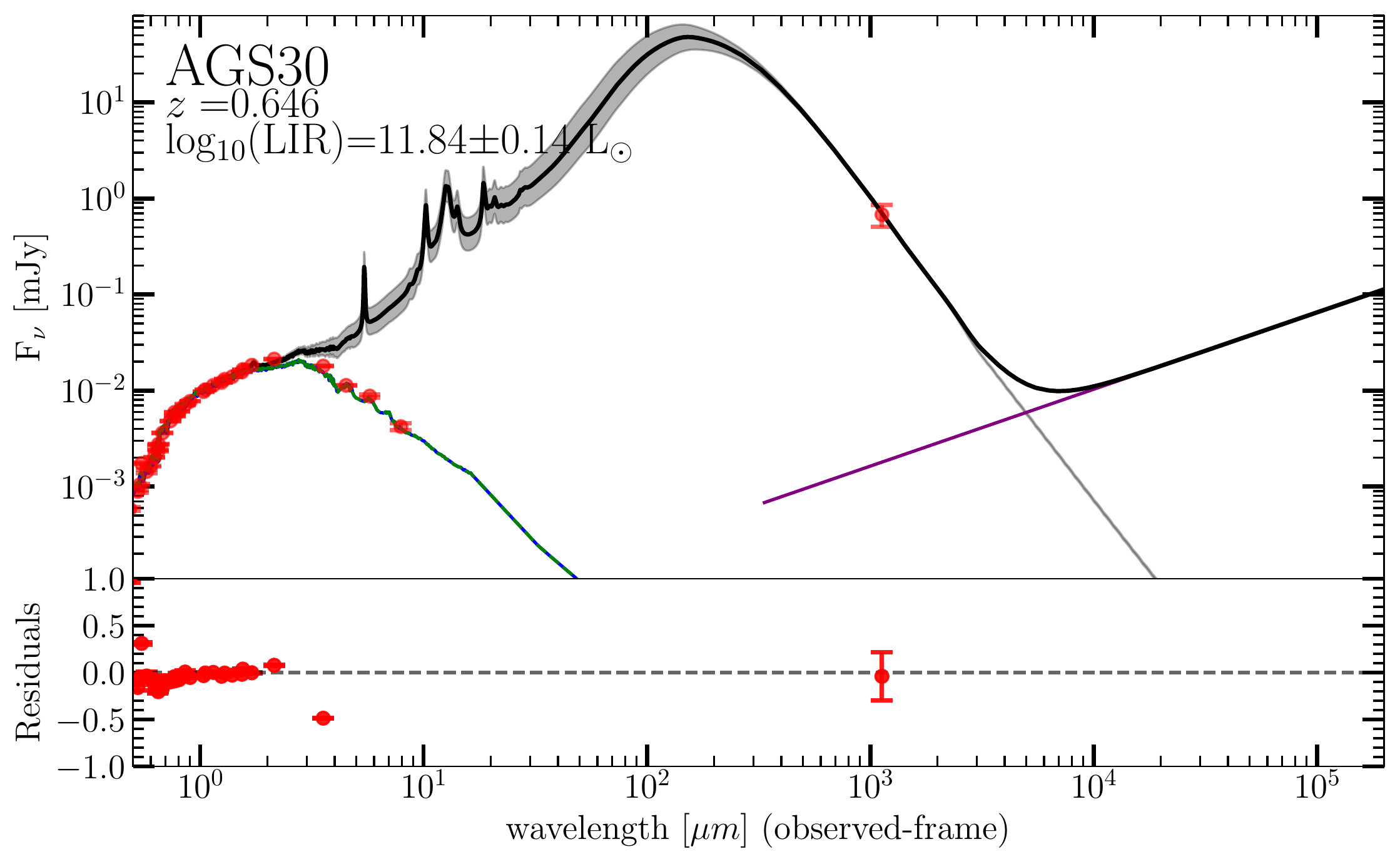} 
\includegraphics[width=3cm,clip]{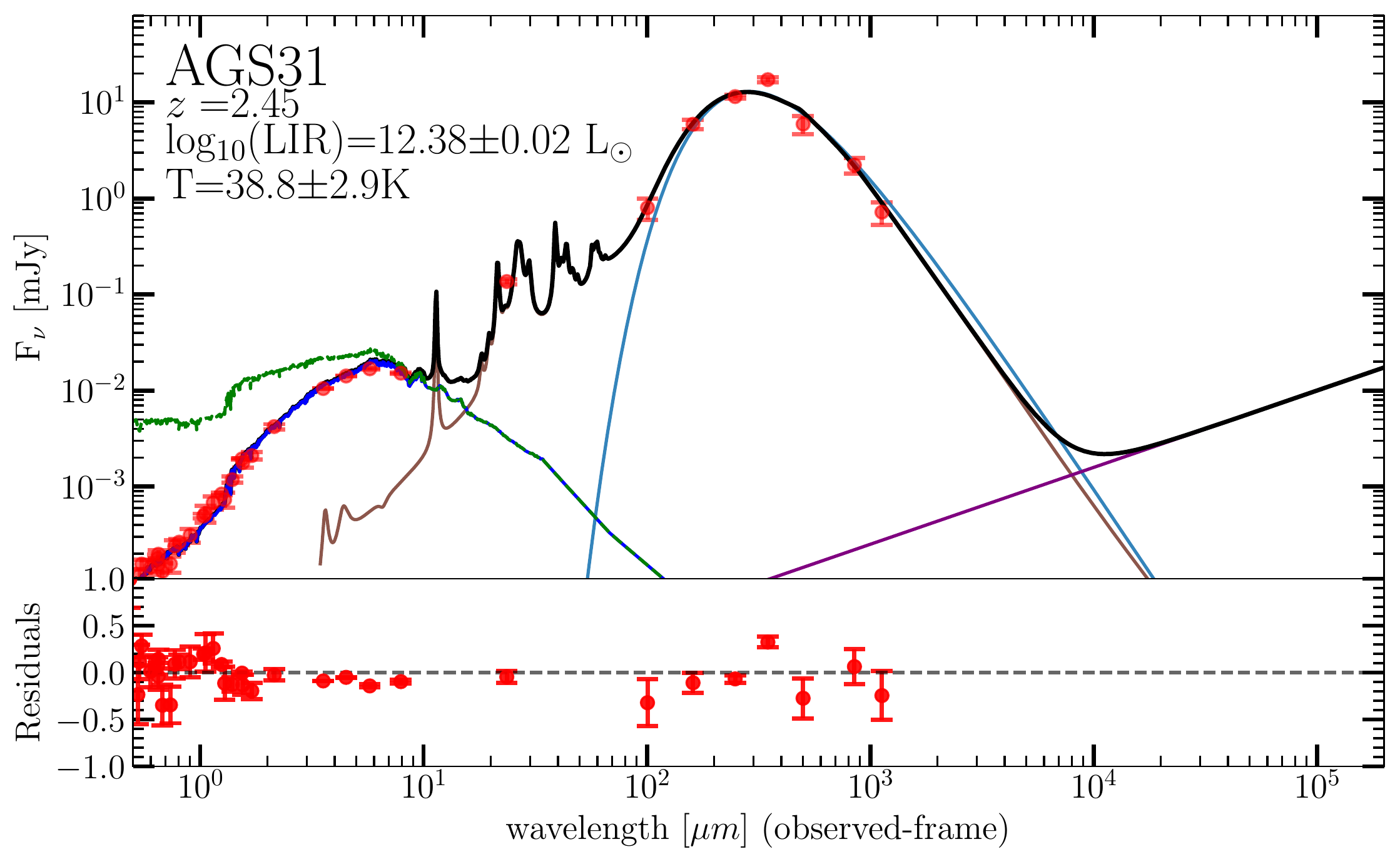} 
}
\end{minipage}
\begin{minipage}[t]{.96\textwidth}
\resizebox{\hsize}{!} { 
\includegraphics[width=3cm,clip]{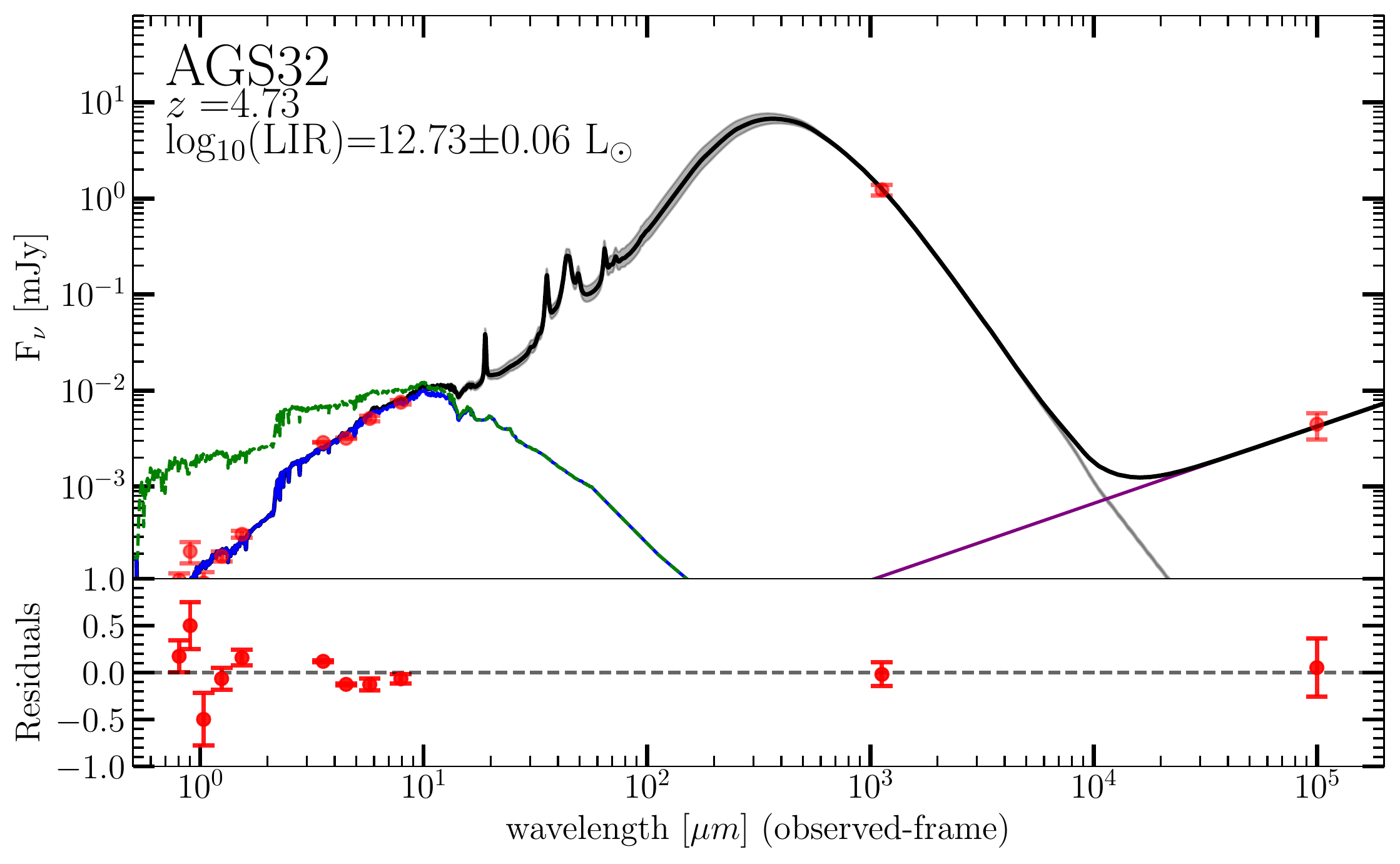} 
\includegraphics[width=3cm,clip]{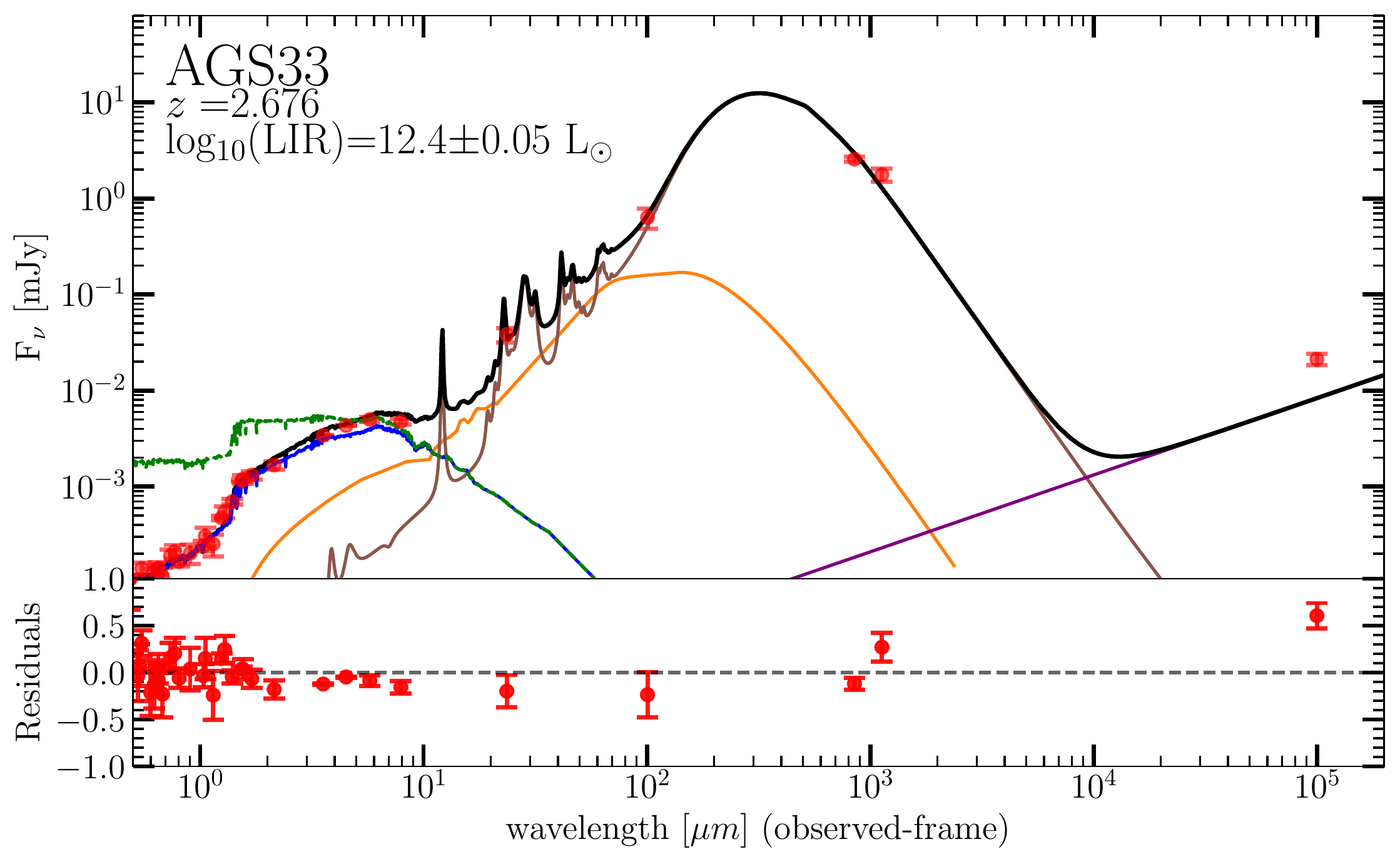} 
\includegraphics[width=3cm,clip]{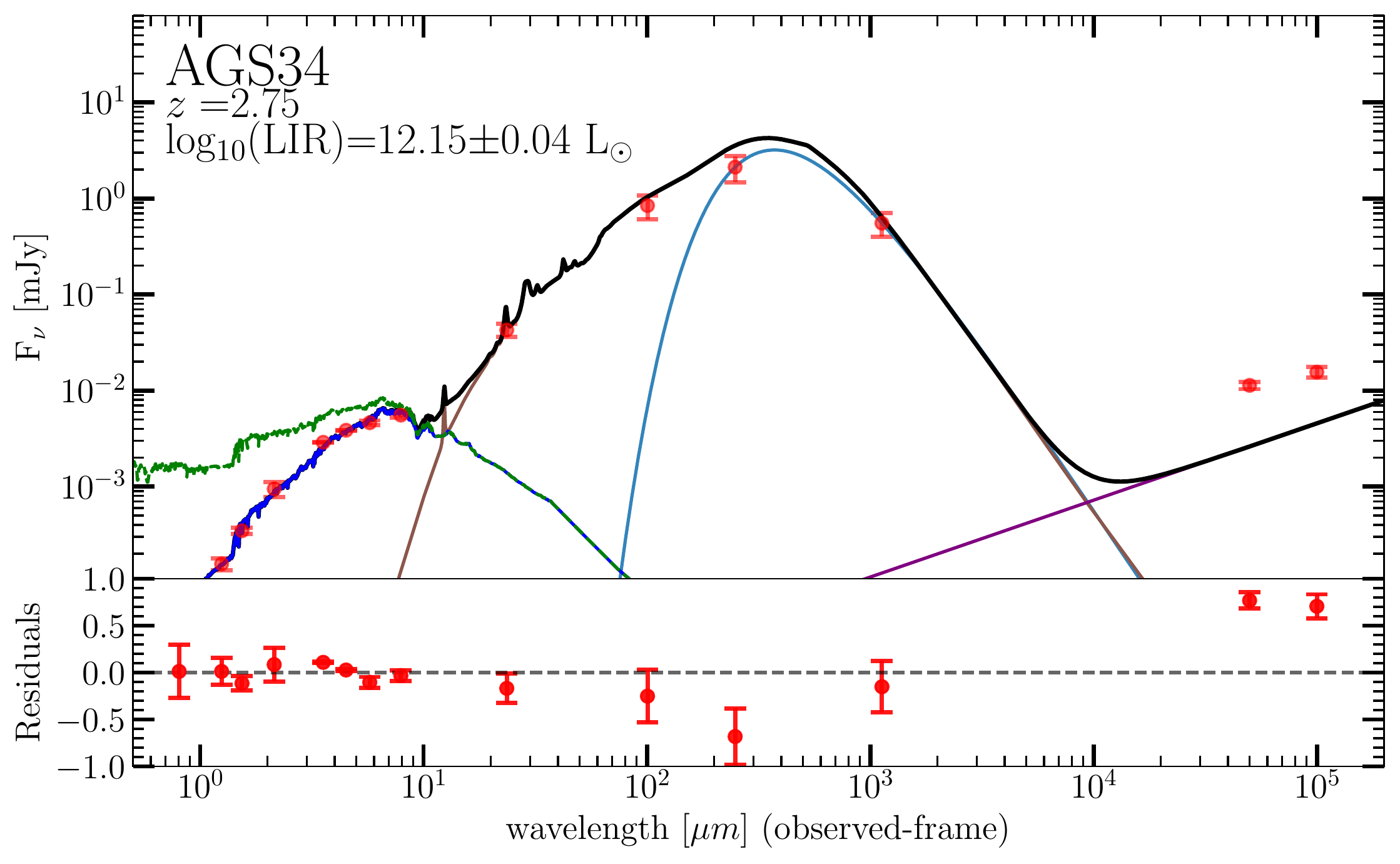} 
}
\end{minipage}
\begin{minipage}[t]{.96\textwidth}
\resizebox{\hsize}{!} { 
\includegraphics[width=3cm,clip]{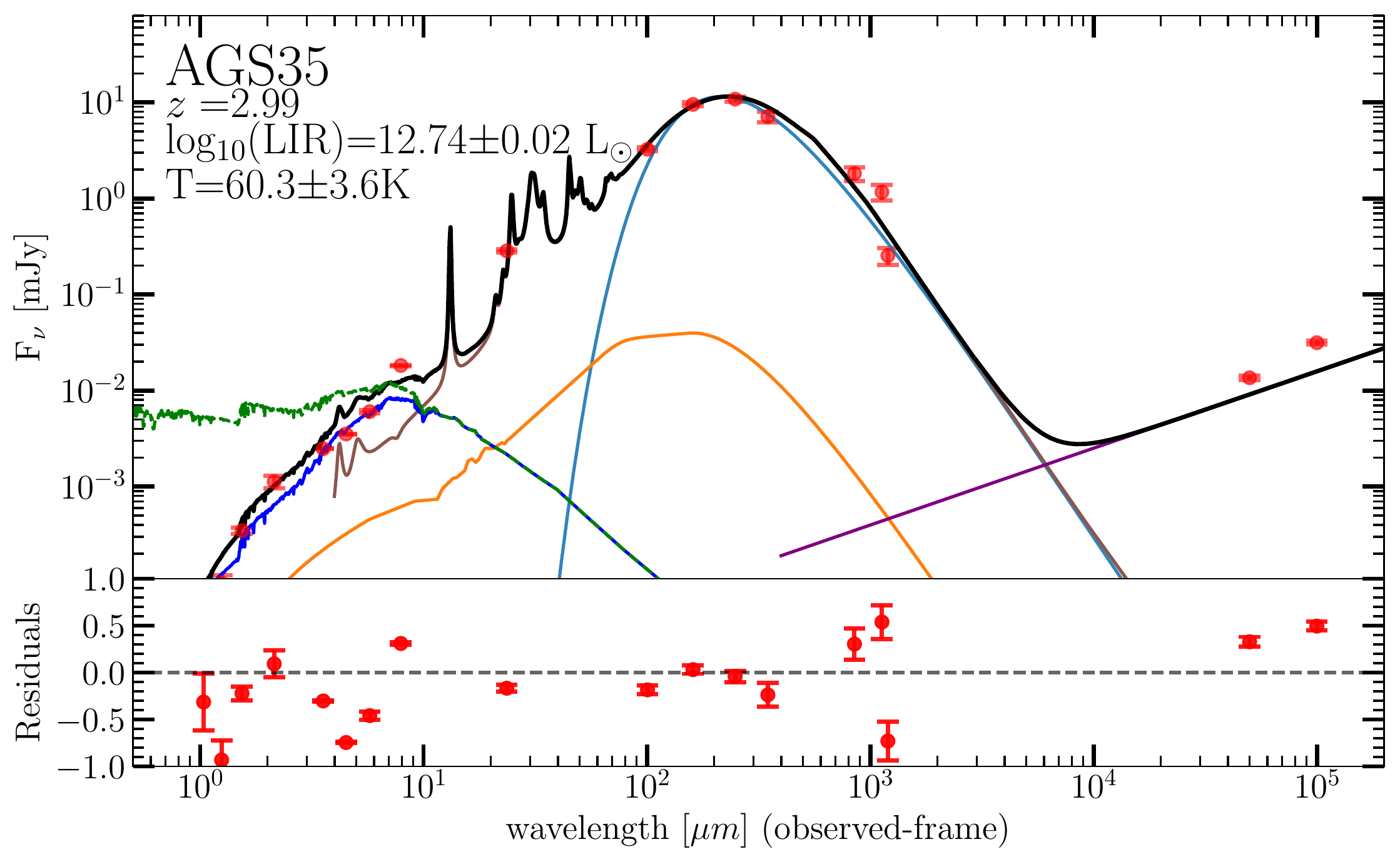} 
\includegraphics[width=3cm,clip]{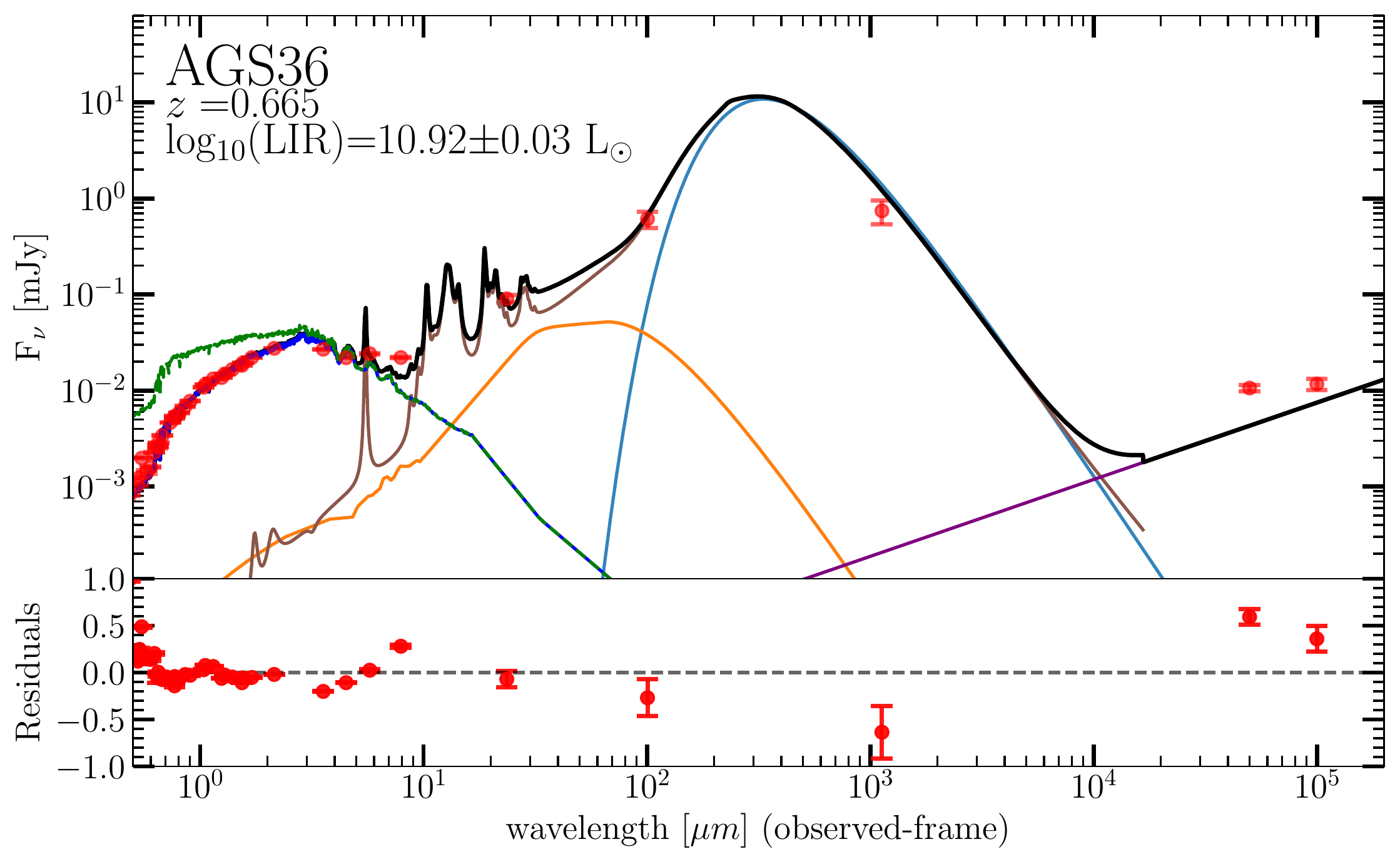} 
\includegraphics[width=3cm,clip]{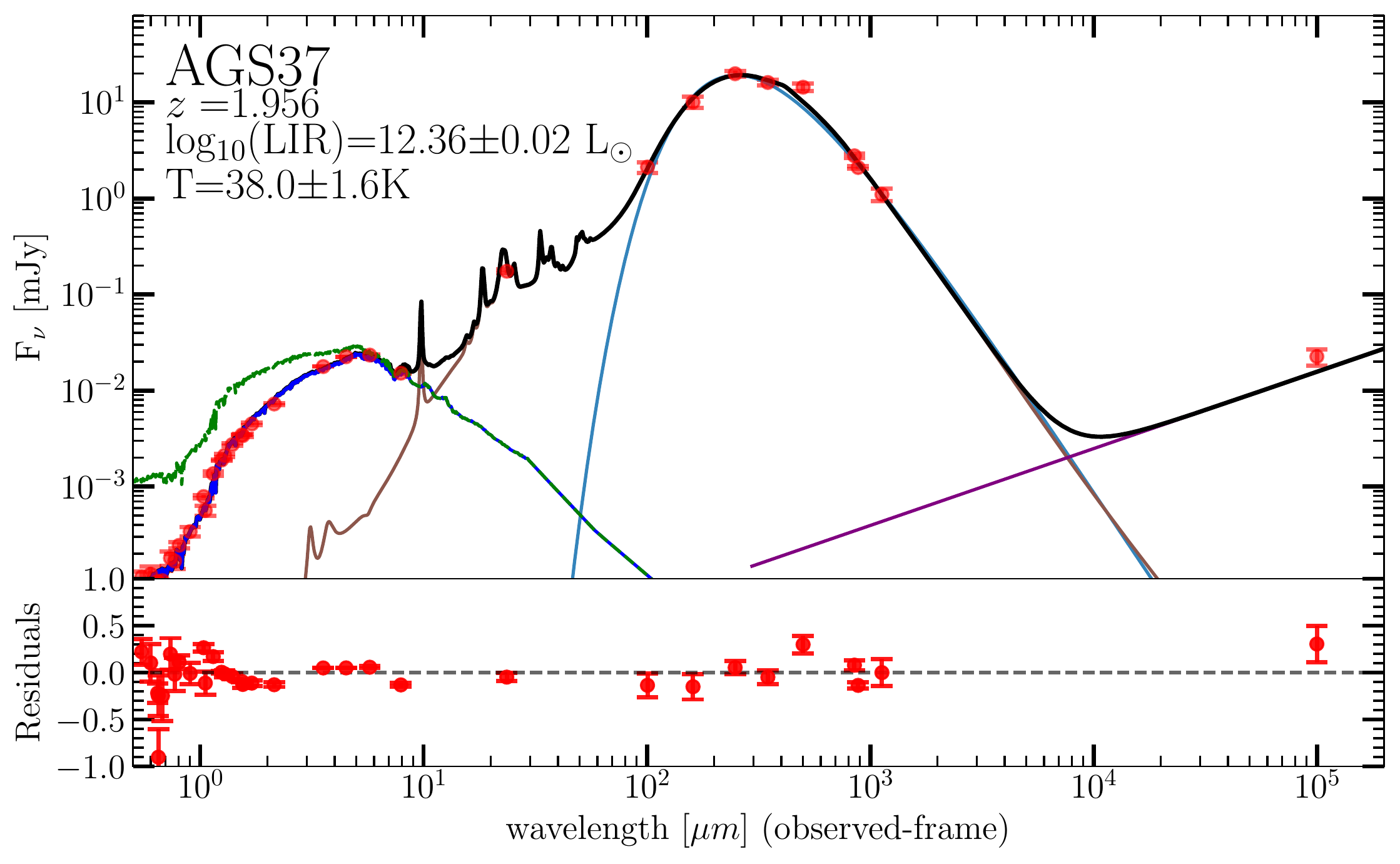} 

}
\end{minipage}
\begin{minipage}[t]{.96\textwidth}
\resizebox{\hsize}{!} { 
\includegraphics[width=3cm,clip]{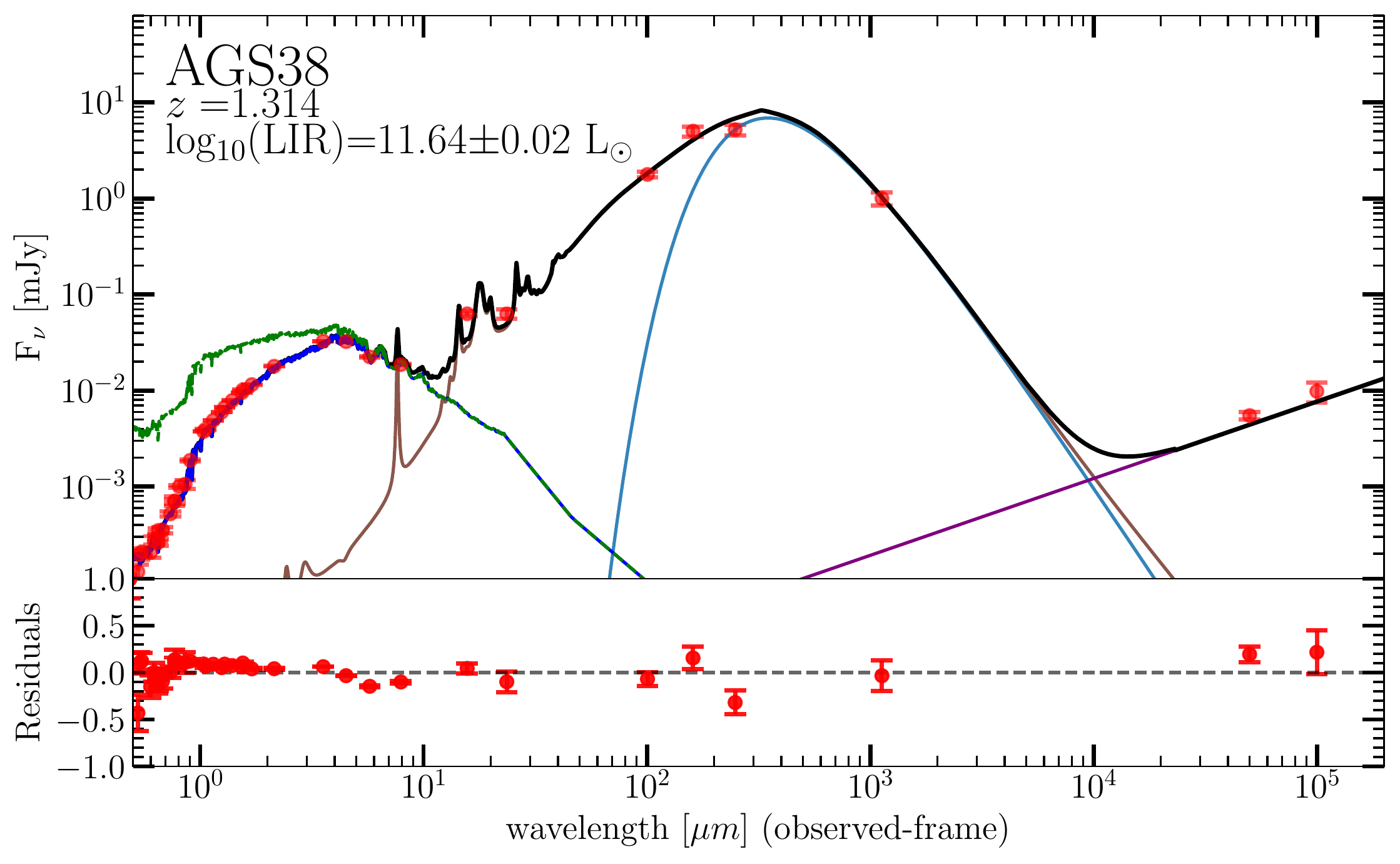} 
\includegraphics[width=3cm,clip]{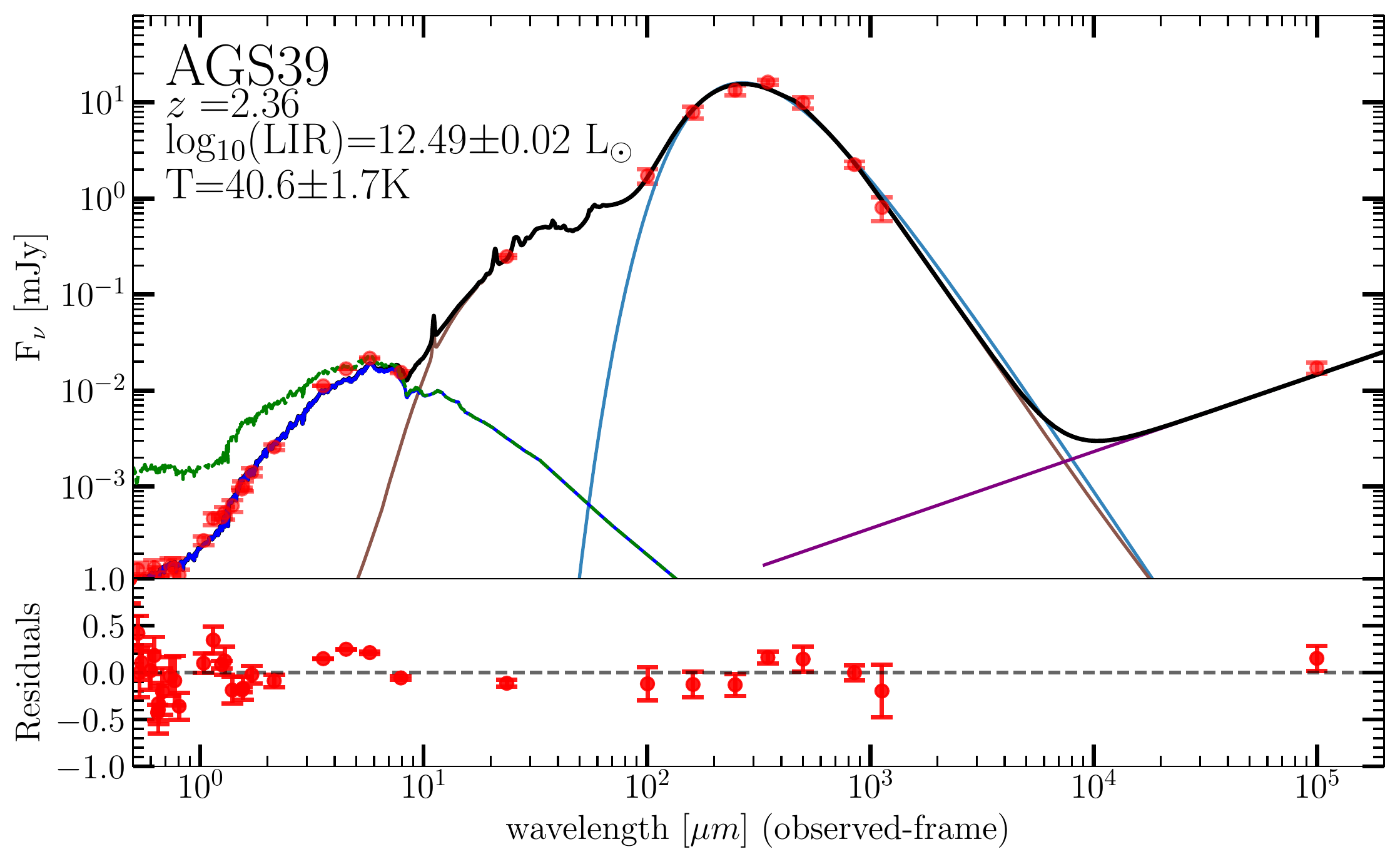}  
\includegraphics[width=3cm,clip]{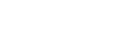}  

}
\end{minipage}
      \caption{(continued).}
\end{figure*}

\begin{table*}\footnotesize
\centering          
\begin{tabular}{l c c c}     
\hline       
 ID & S$_{peak}^{\texttt{Blobcat}}$ & S$_{integrated}^{\texttt{uvmodelfit}}$ & FHWM$^{\texttt{uvmodelfit}}$ \\
    & mJy & mJy & arcsec \\
\hline  
\hline
     AGS1 &  1.90 $\pm$  0.20* &   2.20 $\pm$  0.13     &   0.21 $\pm$  0.02 \\
     AGS2 &  1.99 $\pm$  0.22* &   2.31 $\pm$  0.12     &   0.16 $\pm$  0.01 \\
     AGS3 &  1.84 $\pm$  0.21* &   1.97 $\pm$  0.12     &   0.16 $\pm$  0.16 \\
     AGS4 &  1.72 $\pm$  0.20* &   1.68 $\pm$  0.11     &   0.18 $\pm$  0.02 \\
     AGS5 &  1.56 $\pm$  0.19* &   2.49 $\pm$  0.18     &   0.19 $\pm$  0.02 \\
     AGS6 &  1.27 $\pm$  0.18* &   1.37 $\pm$  0.13     &   0.11 $\pm$  0.03 \\
     AGS7 &  1.15 $\pm$  0.17* &   1.64 $\pm$  0.11     &   0.10 $\pm$  0.02 \\
     AGS8 &  1.43 $\pm$  0.22* &   2.23 $\pm$  0.17     &   0.23 $\pm$  0.02 \\
     AGS9 &  1.25 $\pm$  0.21* &   1.70 $\pm$  0.18     &   0.23 $\pm$  0.03 \\
    AGS10 &  0.88 $\pm$  0.15* &   1.18 $\pm$  0.13     &   0.14 $\pm$  0.03 \\
    AGS11 &  1.34 $\pm$  0.25* &   1.71 $\pm$  0.17     &   0.12 $\pm$  0.02 \\
    AGS12 &  0.93 $\pm$  0.18* &   1.17 $\pm$  0.15     &   0.22 $\pm$  0.04 \\
    AGS13 &  0.78 $\pm$  0.15* &   0.84 $\pm$  0.11     &   0.17 $\pm$  0.03 \\
    AGS15 &  0.80 $\pm$  0.16  &   1.21 $\pm$  0.11*    &   0.07 $\pm$  0.02 \\
    AGS17 &  0.93 $\pm$  0.19  &   2.30 $\pm$  0.20*    &   0.41 $\pm$  0.03 \\
    AGS18 &  0.85 $\pm$  0.18  &   1.70 $\pm$  0.28*    &   0.50 $\pm$  0.08 \\
    AGS20 &  1.11 $\pm$  0.24* &   1.90 $\pm$  0.20     &   0.19 $\pm$  0.02 \\
    AGS21 &  0.64 $\pm$  0.11* &   0.75 $\pm$  0.12     &   0.09 $\pm$  0.04 \\
    AGS23 &  0.98 $\pm$  0.21* &   0.94 $\pm$  0.16     &   0.22 $\pm$  0.04 \\
    AGS24 &  0.88 $\pm$  0.22* &   0.81 $\pm$  0.29     &   0.06 $\pm$  0.55 \\
    AGS25 &  0.81 $\pm$  0.19* &   1.06 $\pm$  0.28     &   0.12 $\pm$  0.24 \\
    AGS26 &  0.74 $\pm$  0.18  &   0.97 $\pm$  0.15*    &   0.30 $\pm$  0.09 \\
    AGS27 &  0.82 $\pm$  0.22  &   1.43 $\pm$  0.28*    &   0.54 $\pm$  0.12 \\
    AGS28 &  0.85 $\pm$  0.21  &   1.56 $\pm$  0.21*    &   0.50 $\pm$  0.07 \\
    AGS29 &  0.61 $\pm$  0.18* &   1.22 $\pm$  0.19   &   ...              \\
    AGS30 &  0.67 $\pm$  0.17* &   0.83 $\pm$  0.23     &   ...              \\
    AGS31 &  0.72 $\pm$  0.19* &   1.01 $\pm$  0.17     &   ...              \\
    AGS32 &  0.63 $\pm$  0.16  &   1.23 $\pm$  0.16*    &   0.33 $\pm$  0.10 \\
    AGS33 &  0.70 $\pm$  0.19  &   1.77 $\pm$  0.27*    &   0.51 $\pm$  0.10 \\
    AGS34 &  0.55 $\pm$  0.15* &   ...                  &    ...             \\
    AGS35 &  0.65 $\pm$  0.18  &   1.16 $\pm$  0.21*    &   0.45 $\pm$  0.12 \\
    AGS36 &  0.74 $\pm$  0.21* &   0.74 $\pm$  0.18     &   0.23 $\pm$  0.13 \\
    AGS37 &  0.63 $\pm$  0.18  &   1.10 $\pm$  0.16*    &   0.28 $\pm$  0.10 \\
    AGS38 &  0.58 $\pm$  0.16  &   1.00 $\pm$  0.16*    &   0.32 $\pm$  0.10 \\
    AGS39 &  0.80 $\pm$  0.23* &   0.98 $\pm$  0.28     &   0.25 $\pm$  0.14 \\
\hline      
\end{tabular}
\caption{Summary of fluxes (peak flux using \texttt{Blobcat} and integrated flux using \texttt{uvmodelfit}), as well as the sizes (measured using \texttt{uvmodelfit}) for the galaxies in the Main and Supplementary catalogs. For each galaxy, the flux used (as explained in \citetalias{Franco2018} and \citetalias{Franco2020}) is indicated by an asterisk. The absence of a size indicates a nonconvergence of \texttt{uvmodelfit}. For AGS15, we obtained new 2\,mm ALMA data for this galaxy, (project 2018.1.01079.S PI: M. Franco; see \citealt{Zhou2020}), which led us to revise our hypothesis that the source was compact and to favor the \texttt{uvmodelfit} flux which is in agreement with the 2mm flux and the flux at 850$\mu$m \citep{Cowie2018}.}
\label{blobcat_uvmodelfit}   
\end{table*}

\end{appendix}
\end{document}